\documentclass[fleqn,usenatbib]{mnras}

\usepackage{newtxtext}

\usepackage[T1]{fontenc}
\usepackage{ae,aecompl}
\usepackage[dvipsnames]{xcolor}

\usepackage{graphicx}	
\usepackage{amsmath}	
\usepackage{amssymb}	
\usepackage{cancel}
\usepackage{ulem}

\usepackage{cuted}

\newcommand{\FT}{\widetilde}
\newcommand{\VEC}{\mathbfit}
\newcommand{\MAT}{\mathbfit}
\newcommand{\PP}{\mathbf{\Psi}}

\newcommand{\hMpc}{\,h^{-1}\,{\rm Mpc}}

\newcommand{\hk}{\,h\,{\rm Mpc^{-1}}}

\newcommand{\Mod}[1]{\textcolor{black}{#1}}

\title[Analysis of anisotropic 3PCFs]{
Towards a self-consistent analysis of the anisotropic galaxy two- and three-point correlation functions on large scales: application to mock galaxy catalogues}

\author[N. S. Sugiyama et al.]{
Naonori S. Sugiyama$^{1}$\thanks{E-mail: nao.s.sugiyama@gmail.com},
Shun Saito$^{2,3}$, Florian Beutler$^{4}$, and Hee-Jong Seo$^{5}$
\\
$^{1}$ National Astronomical Observatory of Japan, Mitaka, Tokyo 181-8588, Japan\\
$^{2}$ Institute for Multi-messenger Astrophysics and Cosmology, Department of Physics,\\
Missouri University of Science and Technology, 1315 N. Pine St., Rolla MO 65409, USA\\
$^{3}$ Kavli Institute for the Physics and Mathematics of the Universe (WPI), \\
Todai Institutes for Advanced Study, The University of Tokyo, Chiba 277-8582, Japan\\
$^4$ Institute for Astronomy, University of Edinburgh, Royal Observatory, Blackford Hill, Edinburgh EH9 3HJ, UK\\
$^5$ Department of Physics and Astronomy, Ohio University, Clippinger Labs, Athens, OH 45701 
}

\date{}

\pubyear{}

\begin{document}
\label{firstpage}
\pagerange{\pageref{firstpage}--\pageref{lastpage}}
\maketitle

\begin{abstract}
We establish a practical method for the joint analysis of anisotropic galaxy two- and three-point correlation functions (2PCF and 3PCF) on the basis of the decomposition formalism of the 3PCF using tri-polar spherical harmonics. We perform such an analysis with MultiDark Patchy mock catalogues to demonstrate and understand the benefit of the anisotropic 3PCF. We focus on scales above $80\hMpc$, and use information from the shape and the baryon acoustic oscillation (BAO) signals of the 2PCF and 3PCF. We also apply density field reconstruction to increase the signal-noise ratio of BAO in the 2PCF measurement, but not in the 3PCF measurement. In particular, we study in detail the constraints on the angular diameter distance and the Hubble parameter. We build a model of the bispectrum or 3PCF that includes the nonlinear damping of the BAO signal in redshift space. We carefully account for various uncertainties in our analysis including theoretical models of the 3PCF, window function corrections, biases in estimated parameters from the fiducial values, the number of mock realizations to estimate the covariance matrix, and bin size. The joint analysis of the 2PCF and 3PCF monopole and quadrupole components shows a $30\%$ and $20\%$ improvement in Hubble parameter constraints before and after reconstruction of the 2PCF measurements, respectively, compared to the 2PCF analysis alone. This study clearly shows that the anisotropic 3PCF increases cosmological information from galaxy surveys and encourages further development of the modeling of the 3PCF on smaller scales than we consider.
\end{abstract}


\section{INTRODUCTION}
\label{Sec:Introduction}

Baryon Acoustic Oscillations (BAOs) are a powerful tool for measuring the expansion history of the Universe. In particular, the anisotropic signal of BAO via the Alcock-Paczy\'{n}ski (AP) effect~\citep{Alcock:1979mp} provides an extremely important means of measuring the angular diameter distance $D_{\rm A}(z)$ and the expansion rate $H(z)$ at each redshift $z$ separately. As dark energy, one of the most mysterious aspects of cosmology, mainly affects the cosmic expansion history of the universe, robust estimates of $D_{\rm A}(z)$ and $H(z)$ via the anisotropic BAO will lead to an accurate probe of dark energy.

\citet{Eisenstein:2005su} reported the first unambiguous detection of the baryon acoustic peak in the galaxy two-point correlation function (2PCF) using spectroscopic samples of luminous red galaxies from the Sloan Digital Sky Survey (SDSS)~\citep{Eisenstein:2001cq}. Since this first detection, much effort has been put into measuring the BAO signature for the samples of galaxies~\citep{Tegmark:2006az,Okumura:2007br,Percival:2009xn,Blake:2011wn,Beutler:2011hx,Blake:2011en,Seo:2012xy,Padmanabhan:2012hf,Anderson:2012sa,Xu:2012fw,Anderson:2013zyy,Tojeiro:2014eea,Ross:2014qpa,Beutler:2016ixs,Zhao:2016das}, galaxy clusters~\citep{Veropalumbo:2015dpi} and quasars~\citep{Bourboux:2017cbm,Zhu:2018edv}. More recently, the accuracy of the BAO data has been further improved by precise measurements in the Baryon Oscillation Spectroscopic Survey (BOSS;~\citealt{Eisenstein:2011sa,Bolton:2012hz,Dawson:2012va}) Data Release 12 (DR12; \citealt{Alam:2015mbd}). Currently, the precision on $D_{\rm A}(z)$ and $H(z)$ obtained from the 2PCF and its Fourier counter part, the power spectrum, is on the order of $2$-$3$ percent and $5$-$7$ percent before reconstruction, and $1$-$2$ percent and $3$-$4$ percent after reconstruction~(e.g., \citealt{Alam:2016hwk}). Future missions such as the Subaru Prime Focus Spectrograph (PFS; ~\citealt{Takada2014PASJ...66R...1T}), the Dark Energy Spectroscopic Instrument (DESI;~\citealt{Levi2013arXiv1308.0847L}) and the space-based \textit{Euclid} mission~\citep{Laureijs:2011gra} are expected to provide unprecedentedly accurate measurements of $D_{\rm A}(z)$ and $H(z)$.

Based on the great success of the analysis of the two-point statistics, there is a growing interest in using measurements of the three-point function (3PCF) or its Fourier counter part, the bispectrum for the cosmological data analysis~\citep[as recent works, see e.g.][]{Hahn:2019zob,Gualdi:2020ymf}. There have been several studies on cosmological analysis using the three-point statistics, but they dealt only with the isotropic component of the three-point statistics, i.e., the \textit{monopole} component. For example, \citet{Slepian:2016kfz} successfully detected a $4.5\sigma$-level BAO signal from BOSS DR12 data using the 3PCF, and \citet{Pearson:2017wtw} also detected a $4.1\sigma$-level BAO signal from the same data using the bispectrum. In the combined analysis of the two- and three-point statistics, \citet{Pearson:2017wtw} reported an improvement of $\sim 10\%$ for the volume-averaged angular diameter distance $D_{\rm V}(z)$; \citet{Gil-Marin:2016wya} reported the results of a joint analysis using the monopole and quadrupole power spectra and the monopole bispectrum. 

Now, it is time to analyze the anisotropic component of the 3PCF or bispectrum, i.e., the \textit{quadrupole} component. In order to do so, we have several problems to solve. The first problem is how to characterize and decompose the anisotropic components. For example, adopting the plane-parallel approximation, the redshift-space 3PCF depends on two relative coordinate vectors and one line-of-sight (LOS) unit vector. With three angular dependences, the choice of coordinate system to characterize the anisotropy of the 3PCF is arbitrary. Although decomposition methods dependent on a specific coordinate system have been proposed by~\citet{Scoccimarro:1999ed,Slepian:2017lpm}, in this paper, we will adopt a coordinate system-independent formalism using the tri-polar spherical harmonics (TripoSH) proposed by~\citet{Sugiyama:2018yzo}. The second issue is the impact of survey geometry on the 3PCF and bispectrum. Since we measure the 3PCF and bispectrum using an estimator based on the Fast Fourier Transform (FFT) proposed by~\citet{Scoccimarro:2015bla,Slepian:2016qwa,Sugiyama:2018yzo}, the measured 3PCF and bispectrum contain the effects of survey geometry. In particular, survey geometry can distort the observed density fluctuations and introduce spurious anisotropic signals. Therefore, we need to compute a theoretical model of the 3PCF or bispectrum corrected for the effects of survey geometry, following the method proposed by~\citet{Sugiyama:2018yzo}. Third, we need to construct a theoretical model to predict the galaxy 3PCF or bispectrum in redshift space. \Mod{There are several empirical models of the monopole bispectrum (3PCF) applied to the actual BOSS dataset~\citep{Slepian:2016weg,Gil-Marin:2014sta,Pearson:2017wtw}, which take into account the non-linear damping of BAO by replacing linear power spectra appearing in a tree-level based solution with non-linear power spectra. Since the non-linear effect on BAO is well understood in the power spectrum (2PCF) using cosmological perturbation theory~\citep[e.g.,][]{Crocce:2007dt,Matsubara:2007wj}, it is desirable to have a model that can explain the BAO damping based on the perturbation theory for the bispectrum (3PCF) as well.} Furthermore, it is intrinsically important to be able to quickly compute the theoretical model, as cosmological analysis using e.g. Markov-Chain Monte Carlo (MCMC) algorithm requires tens or hundreds of thousands of iterations of theoretical models with different cosmological parameters. The forth is the relationship between the number of binned measurement data, the number of realizations of the mock simulation that reproduce the observed data, and the number of fitting parameters. To estimate the covariance matrix of the observed data, brute-force methods using a huge number of mock simulations are commonly used. The number of simulations required to compute a reliable covariance matrix depends on the number of data and fitting parameters~\citep{Hartlap:2006kj,Percival:2013sga}. To perform a conservative analysis, one must either reduce the number of data and fitting parameters for a given number of simulations or increase the number of simulations for a given number of data and fitting parameters. Alternatively, one idea is to use an analytic model of the covariance matrix that corresponds to the results obtained from an infinite number of realizations, which is being vigorously studied~\citep{Sugiyama:2019ike,Philcox:2019xzt}.

The aim of this paper is to establish a self-consistent cosmological analysis method using the anisotropic component of the 3PCF by solving all the four problems, for the first time. As the first and second problems have been already addressed by~\citet{Sugiyama:2018yzo}, we focus on the third and forth ones in this paper. We propose a simple template model of the galaxy bispectrum in redshift space as an analogy of the power spectrum template model presented by~\citet{Eisenstein:2006nj}. That is, our bispectrum template model restores a tree-level solution consisting of a smooth version (without BAO) of the linear power spectrum after degrading the BAO signature. We then develop an approximation method to compute the bispectrum template model fast, taking into account the window function correction. The specific form of the bispectrum model is as follows:
\begin{eqnarray}
	&&B(\VEC{k}_1,\VEC{k}_2,\hat{n}) \nonumber 
	\\&&= 2\, Z^{[1]}(\VEC{k}_1,\hat{n})Z^{[1]}(\VEC{k}_2,\hat{n})Z^{[2]}(\VEC{k}_1,\VEC{k}_2,\hat{n})
	\nonumber \\
	&&\times
	\Big\{
	{\cal D}(\VEC{k}_1,\hat{n}){\cal D}(\VEC{k}_2,\hat{n}){\cal D}(\VEC{k}_{12},\hat{n})
	P_{\rm w}(k_1) P_{\rm w}(k_2) \nonumber \\
	&&\hspace{0.20cm}+ {\cal D}^2(\VEC{k}_1,\hat{n}) P_{\rm w}(k_1) P_{\rm nw}(k_2)
	+  {\cal D}^2(\VEC{k}_2,\hat{n}) P_{\rm nw}(k_1) P_{\rm w}(k_2) \nonumber \\
	&&\hspace{0.20cm}+ P_{\rm nw}(k_1)P_{\rm nw}(k_2)\Big\} + \mbox{2 cyc.}.
	\label{Eq:MainResultIntro}
\end{eqnarray}
where $\VEC{k}_1$, $\VEC{k}_2$ and $\VEC{k}_{12}=\VEC{k}_1+\VEC{k}_2$ denote wavevectors, and $\hat{n}$ represents the LOS unit vector. The first order kernel function $Z^{[1]} = b_1 + f (\hat{k}\cdot\hat{n})^2$ in the standard perturbation theory~\citep[for a review, see][]{Bernardeau:2001qr} is the so-called Kaiser formula of linear redshift space distortions~\citep[RSD;][]{Kaiser:1987qv,Hamilton:1997zq}, where $b_1$ and $f$ are the linear bias parameter $b_1$ and the logarithmic growth rate function $f$, respectively. The second order kernel function $Z^{[2]}$ depends on the non-linear gravity effect, non-linear redshift-space distortion effect, and the non-linear bias effect. For the BAO signal in the linear matter power spectrum $P_{\rm lin}$, the ``no-wiggle (nw)'' part $P_{\rm nw}$ is a smooth version of $P_{\rm lin}$ with the baryon oscillations removed~\citep{Eisenstein:1997ik}, and the ``wiggle (w)'' part is defined as $P_{\rm w}= P_{\rm lin}-P_{\rm nw}$. The non-linear BAO degradation is represented by the two-dimensional Gaussian damping factor derived from an differential motions of Lagrangian displacements~\citep{Eisenstein:1997ik,Crocce:2005xy,Matsubara:2007wj}: ${\cal D}(\VEC{k},\hat{n}) = \exp[-k^2 ((1-\mu^2)\Sigma_{\perp}^2 + \mu^2\Sigma_{\parallel})/4]$, where $\mu=\hat{k}\cdot\hat{n}$, and $\Sigma_{\parallel}$ and $\Sigma_{\perp}$ are the radial and transverse components of smoothing parameters.

We perform data analysis on the $2048$ MultiDark-Patchy mock catalogues~\cite[MD-Patchy mocks;][]{Klypin:2014kpa,Kitaura:2015uqa} using the 2PCF and 3PCF. The reason of this choice is because the MultiDark-Patchy mock catalogues and the cosmological constraints on 2PCF or the power spectrum are well studied in previous works~\citep{Ross:2016gvb,Satpathy:2016tct}. We limit the scale of interest to $80\hMpc$ and above. This restriction allows us to capture all the BAO signals appearing around $100\hMpc$, while still maintaining the validity of the template model of the 2PCF and 3PCF based on tree-level solutions (linear theory for the 2PCF and the second-order perturbation theory for the 3PCF) thanks to small non-linearities. Note that our analysis in principle can serve as a RSD analysis, since the theoretical model we use does not include any unphysical nuisance parameters, and we can also constrain the growth rate parameter. However, since we use only large scales, our RSD constraints are not expected to be competitive with previous studies~\citep{Satpathy:2016tct,Beutler:2016arn}, and hence are not main focus of this paper.

The methods for reducing the number of data bins are as follows. First, we use the 2PCF and 3PCF only around the BAO scale ($80\leq r \leq 150\hMpc$). For example, in the case of the power spectrum, the BAO signal appears as an oscillation function up to $k\sim0.4\hk$. Therefore, in order to analyze BAO in the power spectrum, it is common to fit the broadband shape up to small scales, using some nuisance parameters. A similar analysis can be performed in the bispectrum, but the number of data bins required increases dramatically because the bispectrum depends on two scales. Thus, the 3PCF is more useful than the bispectrum in terms of reducing the number of data bins. Second, \Mod{we adopt a wider bin size for the 3PCF than for the 2PCF in order for the compression of data. Specifically,} we use a bin width of $5\hMpc$ for the 2PCF, which is considered by e.g.~\citet{Ross:2016gvb} to be a fiducial bin size, but for the 3PCF, we adopt a wider bin width of $10\hMpc$. Finally, when analyzing the data, we find a minimal combination of data by examining which coefficients in the TripoSH decomposition of the 3PCF have the main information on the anisotropic BAO. Through these efforts, we manage to keep the $M_2$ parameter (\ref{Eq:M2}), which corrects for the effect on the variance of the estimated parameters due to the fact that the number of the MD-Patchy mocks is finite ($2048$), to $\sim 1.06$.

We apply the density field reconstruction to the 2PCF measurements. The reconstruction was originally proposed to enhance the BAO signal in 2PCF~\citep{Eisenstein:2006nk}, but it is also known to reduce the non-Gaussian terms of the covariance matrix due to its ability to partially remove nonlinear gravity~\citep{Beutler:2016arn}. Thus, it is expected to reduce the statistical error of the entire parameter of interest, in addition to the BAO signal. The reconstruction employed in this paper is the simplest one and does not remove the linear RSD effect. As a result, at the large scales above $80\hMpc$ that we focus on, the shape of the 2PCF can be evaluated by linear theory, except for the BAO smoothing parameter. In other words, we can use \citet{Eisenstein:2006nj}' template power spectrum model for theoretical predictions of the 2PCF even after reconstruction~\citep[e.g.,][]{White:2015}. Note that our analysis is the first RSD analysis of the post-reconstruction 2PCF and results in a rigorous (again, not competitive, though) constraint on $f\sigma_8$. We do not apply reconstruction to the measurement of the 3PCF because the method for analyzing 3PCFs after reconstruction has not yet been established.

This paper is organized as follows: Section~\ref{Sec:BispectrumMultipoles} briefly reviews the TripoSH decomposition formalism of the 3PCF and bispectrum; Section~\ref{Sec:Model} builds a bispectrum template model for use in data analysis; Section~\ref{Sec:ThreePointCorrelationFunctions} describes how to correct for the effect of survey geometry when calculating the 3PCF, and an approximation method to compute the 3PCF quickly; Section~\ref{Sec:2PCFanalysis} performs parameter estimation using the 2PCFs before and after reconstruction; Section~\ref{Sec:JointAnalysisWith3PCF} performs the joint analysis with the 3PCF while checking for various systematic errors; Section~\ref{Sec:Conclusions} summarizes the discussion and conclusions of this paper. Throughout this paper, we adopt a flat $\Lambda$CDM cosmology that is the same as used in the Patchy mocks: $(\Omega_{\Lambda}, \Omega_{\rm m}, \Omega_{\rm b}, \sigma_8, h) = (0.693,  0.307, 0.048,0.829,0.678)$.

\section{Decomposition formalism}
\label{Sec:BispectrumMultipoles}

We begin with a review of the decomposition formalism of the redshift-space bispectrum introduced by \citet{Sugiyama:2018yzo}. In general, a function depending on an orientational unit vector $\hat{x}$ can be expanded in the basis of spherical harmonic functions $Y_{\ell}^m(\hat{x})$. 
The power spectrum and bispectrum in redshift space are characterized by $P(\VEC{k},\hat{n})$ and $B(\VEC{k}_1,\VEC{k}_2,\hat{n})$, where $\VEC{k}$ and $\hat{n}$ are wave vectors and the unit vector in the line-of-sight (LOS) direction, respectively. 
Since $P$ and $B$ thus depend on the two and three unit vectors, we can expand them using poly-polar spherical harmonics~\citep{Varshalovich}, i.e., bipolar spherical harmonics~\citep[BipoSH; e.g.,][]{Hajian:2003qq,Shiraishi:2017wec,Sugiyama:2017ggb,Chiang:2018mau} for the power spectrum and tri-polar spherical harmonics~\citep[TripoSH; e.g.,][]{Verde:2000xj,Bertacca:2017dzm,Sugiyama:2018yzo} for the bispectrum. 
In particular, under the assumption of the statistical isotropy and parity symmetry of the universe, the $M$-modes appearing in the BipoSH and TripoSH expansions disappear. As a result, we can expand the redshift-space power spectrum using the Legendre polynomial function ${\cal L}_{\ell} = (4\pi/(2\ell+1))\sum_m Y_{\ell}^m Y_{\ell}^{m*}$ as follows~\citep[e.g.,][]{Hamilton:1997zq}:
\begin{eqnarray}
    P(\VEC{k},\hat{n}) = \sum_{\ell} P_{\ell}(k)\, {\cal L}_{\ell}(\hat{k}\cdot\hat{n}).
\end{eqnarray}
For the bispectrum, we define the base function as
\begin{eqnarray}
    {\cal S}_{\ell_1\ell_2\ell}(\hat{k}_1,\hat{k}_2,\hat{n}) 
	\hspace{-0.25cm}&=&\hspace{-0.25cm}
   \frac{4\pi}{h_{\ell_1\ell_2\ell}} 
   \sum_{m_1m_2m} 
   \left( \begin{smallmatrix} \ell_1 & \ell_2 & \ell \\ m_1 & m_2 & m \end{smallmatrix}  \right)\nonumber \\ 
    \hspace{-0.25cm}&\times&\hspace{-0.25cm}
    Y_{\ell_1}^{m_1}(\hat{k}_1) Y_{\ell_2}^{m_2}(\hat{k}_2) Y_{\ell}^m(\hat{n})
    \label{Eq:Slll}
\end{eqnarray} 
with
\begin{eqnarray}
    h_{\ell_1\ell_2\ell}= \sqrt{ \frac{(2\ell_1+1)(2\ell_2+1)(2\ell+1)}{4\pi}}
   \left( \begin{smallmatrix} \ell_1 & \ell_2 & \ell \\ 0 & 0 & 0 \end{smallmatrix}  \right),
\end{eqnarray}
where the circle bracket with $6$ multipole indices, $(\dots)$, denotes the Wigner-3j symbol, and expand the redshift-space bispectrum as~\citep{Sugiyama:2018yzo}
\begin{eqnarray}
    B(\VEC{k}_1,\VEC{k}_2,\hat{n}) &=&\hspace{-0.85cm}  \sum_{\ell_1+\ell_2+\ell={\rm even}} 
    \hspace{-0.40cm}
    B_{\ell_1\ell_2\ell}(k_1,k_2)\, {\cal S}_{\ell_1\ell_2\ell}(\hat{k}_1,\hat{k}_2,\hat{n}).
    \label{Eq:TripoSH}
\end{eqnarray}
We note here that the parity symmetry condition restricts $\ell_1+\ell_2+\ell$ to even numbers. The corresponding coefficients are given by
\begin{eqnarray}
	P_{\ell}(k)
	\hspace{-0.25cm}&=&\hspace{-0.25cm}
    (2\ell+1) \int \frac{d\hat{k}}{4\pi}\int \frac{d^2\hat{n}}{4\pi} {\cal L}_{\ell}(\hat{k}\cdot\hat{n}) P(\VEC{k},\hat{n}) \nonumber \\
	B_{\ell_1\ell_2\ell}(k_1,k_2)
	\hspace{-0.25cm}&=&\hspace{-0.25cm}
    4\pi\, h_{\ell_1\ell_2\ell}^2
	\int \frac{d^2\hat{k}_1}{4\pi}\int \frac{d^2\hat{k}_2}{4\pi}\int \frac{d^2\hat{n}}{4\pi} \nonumber \\
	\hspace{-0.25cm}&\times&\hspace{-0.25cm}
	{\cal S}^*_{\ell_1\ell_2\ell}(\hat{k}_1,\hat{k}_2,\hat{n}) B(\VEC{k}_1,\VEC{k}_2,\hat{n}).
	\label{Eq:PB_multipole}
\end{eqnarray}
In the multipole expansion method described above, the multipole index $\ell$ characterizes the anisotropy of the power and bispectra along the LOS direction induced by the RSD or AP effect. In the Newtonian limit, this anisotropy is axially symmetric around the LOS direction, and the allowed $\ell$ is restricted to even numbers. In the case of the power spectrum, the first three multipole components, i.e., the monopole ($\ell=0$), quadrupole ($\ell=2$), and hexadecapole ($\ell=4$) components, are known to contain almost all the cosmological information~\citep{Taruya:2011tz}. The relativistic effect leads to an odd $\ell$~\citep[e.g.,][]{Clarkson:2019,Maartens:2020,deWeerd:2020}, which we will ignore throughout this paper. 

In the case of the bispectrum, when $h_{\ell_1\ell_2\ell}$ is non-zero, $B_{\ell_1\ell_2\ell}$ is non-zero. For example, acceptable combinations of ($\ell_1,\ell_2,\ell$) are $(0,0,0)$, $(1,1,0)$, $(2,2,0)$, etc. for $\ell=0$, $(2,0,2)$, $(1,1,2)$, $(0,2,2)$, etc. for $\ell=2$, $(4,0,4)$, $(3,1,4)$, $(2,2,4)$, etc. for $\ell=4$, and $(6,0,6)$, $(5,1,6)$, $(4,2,6)$, etc. for $\ell=6$. Note that we define the base function ${\cal S}_{\ell_1\ell_2L}$ to be a Legendre polynomial function when $\ell=0$, i.e., ${\cal S}_{\ell_1 \ell_2 \ell=0}(\hat{k}_1,\hat{k}_2,\hat{n}) = \delta_{\ell_1\ell_2}^{(\rm K)}{\cal L}_{\ell_1}(\hat{k}_1\cdot\hat{k}_2)$ with $\delta_{\ell_1\ell_2}^{(\rm K)}$ being the Kronecker delta. Thus, we can regard the bispectrum decomposition formalism considered here as a natural extension of the Legendre expansion of the monopole bispectrum~\citep{Szapudi:2004gg,Pan:2005ym,Slepian:2015qza,Slepian:2016qwa}. Since this decomposition formalism does not depend on the choice of coordinate system, we can choose any coordinate system that is convenient for numerical calculations, such as a coordinate system with $\hat{k}_1$ or $\hat{n}$ as the $z$-axis~\citep{Scoccimarro:1999ed,Slepian:2017lpm}. In addition, this coordinate independence facilitates comparison with the observed bispectrum because it does not matter if the coordinate system used for the measurement is different from that used for the theoretical prediction. In this paper, we choose the following coordinate systems for our theoretical predictions: for the power spectrum,
\begin{eqnarray}
    \hat{k} \hspace{-0.25cm}&=&\hspace{-0.25cm} \{0,0,1\} \nonumber \\
    \hat{n} \hspace{-0.25cm}&=&\hspace{-0.25cm} \{\sqrt{1-\mu^2},0,\mu\},
\end{eqnarray}
and for the bispectrum,
\begin{eqnarray}
	\hat{k}_1 \hspace{-0.25cm}&=&\hspace{-0.25cm} \{0,0,1\} \nonumber \\
    \hat{k}_2 \hspace{-0.25cm}&=&\hspace{-0.25cm} \{\sqrt{1-\mu_k^2},0,\mu_{k}\}\nonumber \\
	\hat{n}   \hspace{-0.25cm}&=&\hspace{-0.25cm} \{\sqrt{1-\mu^2}\cos\varphi,\sqrt{1-\mu^2}\sin\varphi,\mu\},
\end{eqnarray}
where $\mu = \cos  \theta$. Then, the multiple integrals in Eq.~(\ref{Eq:PB_multipole}) are single and triple integrals for the power spectrum and bispectrum, respectively. Measuring the power spectrum and bispectrum from the observed data, we use the Cartesian coordinate and choose the north pole as our $z$-axis.

We can expand the two- and three-point correlation functions (2PCF and 3PCF) in the same way as those used for the power and bispectra. The multipole components of the 2PCF and 3PCF are related to the multipole components of the power and bispectra via the Hankel transform as follows:
\begin{eqnarray}
	\xi_{\ell}(r) = i^{\ell} \int \frac{dkk^2}{2\pi^2}j_{\ell}(rk)P_{\ell}(k)
    \label{Eq:P_to_xi}
\end{eqnarray}
and
\begin{eqnarray}
	\zeta_{\ell_1\ell_2\ell}(r_1,r_2) 
    \hspace{-0.25cm}&=&\hspace{-0.25cm}
    i^{\ell_1+\ell_2} \int \frac{dk_1k_1^2}{2\pi^2} \int \frac{dk_2k_2^2}{2\pi^2} \nonumber \\
	\hspace{-0.25cm}&\times&\hspace{-0.25cm}
    j_{\ell_1}(r_1k_1) j_{\ell_2}(r_2k_2) B_{\ell_1\ell_2\ell}(k_1,k_2),
    \label{Eq:B_to_zeta}
\end{eqnarray}
where $j_{\ell}$ is the spherical Bessel function at the $\ell$-th order. These relations mean that $\xi_{\ell}$ and $\zeta_{\ell_1\ell_2\ell}$ have the same information as $P_{\ell}$ and $B_{\ell_1\ell_2\ell}$, respectively, facilitating the comparison of the results of the configuration-space and the Fourier-space analyses.

\section{Model}
\label{Sec:Model}

In this section, we present a model of the galaxy bispectrum in redshift space that describes by construction the nonlinear damping of BAO in the framework of the Lagrangian perturbation theory. It is known that the nonlinear damping of the BAO signal in density fields such as dark matter particles, halos, and galaxies can be well explained by large-scale coherent flows of objects~\citep[e.g.,][]{Eisenstein:2006nj}. In order to mathematically describe such large-scale flows, we need to consider flows with wavelength modes much larger than the scale of interest, i.e., infrared (IR) modes. For this purpose, the Lagrangian view is useful, because the long-wavelength mode of the displacement vector, which is a variable in the Lagrangian picture, corresponds to the IR flow of direct interest~\citep[e.g.,][]{Matsubara:2007wj}. The IR mode can be manipulated to add up to the infinite order of perturbation theory. Namely, it can be treated as non-perturbative, called IR re-summation. On the other hand, in the Eulerian approach, this IR re-summation can be treated by focusing on perturbation solutions of the density field in the high-$k$ limit~\citep[e.g.,][]{Crocce:2007dt}. This is because the high-$k$ limit means that the nonlinear modes affecting the scale $k$ of interest are much larger than $k$. We build our bispectrum model based on this idea of IR re-summation~\citep[e.g.,][]{Blas:2016sfa}.

To simplify our notation, we omit the angular dependence of the LOS direction due to RSD effects in density fields, statistics, and etc., unless we specify otherwise.
We calculate all the linear power spectra used in this paper using {\sc CLASS}~\citep{Blas:2011}.

\subsection{Lagrangian perturbation theory}
\label{Sec:LPT}

In the Lagrangian picture, the displacement field $\PP(\VEC{q})$ maps the galaxies from the initial Lagrangian coordinate $\VEC{q}$ to the final Eulerian coordinate $\VEC{x}$ via the relation
\begin{eqnarray}
	\VEC{x} = \VEC{q} + \PP(\VEC{q}).
\end{eqnarray}
The observed positions of the galaxies are displaced from their real-space coordinates along the direction of the LOS due to the RSD effect. In this paper, we use the distant observer approximation for the theoretical calculation of the power and bispectra and represent the direction of the LOS as a global direction $\hat{n}$. The displacement vector including the RSD effect is then given by~\citep[e.g.,][]{Taylor:1996ne}
\begin{eqnarray}
    \PP(\VEC{q}) = \PP_{\rm real}(\VEC{q}) + \frac{\dot{\PP}_{\rm real}(\VEC{q})\cdot\hat{n}}{H}\, \hat{n},
    \label{Eq:PP}
\end{eqnarray}
where the subscript ``real'' indicates a real space quantity, $\dot{\PP}$ is the time-derivative of the real-space displacement vector, and $H$ is the Hubble parameter. Expanding the displacement field in perturbation theory, the displacement field for each perturbation order in redshift space is given by~\citep{Matsubara:2007wj}
\begin{eqnarray}
	\PP^{[n]}(\VEC{q}) = \PP^{[n]}_{\rm real}(\VEC{q}) 
	+ n\, f\, (\hat{n}\cdot\PP^{[n]}_{\rm real}(\VEC{q}))\,\hat{n},
	\label{Eq:PP2}
\end{eqnarray}
where the superscript $[n]$ indicates the quantity of the $n$-th order in perturbation theory, and the linear logarithmic growth rate function is denoted as $f=(HD)^{-1}\dot{D}$ with $D$ being the growth factor of the linear density perturbation. Note that the left-hand sides of Eqs.~(\ref{Eq:PP}) and (\ref{Eq:PP2}) omit the angle dependence of the dispacement vector, as mentioned at the beginning of this section.

The galaxy density perturbation $\delta(\VEC{x})$ is expressed in the Lagrangian picture as~\citep{Matsubara:2008wx}
\begin{eqnarray}
	\delta(\VEC{x})
	= \int d^3q \left( 1 + \delta_{\rm bias}(\VEC{q}) \right)
	\delta_{\rm D}\left( \VEC{x} - \VEC{q} - \PP(\VEC{q}) \right) - 1.
	\label{Eq:delta_psi}
\end{eqnarray}
In Fourier space, this $\delta(\VEC{x})$ is transformed into
\begin{eqnarray}
	\FT{\delta}(\VEC{k})
	= \int d^3q e^{-i\VEC{k}\cdot\VEC{q}}
	\Big[ \left( 1 + \delta_{\rm bias}(\VEC{q}) \right) e^{-i\VEC{k}\cdot\PP(\VEC{q})} - 1 \Big],
\end{eqnarray}
where the tilde mark over any quantity denotes the Fourier transform of that quantity. The density perturbation $\delta_{\rm bias}(\VEC{q})$ represents the initial distribution of biased objects, such as galaxies, that form later. 

We consider the initial density perturbation of biased objects up to second order~\citep[for a review, see][]{Desjacques:2016bnm}:
\begin{eqnarray}
    \delta_{\rm bias}(\VEC{q}) = b^{\rm L}_{1}\, \delta_{\rm lin}(\VEC{q}) + \frac{1}{2}\, b^{\rm L}_{2}\, [\delta_{\rm lin}(\VEC{q})]^2 + b_{K^2}^{\rm L}\, [K_{ij}]^2(\VEC{q})
\end{eqnarray}
with
\begin{eqnarray}
	K_{ij}(\VEC{q}) = \left(\frac{\partial_{q_i} \partial_{q_j}}{\partial_q^2} - \frac{1}{3}\delta_{ij}  \right)\delta_{\rm lin}(\VEC{q}),
\end{eqnarray}
where $i$ and $j$ run over the three spatial coordinate labels denoted as $1$, $2$, and $3$, and $\delta_{ij}$ represents the Kronecker delta. Throughout this paper, we ignore all relevant stochastic terms and assume that $\delta_{\rm bias}$ can be expanded in terms of the Gaussian linear density perturbation $\delta_{\rm lin}$. 
The Lagrangian bias parameters, $b_1^{\rm L}$, $b_2^{\rm L}$, and $b_{K^2}^{\rm L}$, are related to the Eulerian bias parameters, $b_1$, $b_2$, and $b_{K^2}$ as follows \citep{Baldauf:2012aa,Saito:2014aa}:
\begin{eqnarray}
	b_1 \hspace{-0.25cm}    &=& \hspace{-0.25cm} b_1^{\rm L} + 1, \notag\\
	b_2  \hspace{-0.25cm}   &=& \hspace{-0.25cm} b_2^{\rm L} + \frac{8}{21}\, b_1^{\rm L}\notag\\
	b_{K^2} \hspace{-0.25cm} &=& \hspace{-0.25cm} b_{K^2}^{\rm L} - \frac{2}{7}\, b_{1}^{\rm L}.
    \label{Eq:NLbias}
\end{eqnarray}
We vary all $b_1$, $b_2$ and $b_{\rm K2}$ in our data analysis in Section~\ref{Sec:JointAnalysisWith3PCF}. Note that, for a simple presentation purpose, we set $b_1=2.0$, $b_2=0.0$ and $b_{\rm K^2}=0.0$ in Figures~\ref{fig:pk_IR} and~\ref{fig:bk_IR}.

\subsection{Infra-red flows}
\label{Sec:InfraRedFlow}

Consider decomposing the displacement vector into components at the origin and other components:
\begin{eqnarray}
    \PP(\VEC{q}) = \PP(\VEC{q}\neq\VEC{0}) + \overline{\PP}.
\end{eqnarray}
We refer to this constant vector $\overline{\PP}$ as the infra-red (IR) flow of galaxies, which is defined as
\footnote{There would be nothing wrong with focusing on a certain point other than the origin. However, since we finally compute the quantity taken as an ensemble average, we can consider the origin without loss of generality thanks to statistical translation symmetry.
}
\begin{eqnarray}
	\overline{\PP} = \PP(\VEC{q}=0) = \int \frac{d^3p}{(2\pi)^3}\, \FT{\PP}(\VEC{p}).
\end{eqnarray}
The density perturbation (\ref{Eq:delta_psi}) is then described by
\begin{eqnarray}
	\delta(\VEC{x}) \to \delta(\VEC{x} - \overline{\PP}),
\end{eqnarray}
where we assume that the origin $\VEC{q}=\VEC{0}$ does not contribute to the volume integral in Eq.~(\ref{Eq:delta_psi}). In Fourier space it becomes
\begin{eqnarray}
	\FT{\delta}(\VEC{k}) \to e^{-i\VEC{k}\cdot\overline{\PP}}\, \FT{\delta}(\VEC{k}).
	\label{Eq:delta_high_k}
\end{eqnarray}
Note that assuming the linear IR flow $\overline{\PP}_{\rm lin}$, we can derive the same expression as Eq.~(\ref{Eq:delta_high_k}) in the Eulerian picture by taking the high-$k$ limit~\citep{Sugiyama:2013pwa,Sugiyama:2013gza}. Thus, focusing on the IR flow is equivalent to considering the high-$k$ limit solution in the Eulerian perturbation theory (PT).

If the IR flow is uncorrelated with the density perturbation, it does not affect the $n$-point statistics because of the statistical homogeneity of the universe. For example, the two- and three-point functions, $\xi$ and $\zeta$, are 
\begin{eqnarray}
	\langle \delta(\VEC{x}-\overline{\PP}) \delta(\VEC{y}-\overline{\PP}) \rangle
\hspace{-0.25cm}&=& \hspace{-0.25cm}
	\langle \delta(\VEC{x}) \delta(\VEC{y}) \rangle \nonumber \\
\hspace{-0.25cm}&=& \hspace{-0.25cm}
	\xi(\VEC{x} - \VEC{y}) \nonumber \\
\hspace{-0.25cm}\langle \delta(\VEC{x}-\overline{\PP}) \delta(\VEC{y}-\overline{\PP}) \delta(\VEC{z}-\overline{\PP})\rangle
\hspace{-0.25cm}&=& \hspace{-0.25cm}
\langle \delta(\VEC{x}) \delta(\VEC{y}) \delta(\VEC{z})\rangle \nonumber \\
\hspace{-0.25cm}&=&\hspace{-0.25cm}
\zeta(\VEC{x} - \VEC{z}, \VEC{y}-\VEC{z}),
\end{eqnarray}
where $\langle \cdots \rangle$ means the ensemble average. In Fourier space, the power spectrum and the bispectrum are
\begin{eqnarray}
    && \hspace{-0.25cm} \langle e^{-i\VEC{k}_1\cdot\overline{\PP}} \FT{\delta}(\VEC{k}_1) e^{-i\VEC{k}_2\cdot\overline{\PP}} \FT{\delta}(\VEC{k}_2) \rangle \nonumber \\
	&=& \hspace{-0.25cm} \langle e^{-i\VEC{k}_1\cdot\overline{\PP}} e^{-i\VEC{k}_2\cdot\overline{\PP}} \rangle
	  \langle \FT{\delta}(\VEC{k}_1) \FT{\delta}(\VEC{k}_2) \rangle \nonumber \\
	  &=& \hspace{-0.25cm} (2\pi)^3\delta_{\rm D}\left( \VEC{k}_1+\VEC{k}_2 \right) P(\VEC{k}_1) \nonumber \\\nonumber \\
	&& \hspace{-0.25cm}\langle e^{-i\VEC{k}_1\cdot\overline{\PP}} \FT{\delta}(\VEC{k}_1)
	e^{-i\VEC{k}_2\cdot\overline{\PP}} \FT{\delta}(\VEC{k}_2)
	e^{-i\VEC{k}_3\cdot\overline{\PP}} \FT{\delta}(\VEC{k}_3) \rangle \nonumber \\
	&=& \hspace{-0.25cm} \langle e^{-i\VEC{k}_1\cdot\overline{\PP}} e^{-i\VEC{k}_2\cdot\overline{\PP}} e^{-i\VEC{k}_3\cdot\overline{\PP}}\rangle
	  \langle \FT{\delta}(\VEC{k}_1) \widetilde{\delta}(\VEC{k}_2) \widetilde{\delta}(\VEC{k}_3)\rangle \nonumber \\
	  &=& \hspace{-0.25cm} (2\pi)^3\delta_{\rm D}\left( \VEC{k}_1+\VEC{k}_2 +\VEC{k}_3\right) B(\VEC{k}_1,\VEC{k}_2).
	  \label{Eq:PB_GI}
\end{eqnarray}
This cancellation of the IR flow is known as the high-$k$ limit cancellation in the Eulerian PT and has been shown for the power spectrum at the $1$-loop order~\citep{Vishniac1983,Suto:1990wf,Makino:1991rp}, the $2$-loop order~\citep{Sugiyama:2013gza,Sugiyama:2013mpa,Blas:2013bpa}, and any order~\citep{Jain:1995kx}. This IR flow cancellation is also closely related to the Galilean invariance of the large-scale structure~\citep{Scoccimarro:1995if,Peloso:2013zw,Kehagias:2013yd,Peloso:2016qdr}.

In general, the IR flow is correlated with the density perturbation, so the cancellation of the IR flow in the $n$-point statistics does not occur completely. Therefore, the nonlinear effects arising from $\overline{\PP}$ tend to cancel each other out, but their residual terms are likely to have a physical impact on the $n$-point statistics. In Section~\ref{Sec:TemplateModel}, we show how the breaking of the cancellation of the IR effect gives nonlinear corrections to the BAO signal on the redshift-space bispectrum.

\subsection{$\Gamma$-expansion}
\label{Sec:GammaExpansion}

\begin{figure*}
	\includegraphics[width=\textwidth]{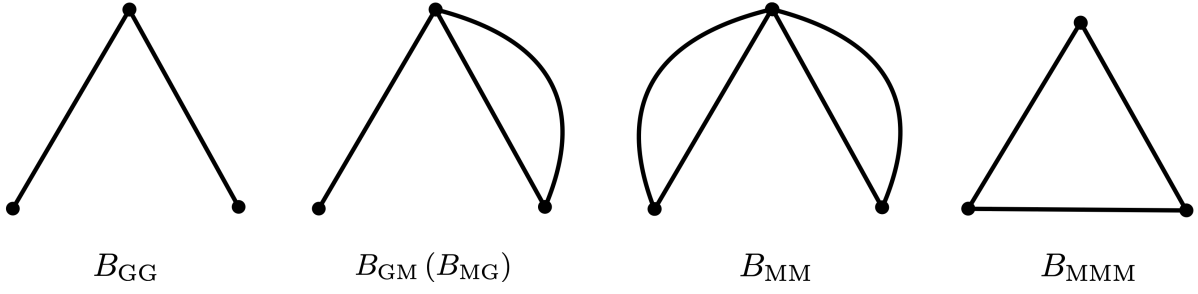}
	\caption{
Schematic diagrams of a nonlinear bispectrum classified by mode-coupling integrals. The straight lines represent the term proportional to the linear power spectrum, and the closed curves are called ``loops'', in analogy to quantum field theory, meaning they are mode-coupling integrals with an infinite number of integration dimensions. The first term from the left, $B_{\rm GG}$, has no modal coupling; the second term, $B_{\rm GM}$ ($B_{\rm MG}$), partially contain the mode-coupling integral. The third and fourth terms consist only of the mode-coupling integrals; therefore, the BAO signal in these two terms should be suppressed.
	}
	\label{fig:diagram}
\end{figure*}

A useful method for extracting most of the information about BAO from the $n$-point statistics is to apply the $\Gamma$-expansion method~\citep{Bernardeau:2008fa,Bernardeau:2011dp}, also known as the multi-point propagator or the Wiener-Hermite expansion~\citep{Matsubara:1995wd,Sugiyama:2012pc}, to density fluctuations. The $\Gamma$-expansion is an expansion method based on the statistical properties of the observed fluctuations such as density perturbations and velocity fields. The first-order term of the $\Gamma$-expansion is defined as a Gaussian distribution, and its higher-order terms describe the departure from the Gaussian distribution. Using the $\Gamma$-expansion, we can decompose the $n$-point statistics into parts with and without the mode-coupling integral. The part without the mode-coupling integral has more information about BAO because the linear power spectrum appears directly, while the mode-coupling integral erases the information about BAO. For example, in the case of the power spectrum, the BAO information on the mode-coupling term is only a few per cent or less~\citep[e.g.,][]{Seo:2008yx}, and such a small effect is negligible in current galaxy surveys such as BOSS. This fact is likely to be true in the case of the bispectrum as well. Therefore, we do not take into account the BAO information in the mode-coupling integral in this paper when we construct the template model for describing the redshift-space bispectrum in Section~\ref{Sec:TemplateModel}.

In the standard PT, the density perturbation is expanded as
\begin{eqnarray}
	\FT{\delta}(\VEC{k}) \hspace{-0.25cm}&=&\hspace{-0.25cm}  \sum_{n=1}^{\infty}\prod_{i=1}^{n}
    \int \frac{d^3p_i}{(2\pi)^3}\,\FT{\delta}_{\rm lin}(\VEC{p}_i) \notag\\
    \hspace{-0.25cm}&\times&\hspace{-0.25cm}
    (2\pi)^3\delta_{\rm D}\big( \VEC{k}-\VEC{p}_{1n} \big) Z^{[n]}(\VEC{p}_1,\dots,\VEC{p}_n),
	\label{Eq:SPT}
\end{eqnarray}
where $\VEC{p}_{1n} = \VEC{p}_1+\cdots + \VEC{p}_n$, $Z^{[1]}(\VEC{k})=(b_1+f(\hat{k}\cdot\hat{n})^2)$ is the so-called Kaiser factor~\citep{Kaiser:1987qv}, and $Z^{[n\geq2]}$ are the non-linear kernel functions including the non-linear gravity effect, the non-linear RSD effect, and the non-linear bias effect. 

Assuming that the linear density perturbation has a Gaussian distribution, i.e., no primordial non-Gaussianity, we define the first-order of the basis function of the $\Gamma$-expansion as $H^{(1)}(\VEC{k})=\FT{\delta}_{\rm lin}(\VEC{k})/\sqrt{P_{\rm lin}(k)}$. Then, the higher-order basis functions are given by
\begin{eqnarray}
	H^{(2)}(\VEC{k}_1,\VEC{k}_2)
	\hspace{-0.25cm}&=&\hspace{-0.25cm}
	H^{(1)}(\VEC{k}_1)H^{(1)}(\VEC{k}_2)
	- (2\pi)^3\delta_{\rm D}\left( \VEC{k}_1+\VEC{k}_2 \right) \nonumber \\
	H^{(3)}(\VEC{k}_1,\VEC{k}_2,\VEC{k}_3) 
	\hspace{-0.25cm}&=&\hspace{-0.25cm}
	H^{(1)}(\VEC{k}_1)H^{(1)}(\VEC{k}_2)H^{(1)}(\VEC{k}_3)  \nonumber \\
	\hspace{-0.25cm}&-&\hspace{-0.25cm} [(2\pi)^3\delta_{\rm D}\left( \VEC{k}_1+\VEC{k}_2 \right) H^{(1)}(\VEC{k}_3) + \mbox{cyc.}]  \nonumber \\
	H^{(4)} \hspace{-0.25cm}&=&\hspace{-0.25cm}  \dots,
\end{eqnarray}
where the superscript $(r)$ means the $r$-th order of the $\Gamma$-expansion. 
Using these basis functions, the density perturbation is expanded as
\begin{eqnarray}
	\hspace{-1.5cm}\widetilde{\delta}(\VEC{k})
	\hspace{-0.25cm}&=&\hspace{-0.25cm}
	\sum_{r=1}^{\infty} \prod_{i=1}^{r} \int \frac{d^3p_i}{(2\pi)^3}\sqrt{P_{\rm lin}(p_i)}\,
    (2\pi)^3\delta_{\rm D}\big( \VEC{k} - \VEC{p}_{1r} \big)\nonumber \\
	\hspace{-0.25cm}&\times&\hspace{-0.25cm} 
	\Gamma^{(r)}(\VEC{p}_1, \dots, \VEC{p}_r)\, H^{(r)}(\VEC{p}_1,\dots,\VEC{p}_r).
\end{eqnarray}
The corresponding coefficients are expressed using the standard PT kernel functions as~\citep{Sugiyama:2012pc,Sugiyama:2013pwa}
\begin{eqnarray}
	\Gamma^{(r)} 
	\hspace{-0.25cm}&=& \hspace{-0.25cm}
	\frac{1}{r!} \sum_{s=0}^{\infty}\frac{(r+2s)!}{2^s s!}
	\prod_{i=1}^s \int \frac{d^3 q_i}{(2\pi)^3}\, P_{\rm lin}(q_i) \nonumber \\
	\hspace{-0.25cm}&\times& \hspace{-0.25cm}
    Z^{[r+2s]}(\VEC{p}_1,\dots,\VEC{p}_r,
	\VEC{q}_1,-\VEC{q}_1,\dots,\VEC{q}_s,-\VEC{q}_s).
	\label{Eq:Gamma_Zn}
\end{eqnarray}
The power spectrum is decomposed into two parts~\citep{Crocce:2005xy,Crocce:2007dt}: 
\begin{eqnarray}
	P(\VEC{k}) = G^2(\VEC{k})\, P_{\rm lin}(k) + P_{\rm MC}(\VEC{k}),
\end{eqnarray}
where $G(\VEC{k})$, known as the ``two-point propagator'', represents how the information in the initial linear power spectrum propagates to the final non-linear power spectrum, and $P_{\rm MC}$, called the ``mode-coupling term'', represents the coupling between different modes. These two terms are expressed using $\Gamma^{(r)}$ as~\citep{Bernardeau:2008fa}
\begin{eqnarray}
	G(\VEC{k}) 
	\hspace{-0.25cm}&=&\hspace{-0.25cm}
	 \Gamma^{(1)}(\VEC{k}) \nonumber \\
	P_{\rm MC}(\VEC{k}) 
	\hspace{-0.25cm}&=&\hspace{-0.25cm}
	\sum_{r=2}^{\infty}\, r!\, \prod_{i=1}^{r}\int \frac{d^3p_i}{(2\pi)^3}\,P_{\rm lin}(p_i) \nonumber \\
	\hspace{-0.25cm}&\times&\hspace{-0.25cm}
	(2\pi)^3\delta_{\rm D}\big( \VEC{k} - \VEC{p}_{1r} \big) 
    \big[ \Gamma^{(r)}(\VEC{p}_1,\dots,\VEC{p}_r) \big]^2.
	\label{Eq:PGamma}
\end{eqnarray}
The expression of the bispectrum using the $\Gamma$-expansion is given by \citet{Bernardeau:2008fa}; we classify that in terms of mode-coupling integrals into the following five parts:
\begin{eqnarray}
    B(\VEC{k}_1,\VEC{k}_2,\VEC{k}_3) 
	\hspace{-0.25cm} &=& \hspace{-0.25cm}
	\left(  B_{\rm GG}(\VEC{k}_1,\VEC{k}_2)\, P_{\rm lin}(k_1)\, P_{\rm lin}(k_2) 
	+  \mbox{2 cyc.} \right) \notag\\
	\hspace{-0.25cm} &+& \hspace{-0.25cm}
	\left(  B_{\rm GM}(\VEC{k}_1,\VEC{k}_2)\, P_{\rm lin}(k_1) +  \mbox{2 cyc.} \right) \notag\\
	\hspace{-0.25cm} &+& \hspace{-0.25cm}
	\left(  B_{\rm MG}(\VEC{k}_1,\VEC{k}_2)\, P_{\rm lin}(k_2) +  \mbox{2 cyc.} \right) \notag\\
	\hspace{-0.25cm} &+& \hspace{-0.25cm}
	\left(  B_{\rm MM}(\VEC{k}_1,\VEC{k}_2) +  \mbox{2 cyc.} \right) \notag\\
	\hspace{-0.25cm} &+& \hspace{-0.25cm}
	\left(  B_{\rm MMM}(\VEC{k}_1,\VEC{k}_2) +  \mbox{2 cyc.} \right),
	\label{Eq:BispectrumGammaExp}
\end{eqnarray}
where
\begin{eqnarray}
	B_{\rm GG}(\VEC{k}_1,\VEC{k}_2) \hspace{-0.25cm}&=& \hspace{-0.25cm}2\, \Gamma^{(2)}\left( \VEC{k}_1, \VEC{k}_2 \right)
	\Gamma^{(1)}\left( \VEC{k}_1\right) \Gamma^{(1)}\left( \VEC{k}_2 \right),  
\end{eqnarray}
\begin{eqnarray}
	B_{\rm GM}(\VEC{k}_1,\VEC{k}_2) 
	\hspace{-0.25cm}&=& \hspace{-0.25cm}
	\Gamma^{(1)}(\VEC{k}_1)\sum_{r=2}^{\infty}(r+1)! 
	\prod_{i=1}^{r}\int \frac{d^3p_i}{(2\pi)^3}P_{\rm lin}(p_i) \nonumber \\
	\hspace{-0.25cm}&\times&\hspace{-0.25cm}
	(2\pi)^3\delta_{\rm D}\big( \VEC{k}_2 - \VEC{p}_{1r}  \big) 
	\Gamma^{(r)}(\VEC{p}_1,\dots,\VEC{p}_r) \nonumber \\
	\hspace{-0.25cm}&\times&\hspace{-0.25cm}
	\Gamma^{(r+1)}(-\VEC{p}_1,\dots,-\VEC{p}_r,-\VEC{k}_1), 
\end{eqnarray}
\begin{eqnarray}
	B_{\rm MG}(\VEC{k}_1,\VEC{k}_2) 
	\hspace{-0.25cm}&=& \hspace{-0.25cm}
	\Gamma^{(1)}(\VEC{k}_2)\sum_{r=2}^{\infty}(r+1)! 
	\prod_{i=1}^{r}\int \frac{d^3p_i}{(2\pi)^3}P_{\rm lin}(p_i) \nonumber \\
	\hspace{-0.25cm}&\times&\hspace{-0.25cm}
	(2\pi)^3\delta_{\rm D}\big( \VEC{k}_1 - \VEC{p}_{1r}  \big) 
	\Gamma^{(r)}(\VEC{p}_1,\dots,\VEC{p}_r) \nonumber \\
	\hspace{-0.25cm}&\times&\hspace{-0.25cm}
	\Gamma^{(r+1)}(-\VEC{p}_1,\dots,-\VEC{p}_r,-\VEC{k}_2), 
\end{eqnarray}
\begin{eqnarray}
	B_{\rm MM}(\VEC{k}_1,\VEC{k}_2) 
	\hspace{-0.25cm}&=& \hspace{-0.25cm}
	\sum_{r=2}^{\infty}\sum_{s=2}^{\infty}(r+s)!  \nonumber \\
	\hspace{-0.25cm}&\times& \hspace{-0.25cm}
	\prod_{i=1}^{r}\int \frac{d^3p_i}{(2\pi)^3}P_{\rm lin}(p_i) 
	\prod_{j=1}^{s}\int \frac{d^3p'_j}{(2\pi)^3}P_{\rm lin}(p'_j) \nonumber \\
	\hspace{-0.25cm}&\times&\hspace{-0.25cm}
	(2\pi)^3\delta_{\rm D}\big( \VEC{k}_1 - \VEC{p}_{1r}  \big)  
	(2\pi)^3\delta_{\rm D}\big( \VEC{k}_2 - \VEC{p}'_{1s} \big) \nonumber \\
	\hspace{-0.25cm}&\times&\hspace{-0.25cm}
	\Gamma^{(r)}(\VEC{p}_1,\dots,\VEC{p}_r)\Gamma^{(s)}(\VEC{p}'_1,\dots,\VEC{p}'_s) \nonumber \\
	\hspace{-0.25cm}&\times&\hspace{-0.25cm}
	\Gamma^{(r+s)}(-\VEC{p}_1,\dots,-\VEC{p}_r,-\VEC{p}'_1,\dots,-\VEC{p}'_s),
\end{eqnarray}
and
\begin{eqnarray}
	B_{\rm MMM}(\VEC{k}_1,\VEC{k}_2) 
	\hspace{-0.25cm}&=& \hspace{-0.25cm}
	\sum_{r=1}^{\infty}\sum_{s=1}^{\infty}\sum_{t=1}^{\infty}
	\frac{(r+t)!(r+s)!(s+t)!}{r!s!t!} \nonumber \\
	\hspace{-0.25cm}&\times& \hspace{-0.25cm}
	\prod_{i=1}^{r}\int \frac{d^3p_i}{(2\pi)^3}
	\prod_{j=1}^{s}\int \frac{d^3p'_j}{(2\pi)^3}
	\prod_{k=1}^{t}\int \frac{d^3p''_k}{(2\pi)^3} \nonumber \\
	\hspace{-0.25cm}&\times&\hspace{-0.25cm}
	(2\pi)^3\delta_{\rm D}\big( \VEC{k}_1 - \VEC{p}_{1r} - \VEC{p}''_{1t} \big)  \nonumber \\
	\hspace{-0.25cm}&\times&\hspace{-0.25cm}
	(2\pi)^3\delta_{\rm D}\big( \VEC{k}_2 - \VEC{p}'_{1s} + \VEC{p}''_{1t} \big) \nonumber \\
	\hspace{-0.25cm}&\times&\hspace{-0.25cm}
	\Gamma^{(r+t)}(\VEC{p}_1,\dots,\VEC{p}_r,\VEC{p}''_1,\dots,\VEC{p}''_t) \nonumber \\
	\hspace{-0.25cm}&\times&\hspace{-0.25cm}
	\Gamma^{(r+s)}(-\VEC{p}_1,\dots,-\VEC{p}_r,-\VEC{p}'_1,\dots,-\VEC{p}'_s) \nonumber \\
	\hspace{-0.25cm}&\times&\hspace{-0.25cm}
	\Gamma^{(s+t)}(\VEC{p}'_1,\dots,\VEC{p}'_s, - \VEC{p}''_1,\dots,-\VEC{p}''_t) \nonumber \\
	\hspace{-0.25cm}&\times&\hspace{-0.25cm}
	\prod_{i=1}^r P_{\rm lin}(p_i) \prod_{j=1}^{s} P_{\rm lin}(p'_j) \prod_{k=1}^{t} P_{\rm lin}(p''_k).
\end{eqnarray}
For intuitive understanding, each of the five terms in Eq.~(\ref{Eq:BispectrumGammaExp}) is schematically illustrated in Figure~\ref{fig:diagram}. The straight line in the figure represents the term proportional to one linear power spectrum, and the closed line represents the mode-coupling effect. The $B_{\rm GG}$ term is proportional to the product of two linear power spectra, the $B_{\rm GM}$ and $B_{\rm MG}$ terms are proportional to the linear power spectrum, and the $B_{\rm MM}$ and $B_{\rm MMM}$ terms are composed only of the mode-coupling effect.

\subsection{IR cancellation}
\label{Sec:Cancellation}

\begin{figure*}
	\scalebox{1.0}{\includegraphics[width=\textwidth]{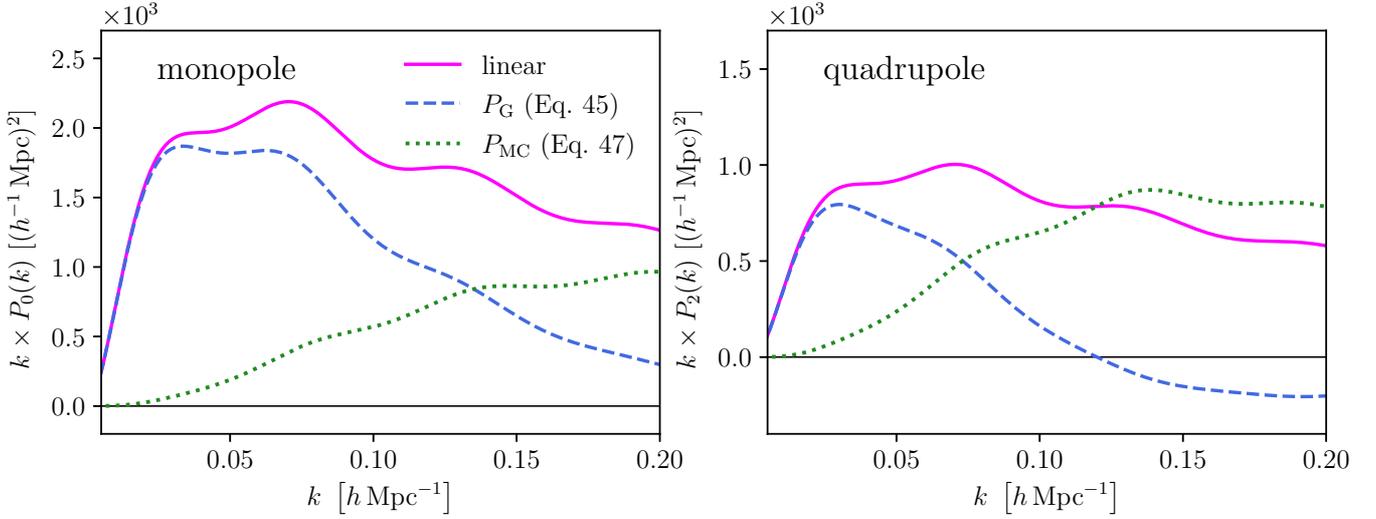}}
	\caption{
Monopole and quadrupole components of the power spectrum, each decomposed into two parts: the propagator term, $P_{\rm G}=G^2 P_{\rm lin}$ (shown as blue lines), and the mode-coupling term, $P_{\rm MC}$ (green). Since these two parts are calculated in the high-$k$ limit, the sum of the two terms, $P = P_{\rm G}+P_{\rm MC}$, is the linear power spectrum (magenta).
}
	\label{fig:pk_IR}
\end{figure*}

\begin{figure*}
	\scalebox{1.0}{\includegraphics[width=\textwidth]{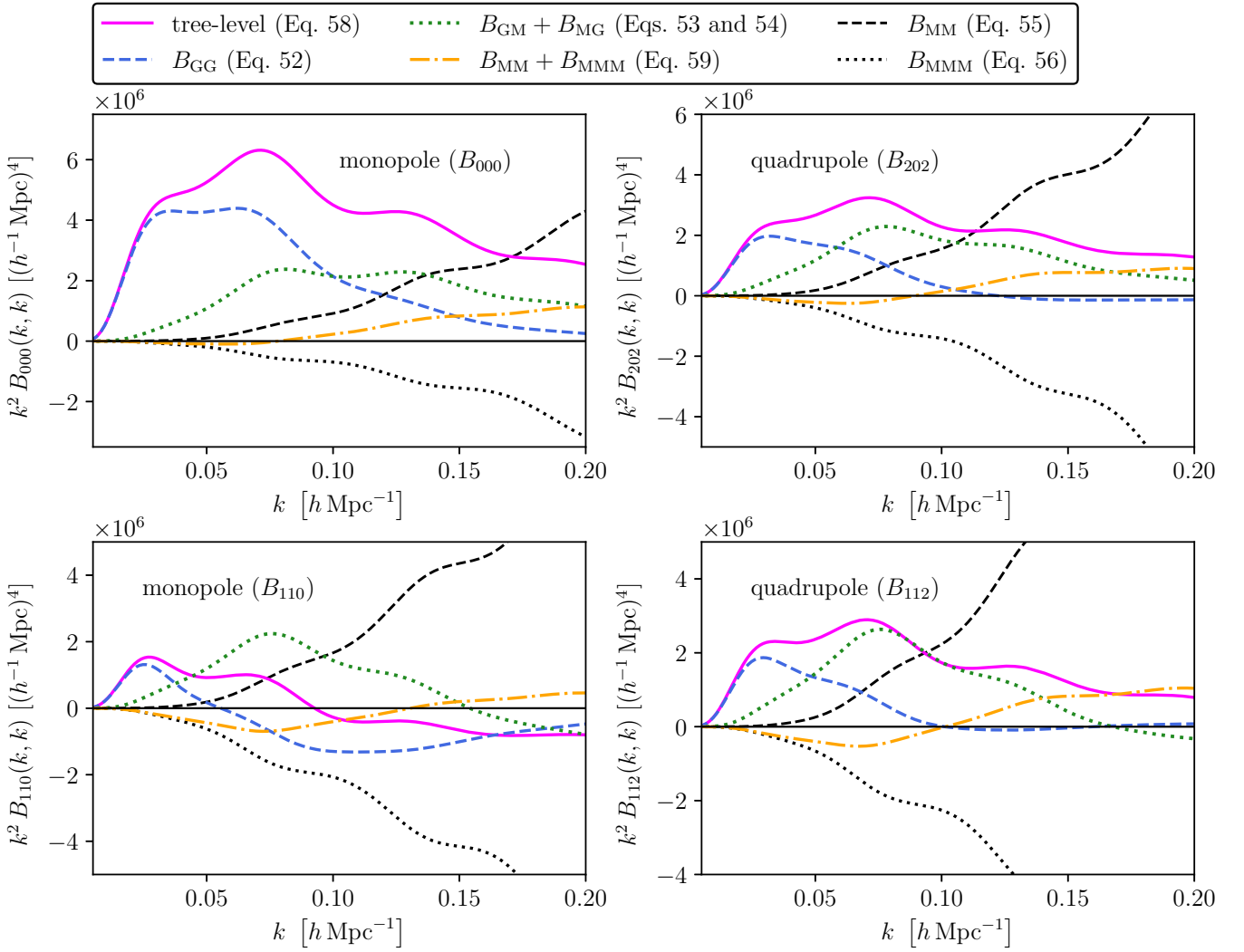}}
	\caption{
First and second terms of the monopole and quadrupole components of the bispectrum: the monopole components are $B_{000}$ (top left) and $B_{110}$ (bottom left), and the quadrupole components are $B_{202}$ (top right) and $B_{112}$ (bottom right). In the $\Gamma$-expansion, the nonlinear bispectrum is decomposed into five parts: $B_{\rm GG}$, $B_{\rm GM}$, $B_{\rm MG}$, $B_{\rm MM}$ and $B_{\rm MMM}$, which are shown schematically in Figure~\ref{fig:diagram}. All of these terms are calculated in the high-$k$ limit, so the bispectrum of the five terms combined is the tree-level solution. 
	}
	\label{fig:bk_IR}
\end{figure*}

In this subsection, we demonstrate the IR cancellation by calculating the full order of the $\Gamma$-expansion for both the power and bispectra. To see this, consider the linear galaxy perturbation with the linear IR flow
\begin{eqnarray}
	\FT{\delta}(\VEC{k}) 
	= e^{-i\VEC{k}\cdot\overline{\PP}_{\rm lin}}\, Z^{[1]}(\VEC{k})\, \FT{\delta}_{\rm lin}(\VEC{k}).
	\label{Eq:delta_psi_lin}
\end{eqnarray}
The linear IR flow is described by
\begin{eqnarray}
	-i \VEC{k}\cdot\overline{\PP}_{\rm lin} = \int \frac{d^3p}{(2\pi)^3} 
    \left( \frac{\VEC{k}_{\rm RSD}\cdot \VEC{p}}{p^2} \right) \FT{\delta}_{\rm lin}(\VEC{p}),
\end{eqnarray}
where
\begin{eqnarray}
    \VEC{k}_{\rm RSD} = \VEC{k} + f\left( \VEC{k}\cdot\hat{n} \right)\hat{n}.
\end{eqnarray}
Eq.~(\ref{Eq:delta_psi_lin}) means that all nonlinear corrections of the density perturbation at the scale $k$ we are interested in are due to long-wavelength (infra-red, IR) modes beyond the scale $k$. 
To explain Equation~(\ref{Eq:delta_psi_lin}) in the standard PT context, we assume in Eq.~(\ref{Eq:SPT}) that the amplitude of one of the $n$ wave vectors is larger than that of the other, i.e., 
$|\VEC{p}_n|\gg |\VEC{p}_{1(n-1)}|$, resulting in $\delta_{\rm D}(\VEC{k}-\VEC{p}_{1n})\to n\,\delta_{\rm D}(\VEC{k}-\VEC{p}_n)$. 
Then, we obtain
\begin{eqnarray}
	\FT{\delta}(\VEC{k}) \hspace{-0.25cm}&=&\hspace{-0.25cm}  \sum_{n=1}^{\infty}\, n\, \prod_{i=1}^{n-1}
    \int \frac{d^3p_i}{(2\pi)^3}\,\FT{\delta}_{\rm lin}(\VEC{p}_i) \notag\\
    \hspace{-0.25cm}&\times&\hspace{-0.25cm}
    Z^{[n]}(\VEC{k},\VEC{p}_1,\dots,\VEC{p}_{n-1})\, \FT{\delta}_{\rm lin}(\VEC{k}).
	\label{Eq:delta_ap}
\end{eqnarray}
Comparing Eq.~(\ref{Eq:delta_psi_lin}) and Eq.~(\ref{Eq:delta_ap}) leads to
\begin{eqnarray}
    Z^{[n]}(\VEC{k}, \VEC{p}_1,\dots,\VEC{p}_{n-1}) = \frac{Z^{[1]}(\VEC{k})}{n!}\prod_{i=1}^{n-1} \left( \frac{\VEC{k}_{{\rm RSD}}\cdot\VEC{p}_i}{p_i^2} \right).
\end{eqnarray}
Substituting the above equation into Eq.~(\ref{Eq:Gamma_Zn}) leads to
\begin{eqnarray}
	\Gamma^{(r)}(\VEC{k}, \VEC{p}_1,\dots,\VEC{p}_{r-1})
	= {\cal D}(\VEC{k})\, \frac{Z^{[1]}(\VEC{k})}{r!} \prod_{i=1}^{r-1} \left( \frac{\VEC{k}_{{\rm RSD}}\cdot\VEC{p}_i}{p_i^2} \right).
	\label{Eq:Gamma_P_highK}
\end{eqnarray}
Here, the damping factor ${\cal D}(\VEC{k})$ arising from the linear IR flow is given by~\citep{Matsubara:2007wj}
\begin{eqnarray}
    {\cal D}(\VEC{k}) \hspace{-0.25cm} &=& \hspace{-0.25cm} 
    \sum_{s=0}^{\infty} \frac{1}{s!}\left[ -\frac{1}{2} \int \frac{d^3p}{(2\pi)^3} \left( \frac{\VEC{k}_{\rm RSD}\cdot\VEC{p}}{p^2} \right)^2 P_{\rm lin}(p) \right]^s \nonumber \\
     \hspace{-0.25cm} &=& \hspace{-0.25cm} \exp\left( - \frac{k^2(1 + 2f\mu^2 + f^2\mu^2)}{2}\sigma_{\rm dd}^2 \right),
     \label{Eq:D}
\end{eqnarray}
where $\sigma_{\rm dd}^2$ is the dispersion of the linear displacement vector, given by
\begin{eqnarray}
    \sigma_{\rm dd}^2 = \frac{1}{3}\int \frac{dp}{2\pi^2} P_{\rm lin}(p).
\end{eqnarray}

The propagator term $G^2(\VEC{k})P_{\rm lin}(k)$ is then expressed as
\begin{eqnarray}
	G^2(\VEC{k})P_{\rm lin}(k) \to {\cal D}^2(\VEC{k})[ Z^{[1]}(\VEC{k}) ]^2 P_{\rm lin}(k).
	\label{Eq:PG_high_K}
\end{eqnarray}
To compute the mode-coupling term, we assume $\delta_{\rm D}\left( \VEC{k}-\VEC{k}_{1r} \right)\to r\, \delta_{\rm D}\left( \VEC{k}-\VEC{k}_{r} \right)$ in the limit $|\VEC{k}_r|\gg |\VEC{k}_{1(r-1)}|$ in Eq.~(\ref{Eq:PGamma}), resulting in 
\begin{eqnarray}
	P_{\rm MC}(\VEC{k}) 
	\hspace{-0.25cm}&=&\hspace{-0.25cm}
	P_{\rm lin}(k)\, \sum_{r=2}^{\infty} r!\, r\,
	\prod_{i=1}^{r-1}\int \frac{d^3p_i}{(2\pi)^3}P_{\rm lin}(p_i) \nonumber \\
	\hspace{-0.25cm}&\times&\hspace{-0.25cm}
	\big[ \Gamma^{[r]}(\VEC{k},\VEC{p}_1,\dots,\VEC{p}_{r-1}) \big]^2.
    \label{Eq:PMC_appro}
\end{eqnarray}
Inserting Eq.~(\ref{Eq:Gamma_P_highK}) in the above expression, we finally derive
\begin{eqnarray}
	P_{\rm MC}(\VEC{k}) 
	\hspace{-0.25cm}&\to&\hspace{-0.25cm}
	\left[ 1 - {\cal D}^2(\VEC{k}) \right][Z^{[1]}(\VEC{k})]^2  P_{\rm lin}(k),
	\label{Eq:PMC_high_K}
\end{eqnarray}
and therefore,
\begin{eqnarray}
    P(\VEC{k}) \hspace{-0.25cm} &=&   \hspace{-0.25cm} G^2(\VEC{k})P_{\rm lin}(k) + P_{\rm MC}(\VEC{k}) \nonumber \\
               \hspace{-0.25cm} &\to& \hspace{-0.25cm} [Z^{[1]}(\VEC{k})]^2\,P_{\rm lin}(k).
    \label{P_high_k}
\end{eqnarray}
This result is as predicted in Eq.~(\ref{Eq:PB_GI}). The effect of the IR flow on the power spectrum completely is canceled out by summing up all orders of the $\Gamma$-expansion, i.e., up to all orders in standard perturbation theory. In other words, in the high-$k$ limit, the nonlinear power spectrum becomes just the linear power spectrum.

Figure~\ref{fig:pk_IR} plots the theoretical predictions for the monopole and quadrupole components of the power spectrum in the high-$k$ limit, where the sum of the propagator term $P_{\rm G}$ and the mode-coupling term $P_{\rm MC}$ is the linear theory prediction, as shown in Eq.~(\ref{P_high_k}). The propagator of the quadrupole component is negative on small scales. The scale at which the mode-coupling term dominates, i.e., the point at which the theoretical curves of the propagator term and the mode-coupling term intersect, is $k=0.14\hk$ for the monopole component and $k=0.07\hk$ for the quadrupole component. This fact shows the intrinsic difficulty of theoretical prediction of the quadrupole component. That is, when focusing on a certain scale $k$, the quadrupole component requires to compute higher-order $\Gamma$-expansion terms than the monopole component.

We next turn to the IR cancellation in the bispectrum. As in the case of the power spectrum, consider the second-order density perturbation with the linear IR flow:
\begin{eqnarray}
	\FT{\delta}(\VEC{k}) = e^{-i\VEC{k}\cdot\overline{\PP}_{\rm lin}}\, \FT{\delta}^{[2]}(\VEC{k}).
\end{eqnarray}
The corresponding $\Gamma^{(r)}$ is given by
\begin{eqnarray}
	\hspace{-3.0cm}&& \Gamma^{(r)}(\VEC{k}_1,\VEC{k}_2,\VEC{p}_1,\dots,\VEC{p}_{r-2}) \nonumber \\
	\hspace{-0.55cm}&=&\hspace{-0.25cm}
	{\cal D}(\VEC{k}_{[12]}) \frac{2!}{r!}\prod_{i=1}^{r-2} 
	\left( \frac{\VEC{k}_{[12]\, {\rm RSD}}\cdot\VEC{p}_i}{p_i^2} \right) Z^{[2]}(\VEC{k}_1,\VEC{k}_2),
	\label{Eq:Gamma_B_highK}
\end{eqnarray}
where we used the relation
\begin{eqnarray}
	\delta_{\rm D}\left( \VEC{k} - \VEC{p}_{1n} \right) \to \frac{n!}{2!(n-2)!} \delta_{\rm D}\left( \VEC{k} - \VEC{p}_{n-1} - \VEC{p}_n \right)
	\label{Eq:Delta_highK}
\end{eqnarray}
under the condition that the amplitudes of two of $n$ momenta, $\VEC{p}_n$ and $\VEC{p}_{n-1}$, are much larger than those of the others: $|\VEC{p}_n|,\, |\VEC{p}_{n-1}|\gg |\VEC{p}_i|$ for $i=1,\dots,n-2$. Using Eqs.~(\ref{Eq:Gamma_P_highK}) and (\ref{Eq:Gamma_B_highK}), we derive
\begin{eqnarray}
	\hspace{-0.50cm}&& B_{\rm GG}(\VEC{k}_1,\VEC{k}_2) P_{\rm lin}(k_1) P_{\rm lin}(k_2) \nonumber \\
	 &&=
	{\cal D}(\VEC{k}_1){\cal D}(\VEC{k}_2){\cal D}(\VEC{k}_{12}) B_{\rm tree}(\VEC{k}_1,\VEC{k}_2), 
	\label{Eq:GG}
\end{eqnarray}	
\begin{eqnarray}	
	&&B_{\rm GM}(\VEC{k}_1,\VEC{k}_2) P_{\rm lin}(k_1) \nonumber \\
	&& =
	\left( {\cal D}^2(\VEC{k}_1) - {\cal D}(\VEC{k}_1){\cal D}(\VEC{k}_2){\cal D}(\VEC{k}_{12}) \right) B_{\rm tree}(\VEC{k}_1,\VEC{k}_2),
	\label{Eq:GM}
	\end{eqnarray}	
	\begin{eqnarray}
	&&B_{\rm MG}(\VEC{k}_1,\VEC{k}_2) P_{\rm lin}(k_2)\nonumber\\
	&&= \left( D^2(\VEC{k}_2) - {\cal D}(\VEC{k}_1){\cal D}(\VEC{k}_2){\cal D}(\VEC{k}_{12}) \right) B_{\rm tree}(\VEC{k}_1,\VEC{k}_2),
	\label{Eq:MG}
	\end{eqnarray}
	\begin{eqnarray}	
	&& B_{\rm MM}(\VEC{k}_1,\VEC{k}_2)\nonumber \\
	&&= 
	\Big( {\cal D}(\VEC{k}_1){\cal D}(\VEC{k}_2){\cal D}^{-1}(\VEC{k}_{12}) -
	{\cal D}^2(\VEC{k}_1)- {\cal D}^2(\VEC{k}_2)  \nonumber \\
	&& \hspace{0.25cm}+
	{\cal D}(\VEC{k}_1){\cal D}(\VEC{k}_2){\cal D}(\VEC{k}_{12}) \Big) B_{\rm tree}(\VEC{k}_1,\VEC{k}_2), 
	\label{Eq:MM}
	\end{eqnarray}
	\begin{eqnarray}
	&&B_{\rm MMM}(\VEC{k}_1,\VEC{k}_2)  \nonumber \\
	&&=\Big( 1 - {\cal D}(\VEC{k}_1){\cal D}(\VEC{k}_2){\cal D}^{-1}(\VEC{k}_{12}) \Big) B_{\rm tree}(\VEC{k}_1,\VEC{k}_2),
	\label{Eq:MMM}
\end{eqnarray}
where $B_{\rm tree}$ means the tree-level solution of the bispectrum, given by
\begin{eqnarray}
	\hspace{-0.5cm}&&B_{\rm tree}(\VEC{k}_1,\VEC{k}_2) \nonumber \\\hspace{-0.5cm}&&=  2\, Z^{[1]}(\VEC{k}_1)Z^{[1]}(\VEC{k}_2)Z^{[2]}(\VEC{k}_1,\VEC{k}_2) P_{\rm lin}(k_1) P_{\rm lin}(k_2).
\end{eqnarray}
Obviously, summing up all the five terms $B_{GG}$, $B_{GM}$, $B_{MG}$, $B_{MM}$ and $B_{MMM}$, the nonlinear corrections coming from the IR flow are completely canceled out, and the resulting bispectrum is the same as the tree-level solution.
\begin{eqnarray}
	B(\VEC{k}_1,\VEC{k}_2) \to B_{\rm tree}(\VEC{k}_1,\VEC{k}_2) + \mbox{2 cyc.}.
\end{eqnarray}
This ends the proof.

Figure~\ref{fig:bk_IR} shows the theoretical predictions of the four bispectrum multipoles $B_{000}$, $B_{110}$, $B_{202}$ and $B_{112}$ as a function of $k$, focusing only on the $k_1=k_2$ elements; $B_{\rm GG}$, $B_{\rm GM}$, $B_{\rm MG}$, $B_{\rm MM}$ and $B_{\rm MMM}$ shown in the figure are computed in the high-$k$ limit, and the sum of all of them reproduces the tree-level solution. The scale at which the terms including the mode-coupling integral dominate, i.e., the point at which the theoretical curves of $B_{\rm GG}$ and $B_{\rm GM}+B_{\rm MG}$ intersect, is $k=0.1\hk$ for $B_{000}$, $k=0.035$ for $B_{110}$, $k=0.065\hk$ for $B_{202}$ and $k=0.05\hk$ for $B_{112}$. The contribution of the $B_{\rm MM}+B_{\rm MMM}$, which consists of only the mode-coupling integral, is $\sim5\%$ for $B_{000}$, $\sim80\%$ for $B_{110}$, $\sim8\%$ for $B_{202}$ and $\sim20\%$ for $B_{112}$ up to $k=0.1\hk$. These results indicate that the contributions from the mode-coupling integrals to $B_{000}$ and $B_{202}$ are less than to $B_{110}$ and $B_{112}$, and therefore, $B_{000}$ and $B_{202}$ are easier to predict than $B_{110}$ and $B_{112}$. \citet{Sugiyama:2018yzo} showed that $B_{000}$ and $B_{202}$ have a higher signal-to-noise ratio than $B_{110}$ and $B_{112}$, which suggests that $B_{000}$ and $B_{202}$ should be taken into account first when analyzing the bispectrum (or 3PCF). Note that both the $B_{\rm MM}$ and $B_{\rm MMM}$ terms diverge, because they are proportional to ${\cal D}(\VEC{k}_1){\cal D}(\VEC{k}_2){\cal D}^{-1}(\VEC{k}_{12})$ and $-{\cal D}(\VEC{k}_1){\cal D}(\VEC{k}_2){\cal D}^{-1}(\VEC{k}_{12})$ at small scales, respectively, as shown in Eqs.~(\ref{Eq:MM}) and (\ref{Eq:MMM}), and in Figure~\ref{fig:bk_IR}. However, the sum of them, $B_{\rm MM}+B_{\rm MMM}$, completely cancels out the divergence term (see orange lines in Figure~\ref{fig:bk_IR}.):
\begin{eqnarray}
	&& B_{\rm MM}(\VEC{k}_1,\VEC{k}_2) + B_{\rm MMM}(\VEC{k}_1,\VEC{k}_2)  \nonumber \\
	&&= 
	\Big( 1 -
	{\cal D}^2(\VEC{k}_1)- {\cal D}^2(\VEC{k}_2)  \nonumber \\
	&& \hspace{0.25cm}+
	{\cal D}(\VEC{k}_1){\cal D}(\VEC{k}_2){\cal D}(\VEC{k}_{12}) \Big) B_{\rm tree}(\VEC{k}_1,\VEC{k}_2).
\end{eqnarray}

\subsection{Template model}
\label{Sec:TemplateModel}

Now, we are ready to present a template model for the redshift-space bispectrum that explains the nonlinear degradation of the BAO signal. We start with the power spectrum case and show how to reproduce the power spectrum template model proposed by \citet{Eisenstein:2006nj} in the context of the IR cancellation. We then extend it to the bispectrum case.

The mode-coupling term is known to have a suppressed BAO signal compared to the linear power spectrum~\citep{Crocce:2007dt}. Nevertheless, the mode-coupling term arising from the IR flow is proportional to $P_{\rm lin}$ (\ref{Eq:PMC_high_K}) and has the original BAO signal. This discrepancy is due to the assumption $\delta_{\rm D}\left( \VEC{k} - \VEC{p}_{1r} \right)\to r\, \delta_{\rm D}\left( \VEC{k}-\VEC{p}_r \right)$ in Eq.~(\ref{Eq:PMC_appro}), which corresponds to the assumption that the IR flow is uncorrelated with the density perturbation, as discussed in Section~\ref{Sec:InfraRedFlow}.

We propose an empirical method for breaking the IR cancellation and achieving physical effects. It is simply to replace $P_{\rm lin}$ appearing in the mode-coupling term $P_{\rm MC}$ (\ref{Eq:PMC_high_K}) with $P_{\rm nw}$~\citep{Eisenstein:1997ik} that does not have the BAO signal:
\begin{eqnarray}
	P_{\rm MC}(\VEC{k}) \to \left[ 1 - {\cal D}^2(\VEC{k}) \right] [ Z^{[1]}(\VEC{k}) ]^2 P_{\rm nw}(k).
\end{eqnarray}
Then, we obtain \cite{Eisenstein:2006nj}'s template
\begin{eqnarray}
	P^{\rm (temp)}(\VEC{k}) \hspace{-0.25cm}&=&\hspace{-0.25cm} 
	G^2(\VEC{k})P_{\rm lin}(k) + P_{\rm MC}(\VEC{k}) \nonumber \\
	\hspace{-0.25cm}&\to& \hspace{-0.25cm}
	\big[ Z^{[1]}(\VEC{k}) \big]^2
	\big[{\cal D}^2(\VEC{k})\, P_{\rm w}(k) + P_{\rm nw}(k) \big],
	\label{Eq:Eisenstein2007}
\end{eqnarray}
where
\begin{eqnarray}
	P_{\rm w}(k) = P_{\rm lin}(k) - P_{\rm nw}(k).
\end{eqnarray}
Despite its simplicity, Eq.~(\ref{Eq:Eisenstein2007}) provides highly unbiased constraints on the BAO signal and has passed various tests using high-precision $N$-body simulations~\citep[e.g.,][]{Seo:2008yx,Kim:2008kf,Seo:2009fp,Ross:2016gvb}.

In a similar manner, we replace $P_{\rm lin}$ appearing in $B_{\rm GM}$ (\ref{Eq:GM}), $B_{\rm MG}$ (\ref{Eq:MG}), $B_{\rm MM}$ (\ref{Eq:MM})and $B_{\rm MMM}$ (\ref{Eq:MMM}) with $P_{\rm nw}$, resulting in 
\begin{eqnarray}
	&&B^{(\rm temp)}(\VEC{k}_1,\VEC{k}_2) \nonumber \\&&= 2\, Z^{[1]}(\VEC{k}_1)Z^{[1]}(\VEC{k}_2)Z^{[2]}(\VEC{k}_1,\VEC{k}_2)\nonumber \\
	&&\times
	\Big\{
	{\cal D}(\VEC{k}_1){\cal D}(\VEC{k}_2){\cal D}(\VEC{k}_{12}) P_{\rm w}(k_1) P_{\rm w}(k_2) \nonumber \\
	&&\hspace{0.20cm}+ {\cal D}^2(\VEC{k}_1) P_{\rm w}(k_1) P_{\rm nw}(k_2)+ {\cal D}^2(\VEC{k}_2) P_{\rm nw}(k_1) P_{\rm w}(k_2) \nonumber \\
	&&\hspace{0.20cm}+ P_{\rm nw}(k_1)P_{\rm nw}(k_2)\Big\} + \mbox{2 cyc.}.
	\label{Eq:MainResult}
\end{eqnarray}
The above expression is one of the main results in this paper~\footnote{\citet{Blas:2016sfa} presented a bispectrum model for explaining the nonlinear BAO damping using an IR re-summation method in the context of the time-sliced perturbation theory. They performed the calculations for the case of dark matter in real space, and derived the term consisting of a product of $P_{\rm w}$ and $P_{\rm nw}$ in the third line of Eq.~(\ref{Eq:MainResult}).}. The validity of this model will be examined in detail in Section~\ref{Sec:JointAnalysisWith3PCF} by comparing the theoretical predictions of the 3PCF using this model to the corresponding measurements from the mock catalogues.

\begin{figure*}
    \includegraphics[width=\textwidth]{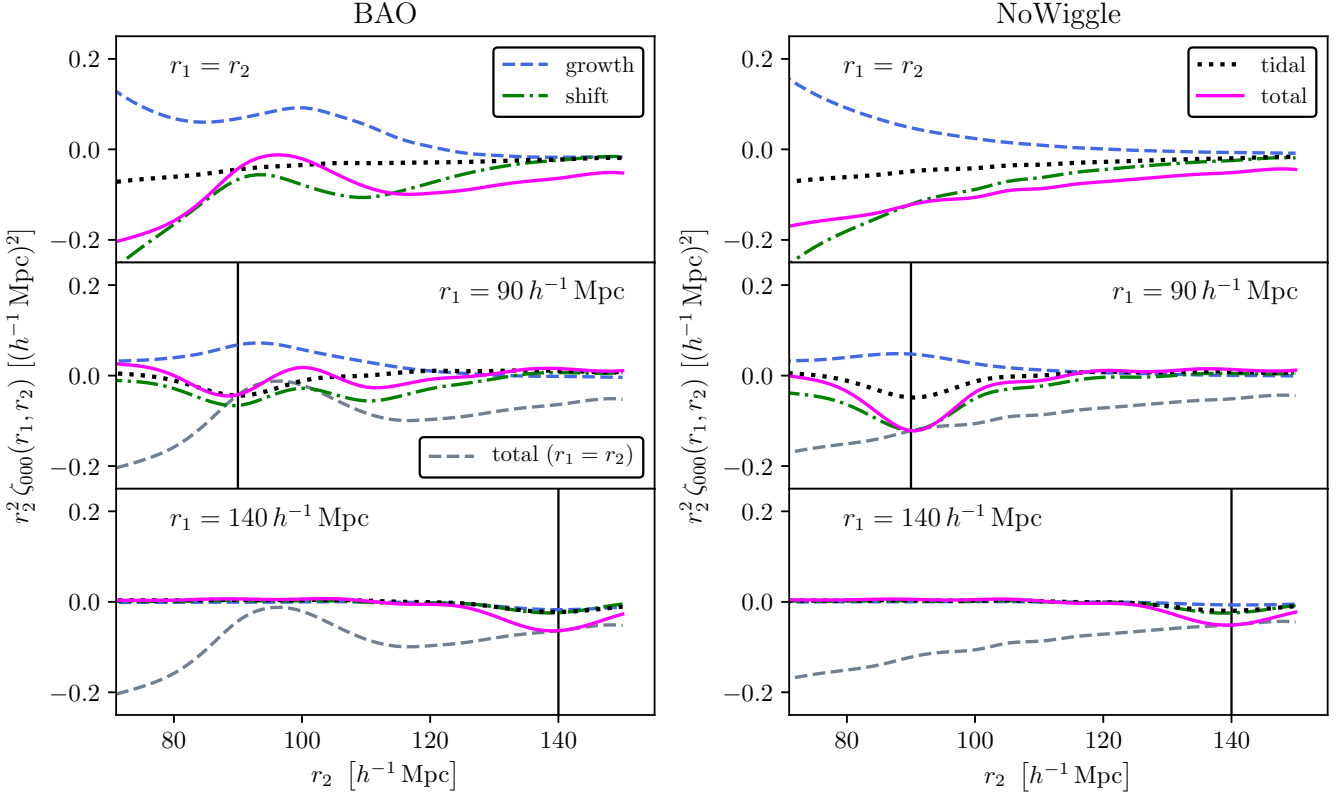}
	\caption{
Theoretical predictions of the lowest order of the monopole component of the 3PCF in real space for dark matter. The non-linear growth term, the shift term, the tidal force term, and their combined total $\zeta_{000}$ are shown; the predictions from the template model (\ref{Eq:MainResult}) with BAO and the model without BAO are plotted in the left and right panels. \Mod{For both the left and right panels, we plot $\zeta_{000}$ as a function of $r_2$ when $r_1$ is fixed to $r_1=r_2$, $r_1=90\hMpc$ and $r_1=140\hMpc$, from top to bottom. The vertical black lines in the middle and bottom panels show $r_2$ as $90\hMpc$ and $140\hMpc$, respectively.}
	}
	\label{fig:3PCF_RealSpace}
\end{figure*}

\section{Three-point correlation functions}
\label{Sec:ThreePointCorrelationFunctions}

\subsection{Predictions of 3PCFs}
\label{Sec:PredictionsOf3PCF}

As explained in the introduction, we adopt the 3PCF to analyze the data in order to extract all of the information from the BAO while keeping the number of data bins under control. 
The template model of the 3PCF for the data analysis is computed through the 2D Hankel transform (\ref{Eq:B_to_zeta}) of the bispectrum template model given in Eq.~(\ref{Eq:MainResult}).

Here let us discuss the BAO information on the 3PCF, focusing on the impact of nonlinear gravity on the BAO signal. To this end, we focus on $\zeta_{000}$ for dark matter in real space as the simplest example. We compare the theoretical predictions of our template model including BAO with those of the model without BAO, which consists of only $P_{\rm nw}$, in Figure~\ref{fig:3PCF_RealSpace}. In the no-wiggle case, the non-linearity arises from the second-order density perturbation of dark matter in real space only, and it can be decomposed into three sources, i.e., non-linear growth, shift effects, and tidal forces, as follows
\begin{eqnarray}
    \delta^{[2]}(\VEC{x})\hspace{-0.25cm} &=&\hspace{-0.25cm}  \frac{17}{21}[\delta_{\rm lin}(\VEC{x})]^2 - \PP_{\rm lin}(\VEC{x}) \cdot \nabla \delta_{\rm lin}(\VEC{x}) \nonumber \\
    \hspace{-0.25cm}&+&\hspace{-0.25cm} \frac{2}{7}\Bigg[\left( \frac{\partial_i\partial_j}{\partial^2}-\frac{1}{3}\delta_{ij} \right)\delta_{\rm lin}(\VEC{x})\Bigg]^2
    \label{Eq:2ndOrder}
\end{eqnarray}
where the first, second and third terms represent the nonlinear growth, shift, and tidal force effects. 
In the figure, we investigate how those three components contribute to the final 3PCF. From the right panel of the figure, we find that in $\zeta_{000}$, in the range $80<r<150 \hMpc$, the non-linear growth term is positive, and the shift and tidal force terms are negative. Since the total $\zeta_{000}$ is negative, we can conclude that in this range, the shift and tidal force terms give the main contribution. In particular, from the middle and bottom right panels of the figure, we notice that at the point where $r_1=r_2$, the shift and tidal force terms have a trough. We attribute this trough to the fact that the shift and tidal force terms arise from the spatial derivative of the density field, i.e., the relation between galaxies at different positions, and that thus the probability of finding a triplet of galaxies comprising the 3PCF at the same scale $r_1=r_2$ is reduced compared to the $r_1\neq r_2$ case. \Mod{For the non-linear growth term, when the value at $r_1=r_2$ is positive or negative, it has a peak or trough, respectively. Thus, when we fix $r_1$ to $90\hMpc$, we find a peak; when we fix $r_1$ to $140\hMpc$, a trough appears, but this effect is too small to be visible in the figure.}

In the case with BAO, we find the BAO signal around $100\hMpc$ in the top left panel where $r_1=r_2$ is satisfied. However, in the middle left panel, the BAO peak at $\sim 100\hMpc$ and the trough at $r_1=r_2=90\hMpc$ cancel each other out and neither signal is clearly visible. On the scale where the BAO signal does not appear, i.e., in the bottom left panel ($r_1=140\hMpc$), we find an trough at $r_1=r_2$ as in the no-wiggle case. 

\subsection{Window function effects}
\label{Sec:WindowFunctionEffects}

Our estimator for measuring the 3PCF as shown in Section~\ref{Sec:MethodologyFor3PCFAnalysis}
requires to take into account how the survey geometry affects the measurement of 3PCFs. Following \citet{Sugiyama:2018yzo}, we include the window function effect into the 3PCF model as follows
\begin{eqnarray}
     \zeta^{(\rm obs)}_{\ell_1\ell_2\ell}(r_1,r_2)
	\hspace{-0.25cm}&=&\hspace{-0.25cm} 
    (4\pi) \sum_{\ell'_1+\ell'_2+\ell'={\rm even}}\ \ \sum_{\ell''_1+\ell''_2+\ell''={\rm even} } \nonumber \\
	\hspace{-0.25cm}&\times&\hspace{-0.25cm} 
	\left\{ \begin{smallmatrix} \ell''_1 & \ell''_2 & \ell'' \\   \ell'_1 & \ell'_2 & \ell' \\   \ell_1 & \ell_2 & \ell \end{smallmatrix}  \right\}
	\left[\frac{h_{\ell_1\ell_2\ell}h_{\ell_1\ell'_1\ell''_1}h_{\ell_2\ell'_2\ell''_2}h_{\ell \ell'\ell''}}{h_{\ell'_1\ell'_2\ell'}h_{\ell''_1\ell''_2\ell''}} \right]
	\nonumber \\
	\hspace{-0.25cm}&\times&\hspace{-0.25cm} 
			Q_{\ell''_1\ell''_2\ell''}(r_1,r_2)\, \zeta_{\ell'_1\ell'_2\ell'}(r_1,r_2),
	\label{Eq:zetaMask}
\end{eqnarray}
where $\zeta^{(\rm obs)}_{\ell_1\ell_2\ell}$ denotes the theoretical model corresponding the observed 3PCF multipoles, the bracket with $9$ multipole indices, $\{\dots\}$, denotes the Wigner-$9$j symbol, and $Q_{\ell_1\ell_2\ell}$ is the 3PCF of the window function expanded in the TripoSH formalism. Throughout this paper, we ignore the contribution from the integral constraint~\citep{Peacock1991}. This relation describes simply mixing of different multipole components, ($\ell_1,\ell_2,\ell$), due to the survey window function. In practice we have one issue in calculating a 3PCF model that includes the effects of the window function in this method. As for the indexes corresponding to the LOS direction, $\ell'$ and $\ell''$, due to RSD, one would expect that their contributions become smaller for larger $\ell'$ and $\ell''$. However, for $\ell_1'$, $\ell_1''$, $\ell_2'$ and $\ell_2''$ corresponding to expansion w.r.t. $\VEC{r}_1$ and $\VEC{r}_2$, it is not clear at which multipoles we should truncate to obtain a converged 3PCF. For this reason, let us examine below in detail the contributions of each multipole component. As a caveat, the results derived here correspond to the BOSS survey region, and a similar study is needed for each survey region in which the 3PCF is measured.

One may wonder why the survey window function correction is required for 3PCF measurements, i.e., in configuration space rather than Fourier space. We summarize key points here, leaving more elaborated discussions with some equations in Appendix~\ref{Ap:Window}. In general, as long as we measure the multipole components of the 3PCF of galaxies, we have to take the survey window function of the 3PCF into account. Note that this fact is independent of whether we use FFTs to measure the 3PCF~\citep{Scoccimarro:2015bla,Slepian:2016qwa,Sugiyama:2018yzo}, or not~\citep{Slepian:2015qza}. We use FFTs for measuring the 3PCF, but if we only consider the monopole components of $\zeta_{\ell_1\ell_2\ell}$ and $Q_{\ell_1\ell_2\ell}$, Eq.~(\ref{Eq:zetaMask}) matches Eq.~(32) in~\citet{Slepian:2015qza}, which is a measurement of 3PCF without FFTs, as already pointed out by~\citet{Sugiyama:2018yzo}. The difference from the treatment of the window function in \citet{Slepian:2015qza} is that we convolve the window function in our theoretical model of the 3PCF, whereas \citet{Slepian:2015qza} treats the multipole components of the 3PCF of the window function as a single matrix $\MAT{A}$ and multiplies the measured multipole components of the 3PCF by its inverse matrix $\MAT{A}^{-1}$.

We propose one way to quantitatively assess the effect of the survey geometry. 
As can be seen from Figures~\ref{fig:3PCFMeasured_000202} and \ref{fig:3PCFMeasured_110112404}, 3PCFs may change sign at large scales, i.e., 3PCFs may pass through a point with zero. In this case, for example, the calculation of the relative change diverges, making it difficult to interpret the results. Therefore, instead of checking for convergence in each ($r_1$, $r_2$)-bin, we calculate the 3PCFs averaged over a range of $80\leq r \leq 150\hMpc$, which is used in the analysis of this paper (see Sections~\ref{Sec:2PCFanalysis} and \ref{Sec:JointAnalysisWith3PCF}), and check whether the average values are converged: we compute 
\begin{eqnarray}
    \bar{\zeta}_{\ell_1\ell_2\ell}^{(\rm obs)} = 
    \begin{cases}
    \frac{1}{N_{\rm b}} \sum_{r_1\geq r_2} \zeta^{(\rm obs)}_{\ell_1\ell_2\ell}(r_1,r_2) \quad \mbox{for $\ell_1=\ell_2$}\\
    \frac{1}{N_{\rm b}} \sum_{r_1, r_2}\  \zeta^{(\rm obs)}_{\ell_1\ell_2\ell}(r_1,r_2) \quad \mbox{for $\ell_1\neq \ell_2$},
    \end{cases}
    \label{Eq:zeta_mean}
\end{eqnarray}
where $N_{\rm b}$ is the number of data bins, which is $N_{\rm b}=(8\times 9)/2=36$ for $\ell_1=\ell_2$ and $N_{\rm b}=8^2=64$ for $\ell_1\neq \ell_2$ because we adopt a bin width of $\Delta r =10\hMpc$.

To estimate the extent to which other multipole components $\zeta_{\ell_1'\ell_2'\ell'}$ affect the multipole component of interest $\zeta_{\ell_1\ell_2\ell}^{(\rm obs)}$, we define the following quantities from Eq.~(\ref{Eq:zetaMask}):
\begin{eqnarray}
    \Delta \bar{\zeta}_{\ell'_1\ell'_2\ell'}
    \hspace{-0.25cm}&=& \hspace{-0.25cm}
    \frac{\mbox{Average}\Big[\Delta \zeta_{\ell'_1\ell'_2\ell'}^{\ell_1\ell_2\ell}(r_1,r_2) \Big]}{\bar{\zeta}_{\ell_1\ell_2\ell}^{(\rm obs)}}\nonumber \\
    \Delta \zeta_{\ell'_1\ell'_2\ell'}^{\ell_1\ell_2\ell}(r_1,r_2)
    \hspace{-0.25cm}&=&\hspace{-0.25cm}
    (4\pi) \ \ \sum_{\ell''_1+\ell''_2+\ell''={\rm even} }
    \left\{ \begin{smallmatrix} \ell''_1 & \ell''_2 & \ell'' \\   \ell'_1 & \ell'_2 & \ell' \\   \ell_1 & \ell_2 & \ell \end{smallmatrix}  \right\}
    \nonumber \\
	\hspace{-0.25cm}&\times&\hspace{-0.25cm} 
		\left[\frac{h_{\ell_1\ell_2\ell}h_{\ell_1\ell'_1\ell''_1}h_{\ell_2\ell'_2\ell''_2}h_{\ell \ell'\ell''}}{h_{\ell'_1\ell'_2\ell'}h_{\ell''_1\ell''_2\ell''}} \right]
	\nonumber \\
	\hspace{-0.25cm}&\times&\hspace{-0.25cm} 
			Q_{\ell''_1\ell''_2\ell''}(r_1,r_2)\, \zeta_{\ell'_1\ell'_2\ell'}(r_1,r_2),
    \label{Eq:Delta_zeta_mean}
\end{eqnarray}
where $\mbox{Average}[\cdot]$ means the same calculation manner as Eq.~(\ref{Eq:zeta_mean}), and $\Delta \bar{\zeta}_{\ell'_1\ell'_2\ell'}$ satisfies $\sum_{\ell_1'\ell_2'\ell'}\Delta \bar{\zeta}_{\ell'_1\ell'_2\ell'} = 1$. In a similar manner, we also define the following quantities to find out which multipole components of the window function 3PCF are required:
\begin{eqnarray}
    \Delta \bar{Q}_{\ell''_1\ell''_2\ell''}
    \hspace{-0.25cm}&=& \hspace{-0.25cm}
    \frac{\mbox{Average}\Big[\Delta Q_{\ell''_1\ell''_2\ell''}^{\ell_1\ell_2\ell}(r_1,r_2) \Big]}{\bar{\zeta}_{\ell_1\ell_2\ell}^{(\rm obs)}}\nonumber \\
    \Delta Q_{\ell''_1\ell''_2\ell''}^{\ell_1\ell_2\ell}(r_1,r_2)
    \hspace{-0.25cm}&=&\hspace{-0.25cm}
    (4\pi) \ \ \sum_{\ell'_1+\ell'_2+\ell'={\rm even} }
    \left\{ \begin{smallmatrix} \ell''_1 & \ell''_2 & \ell'' \\   \ell'_1 & \ell'_2 & \ell' \\   \ell_1 & \ell_2 & \ell \end{smallmatrix}  \right\}
    \nonumber \\
	\hspace{-0.25cm}&\times&\hspace{-0.25cm} 
		\left[\frac{h_{\ell_1\ell_2\ell}h_{\ell_1\ell'_1\ell''_1}h_{\ell_2\ell'_2\ell''_2}h_{\ell \ell'\ell''}}{h_{\ell'_1\ell'_2\ell'}h_{\ell''_1\ell''_2\ell''}} \right]
	\nonumber \\
	\hspace{-0.25cm}&\times&\hspace{-0.25cm} 
			Q_{\ell''_1\ell''_2\ell''}(r_1,r_2)\, \zeta_{\ell'_1\ell'_2\ell'}(r_1,r_2),
    \label{Eq:Delta_Q_mean}
\end{eqnarray}
where $\sum_{\ell_1''\ell_2''\ell''}\Delta \bar{Q}_{\ell''_1\ell''_2\ell''} = 1$. Note that the multipole indeces $\ell'$ and $\ell''$ associate with the theoretical model of the 3PCF, $\zeta_{\ell'_1\ell'_2\ell'}$, and the window 3PCF, $Q_{\ell''_1\ell''_2\ell''}$, respectively.

To compute $\Delta \bar{\zeta}_{\ell'_1\ell'_2\ell'}$ and $\Delta \bar{Q}_{\ell'_1\ell'_2\ell'}$, we consider $9$ multipole components for monopole $(\ell=0)$, $15$ for quadrupole $(\ell=2)$, $10$ for hexadecapole ($\ell=4$) and $7$ for the tetrahexacontapole ($\ell=6$), for a total of $41$ components. The specific multipole components that we calculate and their values for the NGC and SGC are summarized in Tables~\ref{Table:zeta_window} and~\ref{Table:Q_window}, where the absence or presence of round brackets represents the NGC or SGC results, respectively. The terms that contribute more than $0.5\%$ to the final results are highlighted in bold letters, and in this paper, we will include only such terms in the data analysis. These tables show that the contribution of the higher-order multipole components gradually decreases as the higher order is reached. This result allows us to conclude that we can use the expansion formalism of Eq.~(\ref{Eq:zetaMask}) with multipole components truncated at a finite order. The lower-order components of the quadrupole components such as $\Delta \bar{\zeta}_{202}$ and $\Delta \bar{Q}_{202}$, are more important than the higher-order multipole components of the monopole components such as $\Delta \bar{\zeta}_{880}$ and $\Delta \bar{Q}_{880}$. Unless we focus on the measurement of the hexadecapole, we do not need a theoretical model and a window function of the hexadecapole. As for the tetrahexacontapole ($\ell=6$), we can ignore it for all the cases we are interested in. The contribution of each multipole component other than the multipole component of interest can be positive or negative, and the total contribution is $5$-$15\%$ for $\zeta_{000}^{(\rm obs)}$, $\zeta_{110}^{(\rm obs)}$, $\zeta_{202}^{(\rm obs)}$ and $\zeta_{112}^{(\rm obs)}$ in both the NGC and SGC cases. For $\zeta^{(\rm obs)}_{404}$, the window function has a large effect of roughly $40\%$, so the convergence of Eq.~(\ref{Eq:zetaMask}) is poor.  For this reason, the results of $\zeta_{404}^{(\rm obs)}$ may not have converged with just the $41$ components that we have consider in Tables~\ref{Table:zeta_window} and~\ref{Table:Q_window}. We include $\zeta_{404}^{(\rm obs)}$ in our analysis with this caution; in Section~\ref{Sec:JointAnalysisWith3PCF}, however, we find that $\zeta_{404}^{(\rm obs)}$ does not have a significant contribution to our cosmological constraints at least in our setting.

We have further investigated the window function effect for the case of $\zeta_{606}^{(\rm obs)}$, even though we have not shown the results in Tables~\ref{Table:zeta_window} and~\ref{Table:Q_window}. The results are completely dominated by the effects of lower order multipole components rather than $\ell=6$, and have not converged at all with the $41$ components that we have calculated here. This fact makes it difficult to use $\zeta_{606}^{(\rm obs)}$ for cosmological analysis, but implies at the same time that there is little cosmological information in $\zeta_{606}^{(\rm obs)}$.

\begin{table*}

\centering
\begin{tabular}{cccccc}
\hline\hline
$\Delta \bar{\zeta}_{\ell'_1\ell'_2\ell'}\, [\%]$ & $\zeta^{(\rm obs)}_{000}$ & $\zeta^{(\rm obs)}_{110}$ & $\zeta^{(\rm obs)}_{202}$ & $\zeta^{(\rm obs)}_{112}$ & $\zeta^{(\rm obs)}_{404}$ \\
\hline  
monopole ($\ell=0$)\\
\hline\vspace{0.07cm}
$\Delta \bar{\zeta}_{000}$ & \textbf{94.94 (90.02)} &	\textbf{2.16 (2.99)}    &  \textbf{-3.25 (-2.09)} &	 \textbf{2.75 (1.80)}	& \textbf{-2.74 (-3.88)}\\ \vspace{0.07cm}
$\Delta \bar{\zeta}_{110}$ & \textbf{9.33 (12.60)}  &  \textbf{103.22 (100.94)} & 	\textbf{3.81 (2.65)}  &  \textbf{-8.74 (-5.55)} &	\textbf{1.96 (2.03)} \\ \vspace{0.07cm}
$\Delta \bar{\zeta}_{220}$ & -0.02 (-0.05)          &	\textbf{-0.56 (-0.77)}	&  \textbf{2.31 (1.52)}	  &  0.03 (-0.02)           &   \textbf{0.90 (0.95)}\\ \vspace{0.07cm}
$\Delta \bar{\zeta}_{330}$ & 0.07 (0.11)            &  0.05 (0.07)	            &  0.14 (0.06)	          &  0.02 (0.00)	        & \textbf{-0.68 (-0.79)}\\ \vspace{0.07cm}
$\Delta \bar{\zeta}_{440}$ & 0.01 (0.01)            &  -0.01 (-0.02)	        &  0.05 (0.03)	          &  0.02 (-0.00)           &   \textbf{1.62 (2.29)}\\ \vspace{0.07cm}
$\Delta \bar{\zeta}_{550}$ & -0.02 (-0.03)          &	-0.02 (-0.04)	        &  0.02 (0.02)	          &  0.01 (-0.00)           &   0.18 (0.29)\\ \vspace{0.07cm}
$\Delta \bar{\zeta}_{660}$ & -0.03 (-0.04)          &	-0.03 (-0.05)	        &  0.02 (-0.00)	          &  -0.00 (0.00)           &   0.00 (0.00)\\ \vspace{0.07cm}
$\Delta \bar{\zeta}_{770}$ & -0.02 (-0.03)          &	-0.02 (-0.04)	        &  -0.00 (-0.00)          &	-0.00 (-0.00)           &	0.00 (0.00)\\ \vspace{0.07cm}
$\Delta \bar{\zeta}_{880}$ & -0.02 (-0.03)          &	-0.02 (-0.02)	        &  -0.00 (-0.00)          &	-0.00 (-0.00)           &	0.00 (0.00)\\ 
\hline
quadrupole ($\ell=2$)\\
\hline\vspace{0.07cm}
$\Delta \bar{\zeta}_{022}$ & \textbf{-4.51 (-2.74)} &  \textbf{1.16 (0.78)}	        & \textbf{0.76 (0.83)}	 & \textbf{3.28 (3.61)}	  & \textbf{0.75 (0.81)} \\ \vspace{0.07cm}
$\Delta \bar{\zeta}_{112}$ & \textbf{5.08 (3.33)}   &  \textbf{-3.76 (-2.45)}	    & \textbf{5.02 (5.69)}	 & \textbf{97.83 (93.54)} & \textbf{1.78 (1.36)}\\ \vspace{0.07cm}
$\Delta \bar{\zeta}_{202}$ & \textbf{-4.00 (-2.44)} &  \textbf{0.93 (0.63)}	        & \textbf{84.57 (83.32)} & \textbf{2.64 (2.90)}	  & \textbf{14.49 (9.00)}\\ \vspace{0.07cm}
$\Delta \bar{\zeta}_{132}$ & -0.28 (-0.16)          &  \textbf{-2.02 (-1.29)}	    & 0.10 (0.18)	         & \textbf{0.38 (0.81)}   & \textbf{-0.71 (-0.79)}\\ \vspace{0.07cm}
$\Delta \bar{\zeta}_{222}$ & 0.10 (0.00)            &  0.29 (0.18)	                & 0.04 (0.10)	         & \textbf{0.43 (0.69)}   & -0.27 (-0.32)\\ \vspace{0.07cm}
$\Delta \bar{\zeta}_{312}$ & -0.12 (-0.07)          & \textbf{-0.93 (-0.58)}	    & \textbf{2.81 (3.69)}	 & \textbf{0.23 (0.46)}   & \textbf{-9.34 (-7.13)}\\ \vspace{0.07cm}
$\Delta \bar{\zeta}_{242}$ & -0.10 (-0.08)          & -0.20 (-0.12)	                & \textbf{0.72 (0.90)}	 & \textbf{0.35 (0.52)}   & \textbf{0.55 (0.53)}\\ \vspace{0.07cm}
$\Delta \bar{\zeta}_{332}$ & 0.06 (0.00)            &  0.07 (-0.00)	                & 0.05 (0.12)	         & 0.09 (0.19)	          & 0.12 (0.16)\\ \vspace{0.07cm}
$\Delta \bar{\zeta}_{422}$ & -0.10 (-0.06)          & -0.17 (-0.09)	                & \textbf{0.76 (1.09)}	 & 0.23 (0.34)	          & \textbf{20.00 (15.27)}\\ \vspace{0.07cm}
$\Delta \bar{\zeta}_{352}$ & -0.06 (-0.03)          & -0.05 (-0.04)	                & 0.11 (0.15)	         & 0.17 (0.23)	          & -0.10 (-0.12)\\ \vspace{0.07cm}
$\Delta \bar{\zeta}_{442}$ & 0.02 (-0.01)           &  0.02 (-0.00)	                & 0.02 (0.04)	         & 0.05 (0.09)	          & -0.16 (-0.21)\\ \vspace{0.07cm}
$\Delta \bar{\zeta}_{532}$ & -0.05 (-0.02)          & -0.05 (-0.03)	                & \textbf{0.32 (0.48)}	 & 0.11 (0.16)	          & \textbf{0.85 (0.46)}\\ \vspace{0.07cm}
$\Delta \bar{\zeta}_{462}$ & -0.03 (-0.01)          & -0.02 (-0.02)	                & 0.06 (0.09)	         & 0.11 (0.15)	          & \textbf{0.57 (0.34)}\\ \vspace{0.07cm}
$\Delta \bar{\zeta}_{552}$ & 0.01 (-0.01)           &  0.01 (-0.00)	                & 0.01 (0.02)	         & 0.03 (0.05)	          & 0.03 (0.05)\\ \vspace{0.07cm}
$\Delta \bar{\zeta}_{642}$ & -0.03 (-0.00)          & -0.02 (-0.01)	                & 0.17 (0.24)	         & 0.08 (0.11)	          & 0.33 (0.35)\\ 
\hline
hexadecapole ($\ell=4$) \\
\hline\vspace{0.07cm}
$\Delta \bar{\zeta}_{044}$ & -0.01 (-0.01)	& 0.00 (0.00)	& 0.04 (0.03)	        & 0.00 (0.00)	         & 0.03 (0.05)\\ \vspace{0.07cm}
$\Delta \bar{\zeta}_{134}$ & -0.02 (-0.02)	& 0.03 (0.03)	& 0.03 (0.02)	        & -0.12 (-0.07)	         & 0.05 (0.07)\\ \vspace{0.07cm}
$\Delta \bar{\zeta}_{224}$ & -0.10 (-0.11)	& 0.05 (0.05)	& \textbf{0.78 (0.49)}	& -0.43 (-0.27)          & \textbf{0.51 (0.67)}\\ \vspace{0.07cm}
$\Delta \bar{\zeta}_{314}$ & 0.08 (0.08)	& -0.05 (-0.05)	& -0.14 (-0.09)	        & 0.37 (0.22)	         & \textbf{1.02 (1.39)}\\ \vspace{0.07cm}
$\Delta \bar{\zeta}_{404}$ & -0.17 (-0.21)	& 0.03 (0.03)	& \textbf{0.54 (0.31)}	& 0.06 (0.04)	         & \textbf{62.42 (70.86)}\\ \vspace{0.07cm}
$\Delta \bar{\zeta}_{154}$ & -0.00 (-0.00)	& -0.01 (-0.02)	& -0.00 (-0.01)	        & 0.00 (-0.00)	         & 0.03 (0.05)\\ \vspace{0.07cm}
$\Delta \bar{\zeta}_{244}$ & -0.02 (-0.02)	& -0.00 (0.00)	& 0.06 (0.08)	        & 0.01 (0.02)	         & -0.02 (-0.04)\\ \vspace{0.07cm}
$\Delta \bar{\zeta}_{334}$ & 0.03 (0.04)	& 0.01 (0.01)	& -0.00 (-0.00)	        & 0.00 (-0.01)	         & 0.00 (-0.02)\\ \vspace{0.07cm}
$\Delta \bar{\zeta}_{424}$ & -0.01 (-0.01)	& -0.00 (-0.00)	& 0.04 (0.01)	        & 0.00 (0.01)	         & \textbf{1.97 (1.04)}\\ \vspace{0.07cm}
$\Delta \bar{\zeta}_{514}$ &-0.01 (-0.02)	& -0.07 (-0.09)	& 0.05 (0.02)	        & 0.02 (-0.00)	         & \textbf{3.29 (4.89)}\\ 
\hline
tetrahexacontapole ($\ell=6$)\\
\hline\vspace{0.07cm}
$\Delta \bar{\zeta}_{066}$ & 0.00 (0.00)	& -0.00 (-0.00)	& -0.00 (-0.00)	& -0.00 (-0.00)	& 0.00 (-0.00) \\ \vspace{0.07cm}
$\Delta \bar{\zeta}_{156}$ &-0.00 (-0.00)	& 0.00 (0.00)	& -0.00 (-0.00)	& -0.00 (-0.00)	& -0.00 (-0.00) \\ \vspace{0.07cm}
$\Delta \bar{\zeta}_{246}$ &-0.00 (-0.00)	& 0.00 (0.00)	& 0.01 (0.01)	& -0.00 (-0.00)	& 0.01 (0.01) \\ \vspace{0.07cm}
$\Delta \bar{\zeta}_{336}$ & 0.01 (0.00)	& -0.00 (-0.00)	& -0.00 (-0.01)	& 0.01 (0.01)	& 0.02 (0.01) \\ \vspace{0.07cm}
$\Delta \bar{\zeta}_{426}$ &-0.00 (-0.00)	& 0.00 (0.00)	& 0.01 (0.01)	& -0.00 (-0.00)	& 0.38 (0.28) \\ \vspace{0.07cm}
$\Delta \bar{\zeta}_{516}$ &-0.00 (-0.00)	& 0.00 (0.00)	& 0.00 (0.00)	& -0.00 (-0.00)	& 0.10 (0.07) \\ \vspace{0.07cm}
$\Delta \bar{\zeta}_{606}$ &-0.00 (-0.00)	& 0.00 (0.00)	& 0.00 (0.00)	& 0.00 (0.00)	& 0.05 (0.04) \\ 
\hline
\end{tabular}
\caption{
Contributions of other 3PCF multipole components to the observed 3PCF multipole components, as manifested through the effect of the window function, where the absence or presence of round brackets represents the NGC or SGC results, respectively. When the contribution to the final result exceeds $0.5\%$, it is written in bold. \Mod{ The value of the same multipole component $\Delta \bar{\zeta}_{\ell_1\ell_2\ell}$ as the measured $\zeta^{(\rm obs)}_{\ell_1\ell_2\ell}$ is larger (smaller) than 100\%, when the total contribution from all the other multipole components is negative (positive).}
}
\label{Table:zeta_window}
\end{table*}

\begin{table*}

\centering
\begin{tabular}{cccccc}
\hline\hline
$\Delta \bar{Q}_{\ell''_1\ell''_2\ell''}\, [\%]$ & $\zeta^{(\rm obs)}_{000}$ & $\zeta^{(\rm obs)}_{110}$ & $\zeta^{(\rm obs)}_{202}$ & $\zeta^{(\rm obs)}_{112}$ & $\zeta^{(\rm obs)}_{404}$ \\
\hline  
monopole ($\ell=0$)\\
\hline\vspace{0.07cm}
$\Delta \bar{Q}_{000}$ &  \textbf{94.94 (90.02)} & \textbf{102.60 (99.94)} & \textbf{86.30 (84.07)}  & \textbf{101.29 (95.33)}	& \textbf{64.25 (72.02)} \\ \vspace{0.07cm}
$\Delta \bar{Q}_{110}$ &  \textbf{9.33 (12.60)}	 & \textbf{1.58 (2.19)}	  & \textbf{5.84 (8.14)}	 & \textbf{4.50 (6.08)}	    & \textbf{3.59 (5.72)}\\\vspace{0.07cm}
$\Delta \bar{Q}_{220}$ &  -0.02 (-0.05)	         & \textbf{0.64 (1.03)} & \textbf{1.00 (1.60)}	  & \textbf{0.81 (1.25)}	    & 0.09 (0.16)\\\vspace{0.07cm}
$\Delta \bar{Q}_{330}$ &  0.07 (0.11)	         & -0.00 (-0.01)	 & \textbf{0.49 (0.77)}	  & \textbf{0.58 (0.90)}	    & -0.02 (-0.04)\\\vspace{0.07cm}
$\Delta \bar{Q}_{440}$ &  0.01 (0.01)	         & 0.01 (0.01)	 & \textbf{0.32 (0.51)}	  & \textbf{0.32 (0.49)}	    & -0.02 (-0.04)\\\vspace{0.07cm}
$\Delta \bar{Q}_{550}$ &  -0.02 (-0.03)	         & -0.01 (-0.03)	 & 0.10 (0.16)	  & 0.20 (0.30)	    & 0.03 (0.05)\\\vspace{0.07cm}
$\Delta \bar{Q}_{660}$ &  -0.03 (-0.04)	         & -0.02 (-0.03)	 & 0.06 (0.09)	  & 0.01 (0.01)	    & 0.00 (0.00)\\\vspace{0.07cm}
$\Delta \bar{Q}_{770}$ &  -0.02 (-0.03)	         & -0.02 (-0.03)	 & -0.00 (-0.00)  & -0.00 (-0.00)	& 0.00 (0.00)\\\vspace{0.07cm}
$\Delta \bar{Q}_{880}$ &  -0.02 (-0.03)	         & -0.00 (-0.01)	 & -0.00 (-0.00)  & -0.00 (-0.00)	& 0.00 (0.00)\\
\hline
quadrupole ($\ell=2$) \\
\hline\vspace{0.07cm}
$\Delta \bar{Q}_{022}$ & \textbf{-4.51 (-2.74)}	& \textbf{-3.81 (-2.40)}  & \textbf{3.16 (1.99)}	& \textbf{-6.96 (-4.28)}	& \textbf{23.87 (17.48)}\\  \vspace{0.07cm}
$\Delta \bar{Q}_{112}$ & \textbf{5.08 (3.33)}	& \textbf{2.56 (1.70)}	  & \textbf{6.49 (4.40)}	& \textbf{4.51 (2.92)}  	& \textbf{-8.67 (-6.62)}\\  \vspace{0.07cm}
$\Delta \bar{Q}_{202}$ & \textbf{-4.00 (-2.44)}	& \textbf{-3.00 (-1.89)}  & \textbf{-4.94 (-3.14)}	& \textbf{-6.70 (-4.13)}	& \textbf{13.74 (9.86)}\\  \vspace{0.07cm}
$\Delta \bar{Q}_{132}$ & -0.28 (-0.16)	        & -0.29 (-0.17)	          & 0.10 (0.04)	            & -0.18 (-0.10)	            & \textbf{1.36 (0.84)}\\ \vspace{0.07cm}
$\Delta \bar{Q}_{222}$ & 0.10 (0.00)	        & 0.21 (0.00)	          & 0.12 (0.01)	            & 0.37 (0.01)	            & \textbf{-0.85 (-0.01)}\\ \vspace{0.07cm}
$\Delta \bar{Q}_{312}$ & -0.12 (-0.07)	        & -0.25 (-0.14)	          & -0.43 (-0.24)	        & 0.11 (0.04)	            & \textbf{1.22 (0.78)}\\ \vspace{0.07cm}
$\Delta \bar{Q}_{242}$ & -0.10 (-0.08)	        & -0.07 (-0.05)	          & 0.04 (0.02)	            & -0.06 (-0.03)	            & 0.36 (0.23)\\ \vspace{0.07cm}
$\Delta \bar{Q}_{332}$ & 0.06 (0.00)	        & 0.05 (0.00)	          & 0.04 (-0.00)	        & 0.10 (0.01)	            & -0.35 (-0.04)\\ \vspace{0.07cm}
$\Delta \bar{Q}_{422}$ & -0.10 (-0.06)	        & -0.06 (-0.03)	          & 0.02 (0.00)	            & 0.03 (0.00)	            & 0.25 (0.18)\\ \vspace{0.07cm}
$\Delta \bar{Q}_{352}$ & -0.06 (-0.03)	        & -0.05 (-0.03)	          & 0.03 (0.01)	            & -0.03 (-0.01)	            & 0.00 (0.00)\\ \vspace{0.07cm}
$\Delta \bar{Q}_{442}$ & 0.02 (-0.01)	        & 0.02 (-0.01)	          & 0.02 (-0.01)	        & 0.03 (-0.01)	            & -0.13 (0.05)\\ \vspace{0.07cm}
$\Delta \bar{Q}_{532}$ & -0.05 (-0.02)	        & -0.05 (-0.02)	          & 0.00 (-0.01)	        & 0.01 (-0.01)	            & 0.17 (0.09)\\  \vspace{0.07cm}
$\Delta \bar{Q}_{462}$ & -0.03 (-0.01)	        & -0.01 (-0.00)	          & 0.01 (-0.00)	        & -0.01 (-0.00)	            & 0.00 (0.00)\\  \vspace{0.07cm}
$\Delta \bar{Q}_{552}$ & 0.01 (-0.01)	        & 0.00 (-0.00)	          & 0.00 (-0.00)	        & 0.01 (-0.01)	            & -0.02 (0.02)\\  \vspace{0.07cm}
$\Delta \bar{Q}_{642}$ & -0.03 (-0.00)	        & -0.01 (-0.00)	          & 0.00 (-0.00)	        & 0.01 (-0.01)	            & 0.13 (0.02)\\  
\hline
hexadecapole ($\ell=4$) \\
\hline\vspace{0.07cm}
$\Delta \bar{Q}_{044}$ &-0.01 (-0.01)	& -0.00 (-0.00)	& \textbf{0.72 (0.89)}	& \textbf{0.49 (0.59)}	& \textbf{1.83 (2.62)}\\ \vspace{0.07cm}
$\Delta \bar{Q}_{134}$ &-0.02 (-0.02)	& -0.00 (-0.00)	& -0.12 (-0.10)	        & -0.22 (-0.20)	        & \textbf{-0.64 (-0.86)}\\ \vspace{0.07cm}
$\Delta \bar{Q}_{224}$ &-0.10 (-0.11)	& 0.01 (0.01)	& \textbf{0.40 (0.45)}	& \textbf{0.63 (0.68)}	& \textbf{0.56 (0.59)}\\ \vspace{0.07cm}
$\Delta \bar{Q}_{314}$ &0.08 (0.08)	    & 0.08 (0.09)	& -0.34 (-0.33)	        & -0.15 (-0.14)	        & \textbf{2.89 (3.14)}\\ \vspace{0.07cm}
$\Delta \bar{Q}_{404}$ &-0.17 (-0.21)	& -0.11 (-0.14)	& \textbf{0.48 (0.60)}	& 0.17 (0.21)	        & \textbf{-4.36 (-6.20)}\\ \vspace{0.07cm}
$\Delta \bar{Q}_{154}$ &-0.00 (-0.00)	& 0.00 (0.00)	& 0.02 (0.03)	        & 0.02 (0.03)	        & 0.20 (0.32)\\ \vspace{0.07cm}
$\Delta \bar{Q}_{244}$ &-0.02 (-0.02)	& -0.00 (-0.00)	& -0.05 (-0.07)	        & -0.06 (-0.07)	        & -0.22 (-0.34)\\ \vspace{0.07cm}
$\Delta \bar{Q}_{334}$ &0.03 (0.04)	    & 0.01 (0.01)	& 0.11 (0.15)	        & 0.13 (0.17)	        & -0.22 (-0.34)\\ \vspace{0.07cm}
$\Delta \bar{Q}_{424}$ &-0.01 (-0.01)	& 0.01 (0.01)	& -0.02 (-0.02)	        & -0.02 (-0.03)	        & 0.02 (0.01)\\ \vspace{0.07cm}
$\Delta \bar{Q}_{514}$ &-0.01 (-0.02)	& -0.01 (-0.01)	& -0.01 (-0.01)	        & 0.02 (0.02)	        & -0.12 (-0.25)\\ 
\hline
tetrahexacontapole ($\ell=6$)\\
\hline\vspace{0.07cm}
$\Delta \bar{Q}_{066}$ &0.00 (0.00)	   & 0.00 (0.00)	& -0.00 (-0.00)	& 0.00 (0.00)	& \textbf{0.57 (0.34)}\\ \vspace{0.07cm}
$\Delta \bar{Q}_{156}$ &-0.00 (-0.00)  & 0.00 (0.00)	& -0.01 (-0.01)	& 0.00 (0.00)	& -0.08 (-0.13)\\ \vspace{0.07cm}
$\Delta \bar{Q}_{246}$ &-0.00 (-0.00)  & -0.00 (-0.00)	& 0.01 (0.01)	& -0.00 (-0.00)	& 0.42 (0.43)\\ \vspace{0.07cm}
$\Delta \bar{Q}_{336}$ &0.01 (0.00)	   & 0.00 (0.00)	& -0.00 (-0.00)	& -0.02 (-0.01)	& \textbf{-0.65 (-0.55)}\\ \vspace{0.07cm}
$\Delta \bar{Q}_{426}$ &-0.00 (-0.00)  & -0.00 (-0.00)	& 0.02 (0.02)	& 0.01 (0.01)	& 0.29 (0.28)\\ \vspace{0.07cm}
$\Delta \bar{Q}_{516}$ &-0.00 (-0.00)  & 0.00 (0.00)	& -0.01 (-0.01)	& -0.01 (-0.01)	& -0.10 (-0.17)\\ \vspace{0.07cm}
$\Delta \bar{Q}_{606}$ &-0.00 (-0.00)  & 0.00 (0.00)	& 0.02 (0.01)	& 0.03 (0.02)	& \textbf{0.60 (0.34)}\\ 
\hline
\end{tabular}
\caption{
Contributions of multipole components of the window function 3PCF to the observed 3PCF multipole components, 
where the absence or presence of round brackets represents the NGC or SGC results, respectively.
When the contribution to the final result exceeds $0.5\%$, it is written in bold.
}
\label{Table:Q_window}

\end{table*}

\subsection{Decomposition in terms of parameters}
\label{Sec:ParameterDecomposition}

In this subsection, we describe how to compute practically the theoretical predictions of the 3PCF in a fast manner. Before showing our procedure, let us first discuss the computational cost to calculate the 3PCF from the bispectrum template model given in Eq.~(\ref{Eq:MainResult}) through the two-dimensional (2D) Hankel transformation in Eq.~(\ref{Eq:B_to_zeta}). First of all, a triple integration is needed to calculate the bispectrum multipole component (\ref{Eq:PB_multipole}). In particular, to compute the Hankel transform, we need to compute the model of the bispectrum multipole over a wide range of wavenumbers $k$. In this paper, we compute $150$ bins on the log-scale over the range $10^{-4}\leq k \leq 10\hk$. The 2D Hankel transform itself can be computed fast using the FFTLog algorithm~\citep{Hamilton:1999uv}. Furthermore, as explained in Section~\ref{Sec:WindowFunctionEffects}, when we compute the theoretical prediction of the 3PCF multipole component of interest after taking into account the window function effect, we have to compute about $10$ other multipole components as well. Thus, calculating one multipole component of the 3PCF, e.g., $\zeta_{000}$ or $\zeta_{202}$, requires $150\times 150\times10=225,000$ triple integrals, which is so computationally expensive that it is not well suited for a fitting analysis. 

To speed up the calculation of the theoretical predictions for 3PCFs, we linearize the fitting parameters on which the 3PCF depends by making the following two assumptions. The first is to fix the shape of the linear matter power spectrum $P_{\rm lin}$ and its smooth version $P_{\rm nw}$ contained in the bispectrum template model to their prediction by a fiducial cosmological model introduced in the introduction. Second, we rewrite the two-dimensional damping function describing the nonlinear attenuation of the BAO, ${\cal D}(\VEC{k})$ (\ref{Eq:D}), as
\begin{eqnarray}
    {\cal D}(\VEC{k}) = \exp\left( -\frac{k^2(1-\mu^2)\Sigma_{\perp}^2+k^2\mu^2\Sigma_{\parallel}^2}{4}
    \right),
    \label{Eq:Damping}
\end{eqnarray}
and fix the radial and transverse smoothing parameters, $\Sigma_{\parallel}$ and $\Sigma_{\perp}$, to the linear theory predictions (\ref{Eq:D}) or the best-fitting values (see Section~\ref{Sec:MethodologyFor2PCFAnalysis}). Under these two assumptions, the bispectrum template model depends on five free parameters
\begin{eqnarray}
    \{(b_1\sigma_8), (f\sigma_8), \sigma_8, (b_{2}\sigma_8^2),(b_{\rm K2}\sigma_8^2)\},
\end{eqnarray}
except for the AP parameters, which will be discussed in detail in Section~\ref{Sec:APeffects}. Then, we can represent the model with a linear combination of those parameters as coefficients:
\begin{eqnarray}
    Z_1(\VEC{k}_1)Z_1(\VEC{k}_2)Z_2(\VEC{k}_1,\VEC{k}_2) \sigma_8^4
    \hspace{-0.25cm}&=&\hspace{-0.25cm} \sum_{p=1}^{14} X^{(p)} B^{(p)}(\VEC{k}_1,\VEC{k}_2),
        \label{Eq:Bp}
\end{eqnarray}
where 
\begin{eqnarray}
    X^{(1)}  \hspace{-0.25cm}&=&\hspace{-0.25cm}  (b_1\sigma_8)^3 \sigma_8 \nonumber \\
    X^{(2)}  \hspace{-0.25cm}&=&\hspace{-0.25cm}  (b_1\sigma_8)^2(f\sigma_8) \sigma_8\nonumber \\
    X^{(3)}  \hspace{-0.25cm}&=&\hspace{-0.25cm}  (b_1\sigma_8)(f\sigma_8)^2 \sigma_8\nonumber \\
    X^{(4)}  \hspace{-0.25cm}&=&\hspace{-0.25cm}  (b_1\sigma_8)^2(b_2\sigma_8^2) \nonumber \\
    X^{(5)}  \hspace{-0.25cm}&=&\hspace{-0.25cm}  (b_1\sigma_8)(f\sigma_8)(b_2\sigma_8^2) \nonumber \\
    X^{(6)}  \hspace{-0.25cm}&=&\hspace{-0.25cm}  (f\sigma_8)^2 (b_2\sigma_8^2)\nonumber \\
    X^{(7)}  \hspace{-0.25cm}&=&\hspace{-0.25cm}  (b_1\sigma_8)^2(b_{\rm K2}\sigma_8^2) \nonumber \\
    X^{(8)}  \hspace{-0.25cm}&=&\hspace{-0.25cm}  (b_1\sigma_8)(f\sigma_8)(b_{\rm K2}\sigma_8^2) \nonumber \\
    X^{(9)}  \hspace{-0.25cm}&=&\hspace{-0.25cm}  (f\sigma_8)^2 (b_{\rm K2}\sigma_8^2)\nonumber \\
    X^{(10)} \hspace{-0.25cm}&=&\hspace{-0.25cm}  (b_1\sigma_8)^3 (f\sigma_8)\nonumber \\
    X^{(11)} \hspace{-0.25cm}&=&\hspace{-0.25cm}  (b_1\sigma_8)^2 (f\sigma_8)^2\nonumber \\
    X^{(12)} \hspace{-0.25cm}&=&\hspace{-0.25cm}  (b_1\sigma_8) (f\sigma_8)^3\nonumber \\
    X^{(13)} \hspace{-0.25cm}&=&\hspace{-0.25cm}  (f\sigma_8)^4\nonumber \\
    X^{(14)} \hspace{-0.25cm}&=&\hspace{-0.25cm}  (f\sigma_8)^3\, \sigma_8,
\end{eqnarray}
and
\begin{eqnarray}
    B^{(1)}  \hspace{-0.25cm}&=&\hspace{-0.25cm} F_2(\VEC{k}_1,\VEC{k}_2), \nonumber\\
    B^{(2)}  \hspace{-0.25cm}&=&\hspace{-0.25cm} F_2(\VEC{k}_1,\VEC{k}_2)(V_1(\VEC{k}_1)+V_1(\VEC{k}_2)) + V_2(\VEC{k}_1,\VEC{k}_2),  \nonumber\\
    B^{(3)}  \hspace{-0.25cm}&=&\hspace{-0.25cm} F_2(\VEC{k}_1,\VEC{k}_2) V_1(\VEC{k}_1) V_1(\VEC{k}_2)  
    + V_2(\VEC{k}_1,\VEC{k}_2) ( V_1(\VEC{k}_1) + V_1(\VEC{k}_2) ),  \nonumber\\
    B^{(4)}  \hspace{-0.25cm}&=&\hspace{-0.25cm}  (1/2),  \nonumber\\
    B^{(5)}  \hspace{-0.25cm}&=&\hspace{-0.25cm}  (1/2) (V_1(\VEC{k}_1) + V_1(\VEC{k}_2)),  \nonumber\\
    B^{(6)}  \hspace{-0.25cm}&=&\hspace{-0.25cm}  (1/2)  V_1(\VEC{k}_1) V_1(\VEC{k}_2), \nonumber\\
    B^{(7)}  \hspace{-0.25cm}&=&\hspace{-0.25cm}  K_2(\VEC{k}_1,\VEC{k}_2),  \nonumber\\
    B^{(8)}  \hspace{-0.25cm}&=&\hspace{-0.25cm}  K_2(\VEC{k}_1,\VEC{k}_2) ( V_1(\VEC{k}_1) + V_1(\VEC{k}_2) ), \nonumber\\
    B^{(9)}  \hspace{-0.25cm}&=&\hspace{-0.25cm}  K_2(\VEC{k}_1,\VEC{k}_2) V_1(\VEC{k}_1) V_1(\VEC{k}_2),  \nonumber\\
    B^{(10)} \hspace{-0.25cm}&=&\hspace{-0.25cm}  DV(\VEC{k}_1,\VEC{k}_2),  \nonumber\\
    B^{(11)} \hspace{-0.25cm}&=&\hspace{-0.25cm}  DV(\VEC{k}_1,\VEC{k}_2)  ( V_1(\VEC{k}_1) + V_1(\VEC{k}_2) )
    + V_{11}(\VEC{k}_1,\VEC{k}_2),  \nonumber\\
    B^{(12)} \hspace{-0.25cm}&=&\hspace{-0.25cm}  DV(\VEC{k}_1,\VEC{k}_2) V_1(\VEC{k}_1) V_1(\VEC{k}_2) 
    + V_{11}(\VEC{k}_1,\VEC{k}_2)(V_1(\VEC{k}_1)+V_1(\VEC{k}_2)),  \nonumber\\
    B^{(13)} \hspace{-0.25cm}&=&\hspace{-0.25cm}  V_{11}(\VEC{k}_1,\VEC{k}_2) V_1(\VEC{k}_1) V_1(\VEC{k}_2),  \nonumber\\
    B^{(14)} \hspace{-0.25cm}&=&\hspace{-0.25cm}  V_2(\VEC{k}_1,\VEC{k}_2) V_1(\VEC{k}_1) V_1(\VEC{k}_2), 
\end{eqnarray}
with
\begin{eqnarray}
    F_2(\VEC{k}_1,\VEC{k}_2)\hspace{-0.25cm}
    &=&\hspace{-0.25cm} \frac{5}{7} + \frac{\hat{k}_1\cdot\hat{k}_2}{2}\left( \frac{k_1}{k_2} + \frac{k_2}{k_1} \right) + \frac{2}{7}(\hat{k}_1\cdot\hat{k}_2)^2    \nonumber \\
    G_2(\VEC{k}_1,\VEC{k}_2)\hspace{-0.25cm}
    &=&\hspace{-0.25cm} \frac{3}{7} + \frac{\hat{k}_1\cdot\hat{k}_2}{2}\left( \frac{k_1}{k_2} + \frac{k_2}{k_1} \right) + \frac{4}{7}(\hat{k}_1\cdot\hat{k}_2)^2    \nonumber \\
    V_1(\VEC{k})   \hspace{-0.25cm}
    &=&\hspace{-0.25cm} (\hat{k}\cdot\hat{n})^2\nonumber \\
    V_2(\VEC{k}_1,\VEC{k}_2)   \hspace{-0.25cm}
    &=&\hspace{-0.25cm} (\hat{k}_{12}\cdot\hat{n})^2 G_2(\VEC{k}_1,\VEC{k}_2)\nonumber \\
    V_{11}(\VEC{k}_1,\VEC{k}_2)   \hspace{-0.25cm}
    &=&\hspace{-0.25cm} \frac{1}{2} (\VEC{k}_{12}\cdot\hat{n})^2 \frac{\VEC{k}_1\cdot\hat{n}}{k_1^2} \frac{\VEC{k}_2\cdot\hat{n}}{k_2^2} \nonumber \\
    DV(\VEC{k}_1,\VEC{k}_2)   \hspace{-0.25cm}
    &=&\hspace{-0.25cm} \frac{1}{2} (\VEC{k}_{12}\cdot\hat{n})  \left[ \frac{\VEC{k}_1\cdot\hat{n}}{k_1^2} + \frac{\VEC{k}_2\cdot\hat{n}}{k_2^2}  \right]\nonumber \\
    K_2(\VEC{k}_1,\VEC{k}_2) \hspace{-0.25cm}
    &=&\hspace{-0.25cm} \left( \hat{k}_1\cdot\hat{k}_2 \right)^2 - \frac{1}{3}.
\end{eqnarray}
Finally, we replace $P_{\rm lin}$ and $P_{\rm nw}$ appearing in Eq.~(\ref{Eq:MainResult}) with $\sigma_8^2 P_{\rm lin, fid}/\sigma_{8,\rm fid}^2$ and $\sigma_8^2 P_{\rm nw, fid}/\sigma_{8,\rm fid}^2$, where $P_{\rm lin, fid}$, $P_{\rm nw, fid}$ and $\sigma_{8,\rm fid}^2$ are computed using fixed fiducial cosmological parameters, and substitute Eq.~(\ref{Eq:Bp}) into Eq.~(\ref{Eq:MainResult}). We can then pre-calculate all the other parts of the template model (\ref{Eq:MainResult}) except $X^{(p)}$ and create a table of the resulting data. All that is left to do is to load that table when we perform our cosmological analysis.

\subsection{AP effects}
\label{Sec:APeffects}

\begin{table*}

\centering
\begin{tabular}{cccccc}
\hline\hline
$\Delta \bar{\zeta}^{(\rm AP)}_{\ell'_1\ell'_2\ell'}(\varepsilon=0.02,\ {\rm or}\,\varepsilon=-0.02)\, [\%]$ & $\zeta^{(\rm AP)}_{000}$ & $\zeta^{(\rm AP)}_{110}$ & $\zeta^{(\rm AP)}_{202}$ & $\zeta^{(\rm AP)}_{112}$ & $\zeta^{(\rm AP)}_{404}$ \\
\hline  
monopole ($\ell=0$)\\
\hline
$\Delta \bar{\zeta}^{(\rm AP)}_{000}$ & \textbf{103.17 (98.20)} &	-0.00 (-0.00)	        &\textbf{-3.72 (3.48)}	&-0.00 (-0.00)	        &\textbf{0.33 (0.77)}\\
$\Delta \bar{\zeta}^{(\rm AP)}_{110}$ & -0.00 (-0.00)           &  \textbf{103.20 (98.43)}	&-0.00 (-0.00)	        &\textbf{-4.11 (3.73)}	&0.00 (0.00)\\
$\Delta \bar{\zeta}^{(\rm AP)}_{220}$ & 0.20 (0.21)	            & -0.00 (-0.00)	            &\textbf{0.41 (-0.49)}	&-0.00 (-0.00)	        &\textbf{-1.82 (-4.40)}\\
$\Delta \bar{\zeta}^{(\rm AP)}_{330}$ & -0.00 (-0.00)           &  0.10 (0.10)	            &-0.00 (-0.00)	        &-0.00 (-0.00)	        &0.00 (0.00)\\
$\Delta \bar{\zeta}^{(\rm AP)}_{440}$ & -0.00 (-0.00)           &  -0.00 (-0.00)	        &-0.00 (-0.00)	        &-0.00 (-0.00)	        &\textbf{0.44 (0.93)}\\
$\Delta \bar{\zeta}^{(\rm AP)}_{550}$ & -0.00 (-0.00)           &  -0.00 (-0.00)	        &-0.00 (-0.00)	        &-0.00 (-0.00)	        &0.00 (0.00)\\
$\Delta \bar{\zeta}^{(\rm AP)}_{660}$ & -0.00 (-0.00)           &  -0.00 (-0.00)	        &-0.00 (-0.00)	        &-0.00 (-0.00)	        &0.00 (0.00)\\
$\Delta \bar{\zeta}^{(\rm AP)}_{770}$ & -0.00 (-0.00)           &  -0.00 (-0.00)	        &-0.00 (-0.00)	        &-0.00 (-0.00)	        &0.00 (0.00)\\
$\Delta \bar{\zeta}^{(\rm AP)}_{880}$ & -0.00 (-0.00)           &  -0.00 (-0.00)	        &-0.00 (-0.00)	        &-0.00 (-0.00)	        &0.00 (0.00)\\
\hline
quadrupole ($\ell=2$) \\
\hline
$\Delta \bar{\zeta}^{(\rm AP)}_{022}$ & \textbf{5.42 (-5.73)}	& -0.00 (-0.00)	        &0.18 (0.18)	            &-0.00 (-0.00)	         &0.00 (0.00) \\
$\Delta \bar{\zeta}^{(\rm AP)}_{112}$ & -0.00 (-0.00)	        & \textbf{-1.73 (1.68)}	&-0.00 (-0.00)	            &\textbf{104.59 (96.62)} &0.00 (0.00) \\
$\Delta \bar{\zeta}^{(\rm AP)}_{202}$ & \textbf{-9.30 (9.42)}	& -0.00 (-0.00)	        &\textbf{103.98} (97.26)	&-0.00 (-0.00)	         &\textbf{43.73 (-99.18)} \\
$\Delta \bar{\zeta}^{(\rm AP)}_{132}$ & -0.00 (-0.00)	        & \textbf{2.80 (-3.01)}	&-0.00 (-0.00)	            &\textbf{0.02 (0.64)}	 &0.00 (0.00) \\
$\Delta \bar{\zeta}^{(\rm AP)}_{222}$ & 0.10 (0.11)	            & -0.00 (-0.00)	        &0.26 (-0.31)	            &-0.00 (-0.00)	         &\textbf{-0.97 (-2.33)}\\
$\Delta \bar{\zeta}^{(\rm AP)}_{312}$ & 0.00 (0.00)	            & \textbf{-4.16 (4.25)}	&-0.00 (-0.00)	            &\textbf{-0.84 (0.65)}	 &0.00 (0.00) \\
$\Delta \bar{\zeta}^{(\rm AP)}_{242}$ & 0.00 (0.00)	            & -0.00 (-0.00)	        &-0.11 (-0.11)	            &-0.00 (-0.00)	         &0.00 (0.00) \\
$\Delta \bar{\zeta}^{(\rm AP)}_{332}$ & 0.00 (0.00)	            & 0.04 (0.06)	        &-0.00 (-0.00)	            &0.03 (0.03)	         &0.00 (0.00) \\
$\Delta \bar{\zeta}^{(\rm AP)}_{422}$ & 0.00 (0.00)	            & -0.00 (-0.00)	        &0.13 (0.12)	            &-0.00 (-0.00)	         &\textbf{-13.59 (33.13)} \\
$\Delta \bar{\zeta}^{(\rm AP)}_{352}$ & 0.00 (0.00)	            & -0.00 (-0.00)	        &-0.00 (-0.00)	            &-0.00 (-0.00)	         &0.00 (0.00) \\
$\Delta \bar{\zeta}^{(\rm AP)}_{442}$ & 0.00 (0.00)	            & -0.00 (-0.00)	        &-0.00 (-0.00)	            &-0.00 (-0.00)	         &\textbf{0.27 (0.65)}\\
$\Delta \bar{\zeta}^{(\rm AP)}_{532}$ & 0.00 (0.00)	            & -0.00 (-0.00)	        &-0.00 (-0.00)	            &-0.00 (-0.00)	         &0.00 (0.00) \\
$\Delta \bar{\zeta}^{(\rm AP)}_{462}$ & 0.00 (0.00)	            & -0.00 (-0.00)	        &-0.00 (-0.00)	            &-0.00 (-0.00)	         &0.00 (0.00) \\
$\Delta \bar{\zeta}^{(\rm AP)}_{552}$ & 0.00 (0.00)	            & -0.00 (-0.00)	        &-0.00 (-0.00)	            &-0.00 (-0.00)	         &0.00 (0.00) \\
$\Delta \bar{\zeta}^{(\rm AP)}_{642}$ & 0.00 (0.00)	            & -0.00 (-0.00)	        &-0.00 (-0.00)	            &-0.00 (-0.00)	         &0.01 (0.08)\\
\hline
hexadecapole ($\ell=4$) \\
\hline
$\Delta \bar{\zeta}^{(\rm AP)}_{044}$ & -0.06 (-0.06)&	-0.00 (-0.00)	&-0.00 (-0.00)	        &-0.00 (-0.00)	        &0.00 (0.00)\\
$\Delta \bar{\zeta}^{(\rm AP)}_{134}$ & -0.00 (-0.00)&	0.00 (0.00)  	&-0.00 (-0.00)	        &0.04 (-0.07)	        &0.00 (0.00)\\
$\Delta \bar{\zeta}^{(\rm AP)}_{224}$ & 0.02 (0.02)  &   -0.00 (-0.00)	&-0.19 (0.17)	        &-0.00 (-0.00)	        &\textbf{-0.26 (-0.62)}\\
$\Delta \bar{\zeta}^{(\rm AP)}_{314}$ & -0.00 (-0.00)&	0.01 (0.01)	    &-0.00 (-0.00)	        &\textbf{-1.19 (1.14)}	&0.00 (0.00)\\
$\Delta \bar{\zeta}^{(\rm AP)}_{404}$ & 0.06 (0.06)  &   -0.00 (-0.00)	&\textbf{-1.11 (1.06)}	&-0.00 (-0.00)	        &\textbf{76.00 (161.08)}\\
$\Delta \bar{\zeta}^{(\rm AP)}_{154}$ & -0.00 (-0.00)&	-0.04 (-0.04)	&-0.00 (-0.00)	        &-0.00 (-0.00)	        &0.00 (0.00)\\
$\Delta \bar{\zeta}^{(\rm AP)}_{244}$ & -0.00 (-0.00)&	-0.00 (-0.00)	&-0.01 (-0.01)	        &-0.00 (-0.00)	        &0.00 (0.00)\\
$\Delta \bar{\zeta}^{(\rm AP)}_{334}$ & -0.00 (-0.00)&	0.01 (0.01)	    &-0.00 (-0.00)	        &0.01 (0.01)	        &0.00 (0.00)\\
$\Delta \bar{\zeta}^{(\rm AP)}_{424}$ & -0.00 (-0.00)&	-0.00 (-0.00)	&0.01 (0.01)	        &-0.00 (-0.00)	        &\textbf{-2.17 (5.18)}\\
$\Delta \bar{\zeta}^{(\rm AP)}_{514}$ & -0.00 (-0.00)&	0.03 (0.03)	    &-0.00 (-0.00)	        &-0.00 (-0.00)	        &0.00 (0.00)\\
\hline
tetrahexacontapole ($\ell=6$)\\
\hline
$\Delta \bar{\zeta}^{(\rm AP)}_{066}$ & -0.00 (-0.00)	&-0.00 (-0.00)	&-0.00 (-0.00)	&-0.00 (-0.00)	   & 0.00 (0.00)\\
$\Delta \bar{\zeta}^{(\rm AP)}_{156}$ & -0.00 (-0.00)	&-0.00 (-0.00)	&-0.00 (-0.00)	&-0.00 (-0.00)	   & 0.00 (0.00)\\
$\Delta \bar{\zeta}^{(\rm AP)}_{246}$ & -0.00 (-0.00)	&-0.00 (-0.00)	&-0.00 (-0.00)	&-0.00 (-0.00)	   & 0.00 (0.00)\\
$\Delta \bar{\zeta}^{(\rm AP)}_{336}$ & -0.00 (-0.00)	&-0.00 (-0.00)	&-0.00 (-0.00)	&0.00 (0.00)	   & 0.00 (0.00)\\
$\Delta \bar{\zeta}^{(\rm AP)}_{426}$ & -0.00 (-0.00)	&-0.00 (-0.00)	&0.00 (0.00)	&    -0.00 (-0.00) & \textbf{-0.32 (0.77)}\\
$\Delta \bar{\zeta}^{(\rm AP)}_{516}$ & -0.00 (-0.00)	&-0.00 (-0.00)	&-0.00 (-0.00)	&0.00 (0.00)	   & 0.00 (0.00)\\
$\Delta \bar{\zeta}^{(\rm AP)}_{606}$ & -0.00 (-0.00)	&-0.00 (-0.00)	&0.00 (0.00)	&    -0.00 (-0.00) & -0.17 (0.33)\\
\hline
\end{tabular}
\caption{
Contributions of other 3PCF multipole components to the 3PCF multipole components of interest, as manifested through the AP effect,
where the absence or presence of round brackets represents the results for $\varepsilon=0.02$ and $\varepsilon=-0.02$, respectively.
When the contribution to the final result exceeds $0.5\%$, it is written in bold.
}
\label{Table:zeta_AP}

\end{table*}

In this section, we present a fast method for calculating the 3PCF changes due to the AP effect, which is caused by the use of incorrect cosmological parameters that are different from the true ones when calculating the distance to the galaxy from the measured redshift of the galaxy. In other words, the scale dependence of the theoretical model must be rescaled to compare the theoretical model calculated using the true cosmological parameters with the measurement with the wrong cosmological parameters. In particular, the anisotropic component of the AP effect gives rise to an additional angular dependence of the LOS direction, so the triple integral in Eq.~(\ref{Eq:PB_multipole}) have to be recalculated each time in order to calculate the change of the bispectrum or 3PCF multipole component due to the AP effect. This fact makes it unsuitable for the method proposed in Section~\ref{Sec:ParameterDecomposition}, which involves pre-building a table of the results of the 3PCF calculation. Therefore, we need to come up with a new approximation method to account for the anisotropic component of the AP effect.

To characterize the difference between the true and fiducial values of three-space coordinates (wavevectors), we usually introduce the following two AP parameters:
\begin{eqnarray}
	\alpha_{\perp} \hspace{-0.25cm}&=&\hspace{-0.25cm}
	\frac{D_{\rm A}(z)}{D_{\rm A,\, \rm fid}(z)} \nonumber \\
	\alpha_{\parallel} \hspace{-0.25cm}&=&\hspace{-0.25cm}  \frac{H_{\rm fid}(z)}{H(z)},
\end{eqnarray}
where $D_{\rm A}$ and $H$ are the angular diameter distance and the Hubble parameter estimated at redshift $z$ in the observed region, respectively, and the subscript ``fid'' stands for ``fiducial''. As an alternative parameterization of geometric distortions, we employ isotropic dilation $\alpha$ and anisotropic warping $\varepsilon$ parameters, defined as~\citep{Padmanabhan:2008ag}
\begin{eqnarray}
    \alpha^3 \hspace{-0.25cm} &=& \hspace{-0.25cm} \alpha_{\perp}^2 \alpha_{\parallel}\nonumber \\
	\left(1 + \varepsilon \right)^3 \hspace{-0.25cm} &=& \hspace{-0.25cm} \frac{\alpha_{\parallel}}{\alpha_{\perp}}.
	\label{Eq:AAEE}
\end{eqnarray}
Using $\alpha$ and $\varepsilon$, the true wavevector is represented as
\begin{eqnarray}
	\VEC{k}_{\rm true} = \left( \frac{1+\varepsilon}{\alpha} \right) k
	\left[ \hat{k} + \left( \hat{k}\cdot\hat{n} \right) \left[ \left( 1 + \varepsilon \right)^{-3} - 1 \right]\, \hat{n}\right],
\end{eqnarray}
and therefore, the corresponding wavenumber $k_{\rm true}=|\VEC{k}_{\rm true}|$ is given by
\begin{eqnarray}
	k_{\rm true}
    =
	k \left( \frac{1+\varepsilon}{\alpha} \right) \left[  1 + \left( \hat{k}\cdot\hat{n} \right)^2 \left[ \left( 1 + \varepsilon \right)^{-6}-1 \right] \right]^{\frac{1}{2}}.
\end{eqnarray}
We then obtain the multipole components of the bispectrum, which contains the AP effect:
\begin{eqnarray}
	B^{(\rm AP)}_{\ell_1\ell_2\ell}(k_1,k_2)
	\hspace{-0.25cm}&=&\hspace{-0.25cm}
    \frac{4\pi\, h_{\ell_1\ell_2\ell}^2}{\alpha^6}
	\int \frac{d^2\hat{k}_1}{4\pi}\int \frac{d^2\hat{k}_2}{4\pi}\int \frac{d^2\hat{n}}{4\pi} \nonumber \\
	\hspace{-0.25cm}&\times&\hspace{-0.25cm}
    {\cal S}^*_{\ell_1\ell_2\ell}(\hat{k}_1,\hat{k}_2,\hat{n}) B(\VEC{k}_{\rm true, 1},\VEC{k}_{\rm true, 2}).
	\label{Eq:PB_multipole_APeffect}
\end{eqnarray}
where the pre-factor $1/\alpha^6$ are because the bispectrum have dimensions of the square of the survey volume $V^2$. 

In the following, we explain how to compute the bispectrum multipoles, including the anisotropic warping $\varepsilon$ of the AP effect, in a fast approximation. 
First, we decompose the true wavenumber into 
\begin{eqnarray}
	k_{\rm true} = k' \left[ 1 + \Delta k\right],
\end{eqnarray}
where
\begin{eqnarray}
	\hspace{-0.40cm}k'
    \hspace{-0.25cm}&=&\hspace{-0.25cm} 
	k \left( \frac{1+\varepsilon}{\alpha} \right) \left[  1 + \frac{1}{3}\left[ \left( 1 + \varepsilon \right)^{-6}-1 \right] \right]^{\frac{1}{2}} \nonumber \\
    \hspace{-0.40cm}\Delta k
    \hspace{-0.25cm}&=&\hspace{-0.25cm} 
	\Bigg\{1 +  \frac{\left[  \frac{2}{3} {\cal L}_{2}(\hat{k}\cdot\hat{n})  \left[ \left( 1 + \varepsilon \right)^{-6}-1 \right] \right]}
	{\left[  1 + \frac{1}{3} \left[ \left( 1 + \varepsilon \right)^{-6}-1 \right] \right]} \Bigg\}^{\frac{1}{2}} - 1.
\end{eqnarray}
Thus, $\varepsilon$ as well as $\alpha$ rescale wavenumber in $k'=k'(\alpha,\varepsilon)$, and also generates an additional angular-dependence of the LOS direction through $\Delta k = \Delta k(\varepsilon,\hat{n})$. Second, we expand $B(\VEC{k}_{1,\rm true},\VEC{k}_{2,\rm true})$ in TripoSHs with $\hat{k}_{1,\rm true}$ and $\hat{k}_{2,\rm true}$ as variables,
\begin{eqnarray}
	B(\VEC{k}_{1,\rm true},\VEC{k}_{2,\rm true}) 
    \hspace{-0.25cm}&=&\hspace{-0.25cm}
    \sum_{\ell_1+\ell_2+\ell={\rm even}}
    B_{\ell_1\ell_2\ell}(k_{1, \rm true},k_{2,\rm true}) \nonumber \\
    \hspace{-0.25cm}&\times&\hspace{-0.25cm}
    {\cal S}_{\ell_1\ell_2\ell}(\hat{k}_{1, \rm true},\hat{k}_{2, \rm true},\hat{n}),
    \label{Eq:PB_AP_1}
\end{eqnarray}
where $\hat{k}_{\rm true} = \VEC{k}_{\rm true}/k_{\rm true}$. Third, we expand $B(k_{1,\rm true},k_{2,\rm true})$ around the point $k_{\rm true}=k'$ and obtain
\begin{eqnarray}
	B_{\ell_1\ell_2\ell}(k_{1,\rm true},k_{2,\rm true}) 
    \hspace{-0.25cm}&=&\hspace{-0.25cm}
    \sum_{n,m} \frac{1}{n! m!} B_{\ell_1\ell_2\ell}^{(n)(m)}(k'_1,k'_2)  \nonumber \\ 
	\hspace{-0.25cm}&\times&\hspace{-0.25cm}
	\left[ \Delta k_1 \right]^n  \left[ \Delta k_2 \right]^m,
	\label{Eq:PB_AP_2}
\end{eqnarray}
where
\begin{eqnarray}
   \hspace{-0.15cm} B_{\ell_1\ell_2\ell}^{(n)(m)}(k_1,k_2) = 
    \left( k_1 \frac{\partial}{\partial k_1} \right)^n\hspace{-0.15cm}
    \left( k_2 \frac{\partial}{\partial k_2} \right)^m \hspace{-0.15cm}B_{\ell_1\ell_2\ell}(k_1,k_2).
\end{eqnarray}
Finally, substituting Eqs.~(\ref{Eq:PB_AP_1}) and~(\ref{Eq:PB_AP_2}) into Eq.~(\ref{Eq:PB_multipole_APeffect}) leads to
\begin{eqnarray}
	B_{\ell_1\ell_2\ell}^{(\rm AP)}(k_1,k_2)
    \hspace{-0.25cm}&=& \hspace{-0.25cm}
	\frac{1}{\alpha^6} \sum_{n,m} \sum_{\ell'_1\ell'_2\ell'} 
    T_{\ell_1\ell_2\ell}^{\ell_1'\ell_2'\ell'(n)(m)}(\varepsilon) \nonumber \\
	\hspace{-0.25cm}&\times&\hspace{-0.25cm} 
    B_{\ell'_1\ell'_2\ell'}^{(n)(m)}(k'_1,k'_2),
\end{eqnarray}
where
\begin{eqnarray}
    \hspace{-0.4cm} T_{\ell_1\ell_2\ell}^{\ell_1'\ell_2'\ell'(n)(m)}(\varepsilon)
    \hspace{-0.25cm}&=&\hspace{-0.25cm} 
     \frac{(4\pi) h_{\ell_1\ell_2\ell}^2}{n! m!}
    \int \frac{d^2\hat{k}_1}{4\pi}\int \frac{d^2\hat{k}_2}{4\pi}\int \frac{d^2\hat{n}}{4\pi}  \nonumber \\
	\hspace{-0.25cm}&\times&\hspace{-0.25cm}
	{\cal S}^*_{\ell_1\ell_2\ell}(\hat{k}_1,\hat{k}_2,\hat{n}) 
	{\cal S}_{\ell'_1\ell'_2\ell'}(\hat{k}_{1,\rm true},\hat{k}_{2, \rm true},\hat{n}) 
    \nonumber \\
	\hspace{-0.25cm}&\times&\hspace{-0.25cm}
    \left[ \Delta k_1 \right]^n
    \left[ \Delta k_2 \right]^m.
\end{eqnarray}
The function $T_{\ell_1\ell_2\ell}^{\ell_1'\ell_2'\ell'(n)(m)}$ depends only on $\varepsilon$ and not on any other cosmological parameter. Therefore, it is possible to compute a function $T$ in various values of $\varepsilon$ and create a table of data for it beforehand; $B_{\ell_1\ell_2\ell}^{(n)(m)}$ can be expanded on the cosmological parameters, as described in Section~\ref{Sec:ParameterDecomposition}, and a table of data can also be created by precomputing the parts of the function that do not depend on the cosmological parameters. This approximation method thus allows us to precompute all the functions needed for data analysis. When we actually analyze the data, we read these tables and use a numerical interpolation method to map them to any $\varepsilon$ value. 

Through the 2D Hankel transform (\ref{Eq:B_to_zeta}), we derive the 3PCF multipoles including the AP effect:
\begin{eqnarray}
    \zeta^{(\rm AP)}_{\ell_1\ell_2\ell}(r_1,r_2)
    \hspace{-0.25cm}&=&\hspace{-0.25cm} \frac{1}{E^6(\varepsilon)}\sum_{n,m} \sum_{\ell_1'\ell_2'\ell'} 
    T_{\ell_1\ell_2\ell}^{\ell'_1\ell'_2\ell'(n)(m)}(\varepsilon)
    \nonumber \\
    \hspace{-0.25cm}&\times&\hspace{-0.25cm} 	\zeta^{(n)(m)}_{\ell_1\ell_2;\ell'_1\ell'_2\ell'}(r'_1, r'_2),
    \label{Eq:zeta_approx_AP}
\end{eqnarray}
where
\begin{eqnarray}
	r'(\alpha,\varepsilon) 
    \hspace{-0.25cm}&=&\hspace{-0.25cm}  
    r \left( \frac{\alpha}{1+\varepsilon} \right) \left[  1 + \frac{1}{3}\left[ \left( 1 + \varepsilon \right)^{-6}-1 \right] \right]^{ - \frac{1}{2}}, \\
    E(\varepsilon) 
    \hspace{-0.25cm} &=& \hspace{-0.25cm}
    \left( 1+\varepsilon \right) \left[  1 + \frac{1}{3}\left[ \left( 1 + \varepsilon \right)^{-6}-1 \right] \right]^{\frac{1}{2}},
\end{eqnarray}
and
\begin{eqnarray}
	\hspace{-0.50cm}
    \zeta^{(n)(m)}_{\ell_1\ell_2;\ell'_1\ell'_2\ell'}(r_1,r_2) 
    \hspace{-0.25cm}&=&\hspace{-0.25cm}
    i^{\ell_1+\ell_2} \int \frac{dk_1k_1^2}{2\pi^2} \int \frac{dk_2k_2^2}{2\pi^2} \nonumber \\
	\hspace{-0.25cm}&\times&\hspace{-0.25cm}
    j_{\ell_1}(r_1k_1) j_{\ell_2}(r_2k_2) B^{(n)(m)}_{\ell'_1\ell'_2\ell'}(k_1,k_2).
\end{eqnarray}

\begin{figure*}
	\scalebox{1.0}{\includegraphics[width=\textwidth]{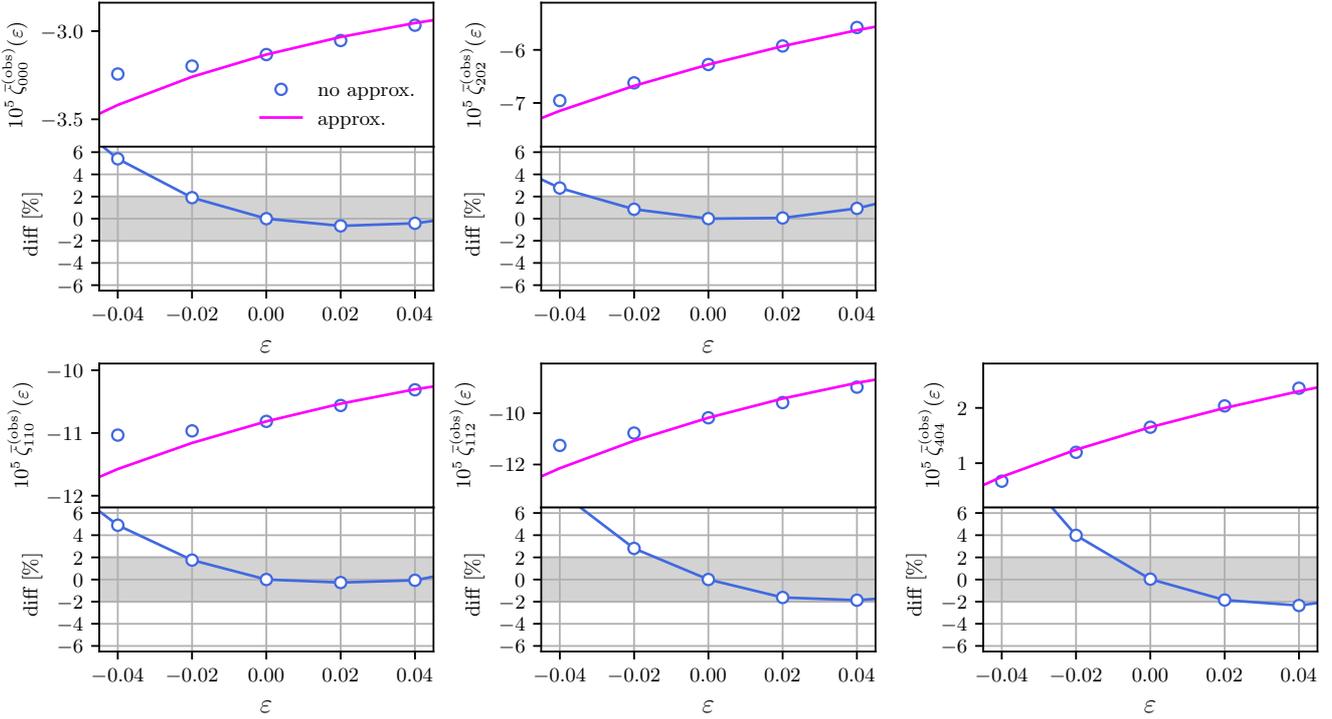}}
    \caption{Averaged 3PCF multipole components $\bar{\zeta}_{\ell_1\ell_2\ell}^{(\rm obs)}$ (\ref{Eq:zeta_mean}) as a function of $\varepsilon$.
    ``no approx.'' means that we calculate Eq.~(\ref{Eq:PB_multipole_APeffect}) directly without using the approximate method of calculating the AP effect described in the text, 
    while ``approx.'' means that we use Eq.~(\ref{Eq:zeta_approx_AP}). 
    ``diff'' represents the relative change in those two theoretical predictions.
	}
	\label{fig:zeta_E}
\end{figure*}

To investigate the convergence of the expansion of Eq.~(\ref{Eq:zeta_approx_AP}), we first calculate the average value of $\zeta^{(\rm AP)}(r_1,r_2)$ in the range $80\leq r \leq 150\hMpc$, as in Eq.~(\ref{Eq:zeta_mean}).
\begin{eqnarray}
    \bar{\zeta}^{(\rm AP)}_{\ell_1\ell_2\ell}
    = \mbox{Average}\Big[\zeta^{(\rm AP)}_{\ell_1\ell_2\ell}(r_1,r_2)\Big].
\end{eqnarray}
Here, $\zeta^{(\rm AP)}(r_1,r_2)$ is obtained by calculating the angle integral in Eq.~(\ref{Eq:PB_multipole_APeffect}) directly and does not use the approximation for the calculation of the AP effect described above. Next, as in Eq.~(\ref{Eq:Delta_zeta_mean}), we define the following quantities
\begin{eqnarray}
    \Delta \bar{\zeta}^{(\rm AP)}_{\ell'_1\ell'_2\ell'}
    \hspace{-0.25cm}&=&\hspace{-0.25cm}
    \frac{\mbox{Average}\Big[\Delta \zeta^{(\rm AP)}_{\ell_1\ell_2\ell;\ell'_1\ell'_2\ell'}(r_1,r_2)\Big]}
    {\bar{\zeta}_{\ell_1\ell_2\ell}^{(\rm AP)}}\nonumber \\
    \Delta \zeta^{(\rm AP)}_{\ell_1\ell_2\ell;\ell'_1\ell'_2\ell'}(r_1,r_2)
    \hspace{-0.25cm}&=&\hspace{-0.25cm} \frac{1}{E^6(\varepsilon)}\sum_{n,m}  
     T_{\ell_1\ell_2\ell}^{\ell'_1\ell'_2\ell'(n)(m)}(\varepsilon)
    \nonumber \\
    \hspace{-0.25cm}&\times&\hspace{-0.25cm}	\zeta^{(n)(m)}_{\ell_1\ell_2;\ell'_1\ell'_2\ell'}(r'_1, r'_2),
\end{eqnarray}
where $\sum_{\ell'_1\ell'_2\ell'} \Delta \bar{\zeta}^{(\rm AP)}_{\ell'_1\ell'_2\ell'}=1$. We restrict the values of $n$ and $m$ that we calculate to $n+m\leq1$. That is, we only calculate up to the first derivative of the bispectrum with respect to $k_1$ or $k_2$, because it is difficult to calculate the higher derivative of the bispectrum with high precision by numerical calculations. Recall that the effect of multipole components other than the measured multipole component, which appears through the effect of the window function, is roughly $5$-$15\%$ (Section~\ref{Sec:WindowFunctionEffects}). Therefore, we will only investigate the AP effect in detail for $\zeta_{000}$, $\zeta_{110}$, $\zeta_{202}$, $\zeta_{112}$ and $\zeta_{404}$, which are the focus of our attention in this paper, and for other components, such as $\zeta_{220}$, $\zeta_{330}$, $\zeta_{022}$, $\zeta_{044}$ and etc., we will simply consider the effect of the isotropic rescaling of the relative distance: $\zeta_{\ell_1\ell_2\ell}^{(\rm AP)}(r_1,r_2) = \zeta_{\ell_1\ell_2\ell}(r'_1,r'_2)/E^6(\varepsilon)$.

Table~\ref{Table:zeta_AP} summarizes the contribution of the other multipolar components appearing through the anisotropic AP effect to the 3PCF multipole component of interest for $\varepsilon=0.02$ and $\varepsilon=-0.02$. Terms that contribute more than $0.5\%$ to the final results are written in bold, and we use only these bolded terms in our data analysis in Section~\ref{Sec:JointAnalysisWith3PCF}. Each multipole component can be positive or negative, and the total contribution to $\zeta_{000}^{(\rm AP)}$, $\zeta_{110}^{(\rm AP)}$, $\zeta_{202}^{(\rm AP)}$ or $\zeta_{112}^{(\rm AP)}$ is less than $\sim 5\%$. $\zeta_{404}^{(\rm AP)}$ is more susceptible to the anisotropic AP effect, with $25\%$ for $\varepsilon=0.02$ and $60\%$ for $\varepsilon=-0.02$.

To check the validity of the approximation of our AP effect calculations in Eq.~(\ref{Eq:zeta_approx_AP}), Figure~\ref{fig:zeta_E} plots $\bar{\zeta}_{\ell_1\ell_2\ell}^{(\rm obs)}$ (see Eq.~(\ref{Eq:zeta_mean})) as a function of $\varepsilon$. Note that, in this figure, the result shown is after accounting for the effect of the window function; therefore, this result is valid for a calculation in our actual data analysis. From this figure, we find that in the range $-0.02<\varepsilon<0.04$, our approximation holds for $\zeta_{000}$, $\zeta_{110}$, $\zeta_{202}$ and $\zeta_{112}$ with a precision of $\lesssim 2\%$. \Mod{Since the standard deviation of $\varepsilon$ estimated by the 2PCF, using the BOSS data we use in this paper, is $\sigma_{\varepsilon}\sim 0.025$ for pre-reconstruction and $\sigma_{\varepsilon}\sim 0.015$ for post-reconstruction~\citep{Ross:2016gvb,Beutler:2016ixs}, it is enough to consider the range $|\varepsilon|\lesssim 0.02$. It is difficult to estimate the extent to which the uncertainty of a few percent in our approximation method will bias the final AP parameter estimates. Therefore, we will test the validity of this approximation by showing that it returns unbiased estimates of $\alpha_{\perp}$ and $\alpha_{\parallel}$ in the data even for the case of $\alpha_{\perp}\neq 1$ and $\alpha_{\parallel}\neq 1$ in Section~\ref{Sec:TestOfAP}.}

Throughout this paper, we do not consider the dependence of $f\sigma_8$ on $\alpha$~\citep{Sanchez:2020vvb,Gil-Marin:2020bct}, because \citet{Gil-Marin:2020bct} have shown that this effect is negligible when $\alpha$ is sufficiently close to $1$.

\section{2PCF analysis}
\label{Sec:2PCFanalysis}

The BAO and AP analysis in 2PCF is well established particularly for the BOSS survey~\citep{Ross:2016gvb,Beutler:2016ixs,Gil-Marin:2020bct}. Nonetheless, in this section we summarize our 2PCF analysis, because we adopt a FFT-based multipole 2PCF estimator for a consistency with the 3PCF estimator. In short, we will show that our 2PCF analysis is consistent with previous studies, and hence readers interested in the joint analysis with 3PCF may skip to the next section.

\subsection{Data}
\label{Sec:Data}

We use the MultiDark-Patchy mock catalogues~\cite[MD-Patchy mocks;][]{Klypin:2014kpa,Kitaura:2015uqa}, which are designed to reproduce the BOSS dataset. These mocks have been calibrated to a $N$-body based reference sample using approximate gravity solvers and analytical-statistical biasing models and incorporate observational effects including the survey geometry, veto mask, and fiber collisions. The BOSS survey spans in two distinct sky regions (North and South Galactic Caps, hereafter NGC and SGC) in three redshift bins ($0.3<z<0.5$, $0.4 < z < 0.6$, and $0.5<z<0.75$). In this paper, we only use the middle redshift range, $0.4<z<0.6$ ($z_{\rm eff}=0.51$), for both NGC and SGC. The fiducial cosmology for these mocks assumes a $\Lambda$CDM cosmology given at the end of the introduction.

\subsection{Methodology for 2PCF analysis}
\label{Sec:MethodologyFor2PCFAnalysis}

The fitting range we use for our analysis is $80 \leq r \leq 150\hMpc$. The bin width is $\Delta r = 5\hMpc$ for the 2PCF, and the number of $r$ bins in each multipole 2PCF is $n_{\rm b}=15$. Specifically, when we use $\xi_0$ and $\xi_2$, the total number of $r$ bins is $2\times n_{\rm b}=30$, and  when we use $\xi_0$, $\xi_2$ and $\xi_4$, the total number of $r$ bins is $3\times n_{\rm b}=45$.

When focusing only on large scales, such as $r\geq 80\hMpc$, we expect \cite{Eisenstein:2006nj}'s template model (\ref{Eq:Eisenstein2007}) to be available to fit the 2PCF measurements, without the need for any nuisance parameters to reproduce the shape of the 2PCF at small scales. The validity of our expectations will be tested in Section~\ref{Sec:2PCFResults}. Then, the theoretical prediction of the 2PCF depends on the Kaiser factor~\citep{Kaiser:1987qv} and the AP effect. Given that the galaxies in the NGC and SGC follow slightly different selections~\citep{Alam:2016hwk}, we use two separate linear bias parameters to describe the clustering amplitude in the two samples: $b_{1, \rm NGC}$ and $b_{1, \rm SGC}$. For the pre-reconstruction case, we compute $\Sigma_{\perp}=6.5\hMpc$ and $\Sigma_{\parallel}=11.3\hMpc$ using the Zel'dovich approximation~\citep{Zeldovich:1970} and fix them. For the post-reconstruction case, we find the best-fitting values of $\Sigma_{\perp}$ and $\Sigma_{\parallel}$ by fixing $\alpha_{\perp}$, $\alpha_{\parallel}$, $f \sigma_8$, $b_{1, \rm NGC}$ and $b_{1, \rm SGC}$ to the best-fit values estimated for the pre-reconstruction case, resulting in $\Sigma_{\perp}=5.7\hMpc$ and $\Sigma_{\parallel}=8.5$ (for details of reconstruction, see Section~\ref{Sec:Reconstruction}). Note that the $\Sigma_{\perp}$ and $\Sigma_{\parallel}$ values for post-reconstruction are larger than the expected values from the PT calculations~\citep{Seo:2015eyw}. \citet{Beutler:2016ixs} pointed out that this excess damping is because the MultiDark Patchy mocks tend to underestimate the BAO signal due to a limitation of the 2LPT approximation used in the mock production (for more details about the MultiDark-Patchy mock production, see~\citet{Kitaura:2015uqa}). As shown by previous studies~\citep[e.g.,][]{Ross:2016gvb,Beutler:2016ixs}, the exact choice for $\Sigma_{\perp}$ and $\Sigma_{\parallel}$ does not affect our analysis. The approach explained above yields $5$ fitting parameters as follows:
\begin{eqnarray}
    \{\alpha_{\perp},\alpha_{\parallel}, f\sigma_8, b_{1}^{\rm NGC}\sigma_8, b_{1}^{\rm SGC}\sigma_8\}. \nonumber
\end{eqnarray}

We measure the 2PCF through the inverse Fourier transform of the power spectrum. We employ the Fast Fourier Transform (FFT)-based estimator suggested by~\citet{Bianchi:2015oia,Scoccimarro:2015bla,Hand:2017irw,Sugiyama:2017ggb}. This method requires multiple FFT operations with the computational complexity of ${\cal O}(N_{\rm grid} \ln N_{\rm grid})$ for each FFT process, where $N_{\rm grid}$ is the number of cells in the three-dimensional (3D) Cartesian grid in which the galaxies are binned. The FFT-based estimator is significantly faster than a straightforward pair counting analysis that would result in ${\cal O}(N_{\rm particle}^2)$ for the 2PCF, where $N_{\rm particle}$ is the number of galaxies or random particles. We use a Triangular Shaped Cloud (TSC) method to assign galaxies to 3D grid cells and correct for the aliasing effect following~\citet{Jing:2004fq}. Each side of our grid is $\sim 5\hMpc$ in size, which is well below the minimum scale we use for analysis ($r_{\rm min}=80\hMpc$). 

Specifically, let $D(\VEC{x})$ and $R(\VEC{x})$ be the numbers of data galaxies and random galaxies at $\VEC{x}$. $D(\VEC{x})$ and $R(\VEC{x})$ include observational systematic weights~\citep{Ross:2012qm,Anderson:2013zyy,Reid:2015gra} and the FKP weight~\citep{Feldman:1993ky}. $N_{\rm gal}=\int d^3x D(\VEC{x})$ and $N_{\rm ran}=\int d^3x R(\VEC{x})$ are the total numbers of weighted data and random galaxies, respectively. The survey volume $V$ is estimated by
\begin{eqnarray}
    V = \frac{N_{\rm ran}^2}{\int d^3x\, [R(\VEC{x})]^2}.
    \label{Eq:V}
\end{eqnarray}
The observed density fluctuation is given by
\begin{eqnarray}
    \delta_{\rm obs}(\VEC{x}) = V\left[  D(\VEC{x})/N_{\rm gal} -  R(\VEC{x})/N_{\rm ran}  \right].
    \label{Eq:delta_obs}
\end{eqnarray}
We compute the Fourier transform of $\delta_{\rm obs}$ weighted by a spherical harmonic function $Y_{\ell}^m$ 
\begin{eqnarray}
    \widetilde{\delta}_{{\rm obs}, \ell m}(\VEC{k}) = \int d^3x e^{-i\VEC{k}\cdot\VEC{x}} Y_{\ell}^{m*}(\hat{x}) \delta_{\rm obs}(\VEC{x}).
\end{eqnarray}
and then, we derive the estimator of the multipole 2PCF as follows:
\begin{eqnarray}
    \widehat{\xi}_{\ell}(r) = 
    \frac{1}{N_r(r)}\sum_{r-\Delta r/2<|\VEC{r}|<r+\Delta r/2}
     \widehat{\xi}_{\ell}(\VEC{r}),
    \label{Eq:xi_hat}
\end{eqnarray}
with 
\begin{eqnarray}
    \widehat{\xi}_{\ell}(\VEC{r})=
    \frac{(4\pi)}{V}
    Y_{\ell}^m(\hat{r})
    \sum_{m}\int \frac{d^3k}{(2\pi)^3}e^{i\VEC{k}\cdot\VEC{r}} 
    \frac{\widetilde{\delta}_{ {\rm obs}, \ell m}(\VEC{k})\, \widetilde{\delta}^*_{ {\rm obs}}(\VEC{k})}{W^2_{\rm mass}(\VEC{k})},
\end{eqnarray}
where $N_r(r)$ is the number of $\widehat{\xi}_{\ell}(\VEC{r})$ included in each $r$ bin, and $W_{\rm mass}$ is the TSC mass assignment function~\citep{HockneyEastwood1981}. We properly subtract the shot-noise term from $\widehat{\xi}_{\ell}$ before transforming to $\widehat{\xi}_{\ell}$~\citep[e.g., see][]{Beutler:2016ixs}. 

For the same reason as we mentioned in our 3PCF case in Section~\ref{Sec:WindowFunctionEffects}, we need to take into account the survey window function in this multipole 2PCF estimator. To this end, we measure the 2PCF multipoles of survey window functions, denoted $Q_{\ell}(r)$, by replacing $\delta_{\rm obs}(\VEC{x})$ by $(V/N_{\rm ran})R(\VEC{x})$ in Eqs.~(\ref{Eq:delta_obs})-(\ref{Eq:xi_hat}). The theoretical model taking the survey geometry effect into account is given by~\citep{Wilson:2015lup}
\begin{eqnarray}
    \xi_{\rm model}(r) = (2\ell+1)\sum_{\ell_1\ell_2} 
     \left( \begin{smallmatrix} \ell_1 & \ell_2 & \ell \\ 0 & 0 & 0 \end{smallmatrix}  \right)^2 Q_{\ell_1}(r)\, \xi_{\ell_2}(r).
     \label{Eq:xi_window}
\end{eqnarray}
where ``model'' means that this masked model will be compared with the measured estimator. 
In the summation on the right-hand-side, we ignore all multipole terms beyond the hexadecapole of their smallness.

We estimate the covariance matrix from the set of $2048$ Patchy mock catalogues described in Section~\ref{Sec:Data}. The mean of each 2PCF multipole is given by
\begin{eqnarray}
    \overline{\xi}_{\ell}(r) = \frac{1}{N_{\rm s}}\sum_{s}^{N_{\rm s}} \widehat{\xi}_{\ell}^{\, (s)}(r)
\end{eqnarray}
with $N_{\rm s}=2048$ being the number of mock catalogues. Then, we derive the covariance matrix of $\widehat{\xi}_{\ell}(r)$ as follows
\begin{eqnarray}
    C_{xy} = \frac{1}{N_{\rm s}-1} \sum_s^{N_{\rm s}}
    \hspace{-0.05cm}
    \left[ \widehat{\xi}_{\ell}^{\, (s)}(r_i) - \overline{\xi}_{\ell}(r_i) \right]
    \hspace{-0.13cm}
    \left[ \widehat{\xi}_{\ell'}^{\, (s)}(r_j) - \overline{\xi}_{\ell'}(r_j) \right].
\end{eqnarray}
The elements of the matrix are given by $(x,y) = (\frac{n_{\rm b}\ell}{2}+i,\frac{n_{\rm b}\ell'}{2}+j)$ with $n_{\rm b}=15$.

The covariance matrix $\MAT{C}$ that is inferred from mock catalogues suffers from the noise due to the finite number of mocks, which propagates directly to increased uncertainties in cosmological parameters~\citep{Hartlap:2006kj,Taylor:2012kz,Dodelson:2013uaa,Percival:2013sga,Taylor:2014ota}. This effect is decomposed into two factors. First, the inverse covariance matrix, $\MAT{C}^{-1}$, provides a biased estimate of the true inverse covariance matrix. To correct for this bias we rescale the inverse covariance matrix as~\citep{Hartlap:2006kj}
\begin{eqnarray}
    \MAT{C}^{-1}_{\rm Hartlap} = \frac{N_{\rm s}-N_{\rm b}-2}{N_{\rm s}-1}\MAT{C}^{-1},
    \label{Eq:Hartlap}
\end{eqnarray}
where $N_{\rm b}$ is the number of data bins. (see also \citet{Sellentin:2016} as a recent work.) Second, we propagate the error in the covariance matrix to the error on the estimated parameters; this is done by scaling the standard deviation for each parameter by~\citep{Percival:2013sga}
\begin{eqnarray}
    M_1 = \sqrt{\frac{1+B(N_{\rm b}-N_{\rm p})}{1+A+B(N_{\rm p}+1)}}
\end{eqnarray}
where $N_{\rm p}$ is the number of parameters, and
\begin{eqnarray}
    A &=& \frac{2}{(N_{\rm s}-N_{\rm b}-1)(N_{\rm s}-N_{\rm b}-4)} \\
    B &=& \frac{N_{\rm s}-N_{\rm b}-2}{(N_{\rm s}-N_{\rm b}-1)(N_{\rm s}-N_{\rm b}-4)}.
\end{eqnarray}

We compute the following quantity in each analysis performed in Sections~\ref{Sec:2PCFanalysis} and~\ref{Sec:JointAnalysisWith3PCF} to estimate how much the fact that the number of mock catalogues is finite affects the final results: 
\begin{eqnarray}
    M_2 = M_1 \sqrt{\frac{N_{\rm s}-1}{N_{\rm s}-N_{\rm b}-2}}.
    \label{Eq:M2}
\end{eqnarray}
This scaling $M_2$ is derived under the assumption of a Gaussian error distribution, which is not strictly true in our dataset. Therefore, it is better to produce many mock catalogues to keep this correction factor small. When using $\xi_0$ and $\xi_2$, $M_2=1.012$, and when using $\xi_0$, $\xi_2$ and $\xi_4$, $M_2 = 1.020$. Thus, as long as we focus on the 2PCF-only analysis, the correction of Eq.~(\ref{Eq:M2}) increases the parameter variance by about $\lesssim2\%$, which is negligible. However, in the joint analysis with the 3PCF, the number of data bins increases significantly, making this correction even more important (see Section~\ref{Sec:JointAnalysisWith3PCF}).

We explore parameter space by performing a Markov-Chain Monte Carlo (MCMC) analysis. We ensure convergence of each MCMC chain, imposing $R-1 < 0.001$ where $R$ is the standard Gelman-Rubin criteria. To this end, we use the $montepython$ package~\citep{Audren:2012wb,Brinckmann:2018cvx}.

\subsection{Reconstruction}
\label{Sec:Reconstruction}

\begin{figure*}
	\includegraphics[width=\textwidth]{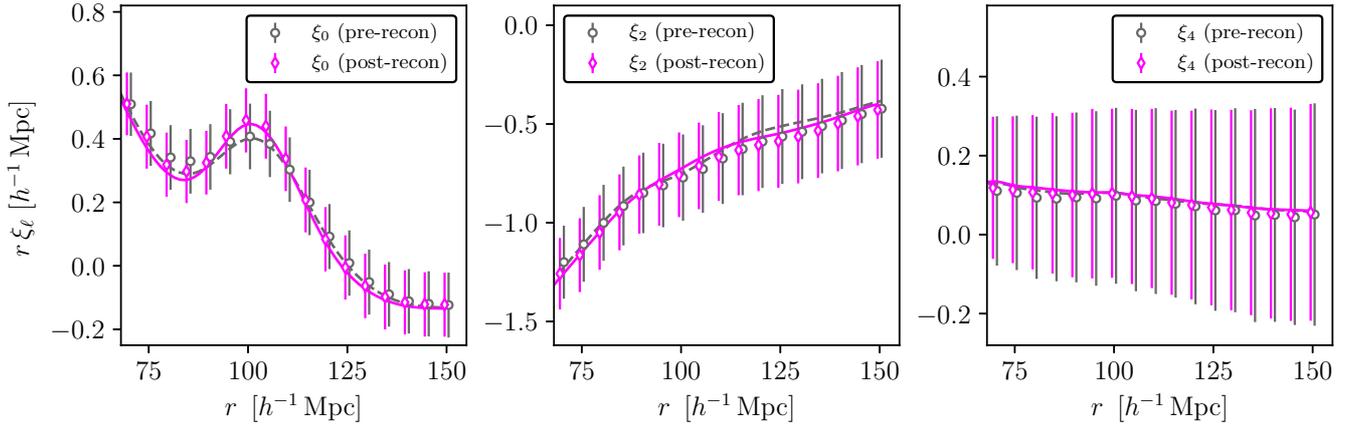}
	\caption{
    Measured monopole, quadrupole and hexadecapole of the 2PCF 
    from the MultiDark-Patchy mock catalogues in the NGC for $0.4<z<0.6$.
    The mean of the $2048$ mock catalogues is shown.
    The errorbars on the data points are computed by the diagonal of the covariance matrix
    \Mod{and are for a single mock, not for the mean of the mocks.}
    The gray and magenta colors indicate the 2PCF multipoles before and after reconstruction, respectively.
    The corresponding gray dashed and magenta solid lines indicate the best fit models.
    }
	\label{fig:2PCFMeasured}
\end{figure*}

Density field reconstruction~\citep{Eisenstein:2006nk,Padmanabhan:2012hf} is designed to improve the signal-to-noise ratio of the BAO signature by partially cancelling the nonlinear effects of structure formation, i.e., by returning the information that leaks to the higher-order statistics of the galaxy distribution back to the two-point statistics~\citep{Schmittfull:2015mja}.

In this paper, we adopt the simplest reconstruction scheme. We compute the displacement vector using the observed density fluctuation,
\begin{eqnarray}
    \VEC{s}(\VEC{x}) =  -\int \frac{d^3k}{(2\pi)^3}\, e^{i\VEC{k}\cdot\VEC{x}}\, \left(\frac{i\VEC{k}}{k^2}  \right)\, W_{\rm G}(k,R_{\rm s}) 
    \frac{\widetilde{\delta}_{{\rm obs}}(\VEC{k})}{b_{1, \rm fid}},
    \label{Eq:disp}
\end{eqnarray}
where $b_{1, \rm fid}=2.0$ is the input linear bias parameter. We smooth the density field with a Gaussian filter of the form, $W_{\rm G}=\exp\left( -k^2 R_{\rm s}^2 /2 \right)$, where we chose $R_{\rm s}=15\hMpc$, which is close to the optimal smoothing scale given by the signal-to-noise ratio of the BOSS data~\citep{Xu:2012hg,Burden:2014cwa,Vargas-Magana:2015rqa,Seo:2015eyw}. The positions of all galaxies are modified based on the estimated displacement field $\VEC{s}(\VEC{x})$. This procedure leads to a shifted galaxy field, $D_{\rm s}(\VEC{x})$, and a shifted random field, $R_{\rm s}(\VEC{x})$:
\begin{eqnarray}
    D_{\rm rec}(\VEC{x}) \hspace{-0.25cm} &=& \hspace{-0.25cm} \int d^3x'\, D(\VEC{x}')\, \delta_{\rm D}(\VEC{x} - \VEC{x}'-\VEC{s}(\VEC{x}')) \nonumber \\
    R_{\rm rec}(\VEC{x}) \hspace{-0.25cm} &=& \hspace{-0.25cm} \int d^3x'\, R(\VEC{x}')\, \delta_{\rm D}(\VEC{x} - \VEC{x}'-\VEC{s}(\VEC{x}')).
\end{eqnarray}
The reconstructed density fluctuation, required for the 2PCF estimate, can be obtained in an analogous way to Eq.~(\ref{Eq:delta_obs}) and is given by
\begin{eqnarray}
    \delta_{\rm rec}(\VEC{x}) = \int d^3x'\, \delta_{\rm obs}(\VEC{x}')\, \delta_{\rm D}(\VEC{x} - \VEC{x}'-\VEC{s}(\VEC{x}')).
    \label{Eq:recon_delta}
\end{eqnarray}
We do not remove linear RSDs. At large scales, our reconstruction does not affect the linear density perturbation except for the damping effect of BAO. Therefore, we do not need to change the theoretical model of the quadrupole and hexadecapole.

We compute the window function multipoles $Q_{\ell}(r)$ using $R_{\rm rec}(\VEC{x})$. In principle, after reconstruction, $R_{\rm rec}(\VEC{x})$ is correlated with $D_{\rm rec}(\VEC{x})$ through $\VEC{s}(\VEC{x})$. In this paper, however, we assume that they are uncorrelated. In other words, we use the same equation (\ref{Eq:xi_window}) as before reconstruction to correct for the window function effect. This assumption is the same as used in~\citet{Beutler:2016ixs}.

\subsection{Results}
\label{Sec:2PCFResults}

Figure~\ref{fig:2PCFMeasured} displays the measured pre- and post-reconstruction $\xi_0$, $\xi_2$ and $\xi_4$, from the MultiDark Patchy mock catalogues in the NGC for $0.4<z<0.6$, and the associated best-fit model. We display the results for BAO fits on the mean of the MD-Patchy 2PCF multipoles in Table~\ref{Table:APlimits}. The expected values of $\alpha_{\perp}$ and $\alpha_{\parallel}$ are unity, and hence, $1-\alpha_{\perp}$ and $1-\alpha_{\parallel}$ show how the estimated $\alpha_{\perp}$ and $\alpha_{\parallel}$ from the analysis are biased. The standard deviations of $\alpha_{\perp}$ and $\alpha_{\parallel}$ are denoted as $\sigma_{\perp}$ and $\sigma_{\parallel}$, respectively. $r_{\alpha_{\perp}\alpha_{\parallel}}$ is the correlation coefficient between $\alpha_{\perp}$ and $\alpha_{\parallel}$. We show $M_2$ (\ref{Eq:M2}) to clarify the impact on the effect due to the finite number of mocks to estimate the covariance matrix. We also fit each of $100$ individual mock catalogues and summarize the mean and standard deviation of the results in Table~\ref{Table:APlimits_mean}. The minimum $\chi^2$ statistics, denoted $\chi^2_{\rm min}$, is estimated for the best-fit parameters for each mock catalogue.

All of the four cases result in the $\langle \chi^2_{\rm min} \rangle/{\rm dof}$ values that are nearly unity and the $p$-values that are larger than $0.15$; therefore, we conclude that the \cite{Eisenstein:2006nj}'s template model fits the measured Patchy 2PCF multipoles well at scales larger than $80\hMpc$. The results of the $\xi_{0,2}$ analysis are $\langle \alpha_{\perp}\rangle=0.997\pm0.026$ and $\langle \alpha_{\parallel}\rangle=1.011\pm0.061$ for pre-reconstruction, and $\langle\alpha_{\perp}\rangle=0.996\pm0.018$ and $\langle \alpha_{\parallel}\rangle=1.000\pm0.035$ for post-reconstruction, which are consistent with the results of the previous studies: e.g., compare our results with Table $4$ in \citealt{Ross:2016gvb} or Table $2$ in \citealt{Beutler:2016ixs}. The addition of the hexadecapole information reduces $\sigma_{\parallel}$ by $\sim20\%$ for pre-reconstruction as shown by~\citet{Taruya:2011tz,Beutler:2016arn}, while in the case of post-reconstruction, the addition of $\xi_4$ does not cause a significant improvement.

Tables~\ref{Table:RSD} and~\ref{Table:RSD_mean} summarize our results of the constraints on $f\sigma_8$.
Since the expected value is $f\sigma_8=0.48$, the bias of the estimated $f\sigma_8$ in our analysis is roughly $1/4$ of the $1$-$\sigma$ error. The $1$-$\sigma$ error in $f\sigma_8$ is $\sim0.085$ before reconstruction and $0.075$ after reconstruction. 
As expected, the density field reconstruction decreases the errors in $f\sigma_8$, with a reduction rate of about $10\%$. In our analysis, the addition of $\xi_4$ does not change the results much.

According to previous studies, the $1\sigma$ error in $f\sigma_8$ using the monopole and quadrupole is $0.060$ for the 2PCF analysis~\citep[see Table 1 in][]{Satpathy:2016tct} and $0.065$ for the power spectrum analysis~\citep[see Table 2 in][]{Beutler:2016arn}. The power spectrum analysis also examines the addition of the hexadecapole, in which case the error in $f\sigma_8$ is reduced to $0.046$. The reason why our constraint is weaker than the results of these previous studies is that our analysis uses only the large scales, $r\geq80\hMpc$, and our setup is such that the 2PCF extracts mostly the BAO-only information with a conservative estimate of $f\sigma_8$.


\section{Joint analysis with 3PCF}
\label{Sec:JointAnalysisWith3PCF}

\subsection{Methodology for 3PCF analysis}
\label{Sec:MethodologyFor3PCFAnalysis}

\begin{figure*}
    \scalebox{0.8}{\includegraphics[width=\textwidth]{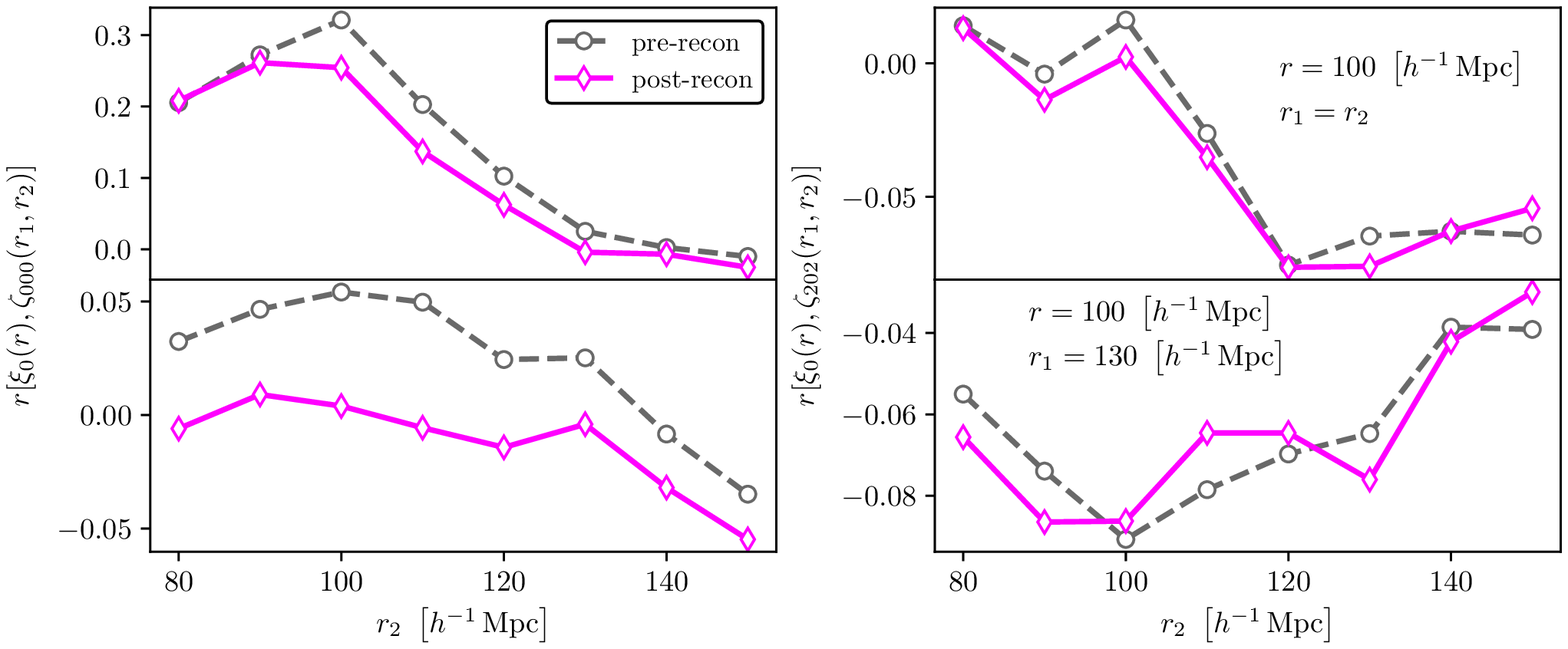}}
    \scalebox{0.8}{\includegraphics[width=\textwidth]{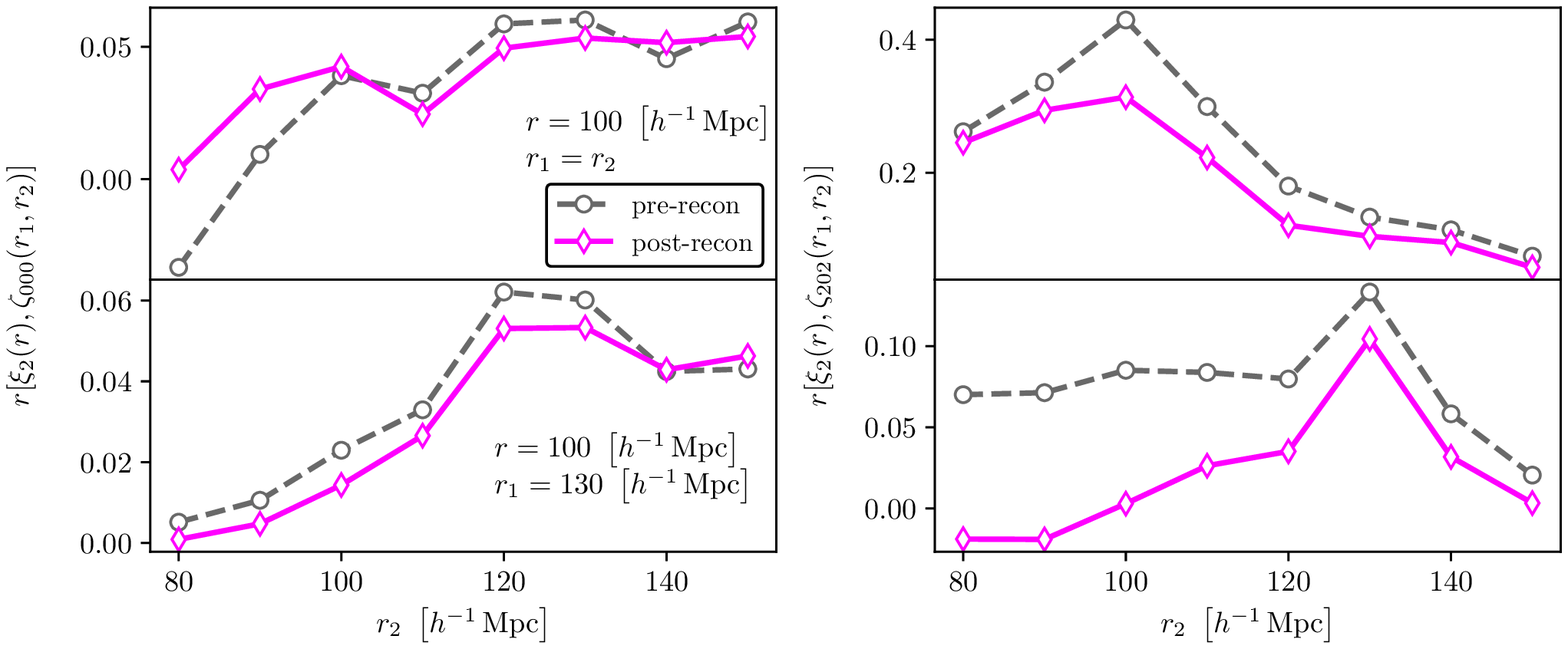}}
	\caption{
	Correlation coefficients between the 2PCF and 3PCF multipoles estimated from the 2048 MultiDark-Patchy mock catalogues
	in the NGC for $0.4<z<0.6$.
	The monopole and quadrupole components of the 2PCF and 3PCF, i.e., $\xi_0$, $\xi_2$, $\zeta_{000}$ and $\zeta_{202}$, are considered. The results before and after reconstruction of the 2PCF are shown by the dashed gray line and the magenta solid line, respectively.
	The 2PCF reconstruction reduces the correlation coefficients between $\xi_0$ and $\zeta_{000}$, and $\xi_2$ and $\zeta_{202}$,
	because they both are dominated by the monopole component and we have performed reconstruction that preserves linear RSDs.
}
	\label{fig:coeff}
\end{figure*}

\begin{figure*}
    \scalebox{0.8}{\includegraphics[width=\textwidth]{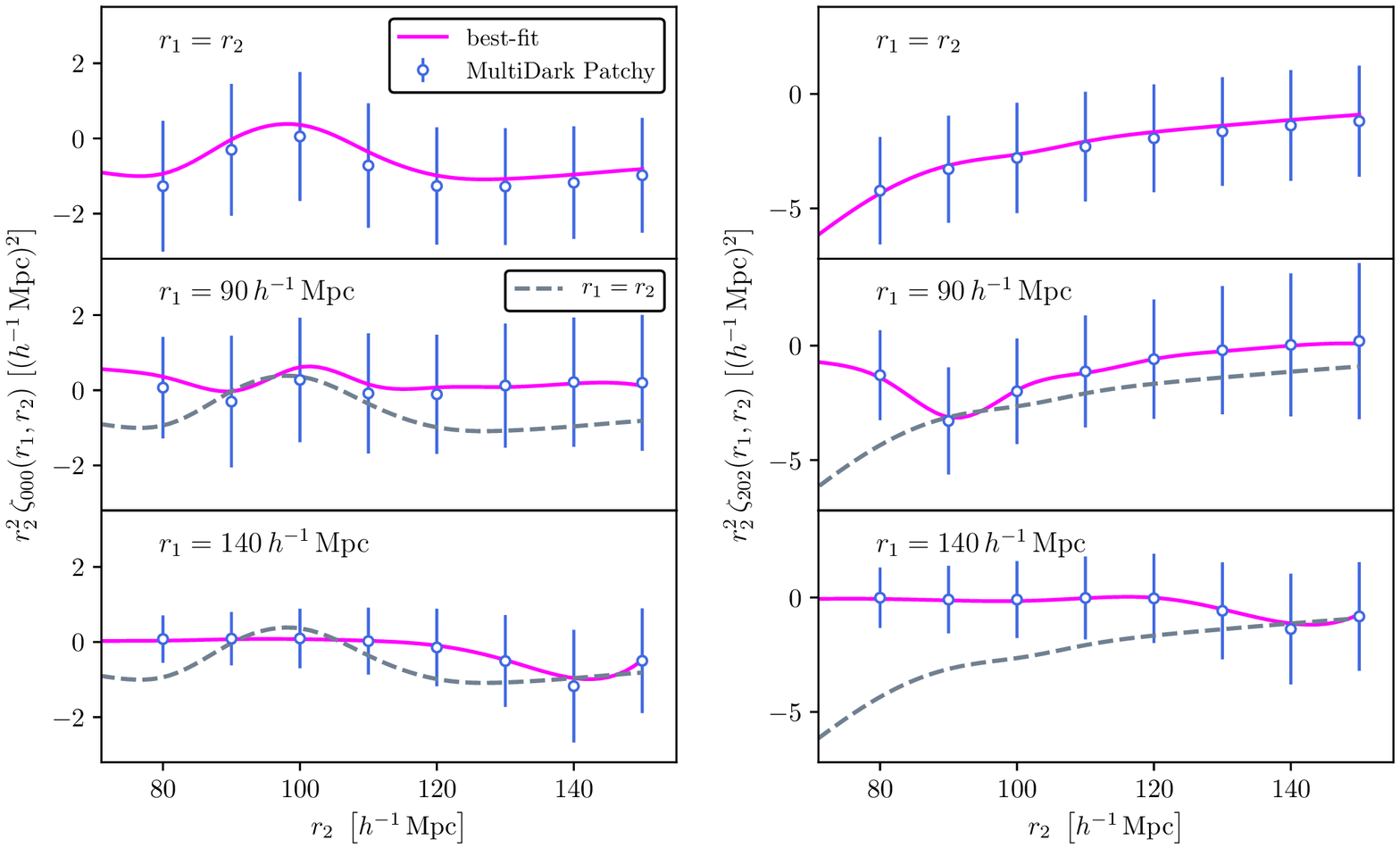}}
	\caption{
    Measured first terms of the monopole and quadrupole 3PCFs from the MultiDark-Patchy mock catalogues in the NGC for $0.4<z<0.6$.
    The mean of the $2048$ mock measurements is shown as a function of $r_2$ when fixing $r_1$ to $r_1=r_2$, $r_1=90\hMpc$ and $r_1=140\hMpc$ from top to bottom panels.
    The errorbars on the data points are computed by the square root of the diagonal of the covariance matrix and
    \Mod{are for a single mock, not for the mean of the mocks.}
    The magenta solid lines indicate the best fit models.
    The gray dashed lines in the middle and bottom panels are the same as the results in the top panels, where the gray and magenta lines intersect when $r_1=r_2$.
	}
	\label{fig:3PCFMeasured_000202}
    \scalebox{0.8}{\includegraphics[width=\textwidth]{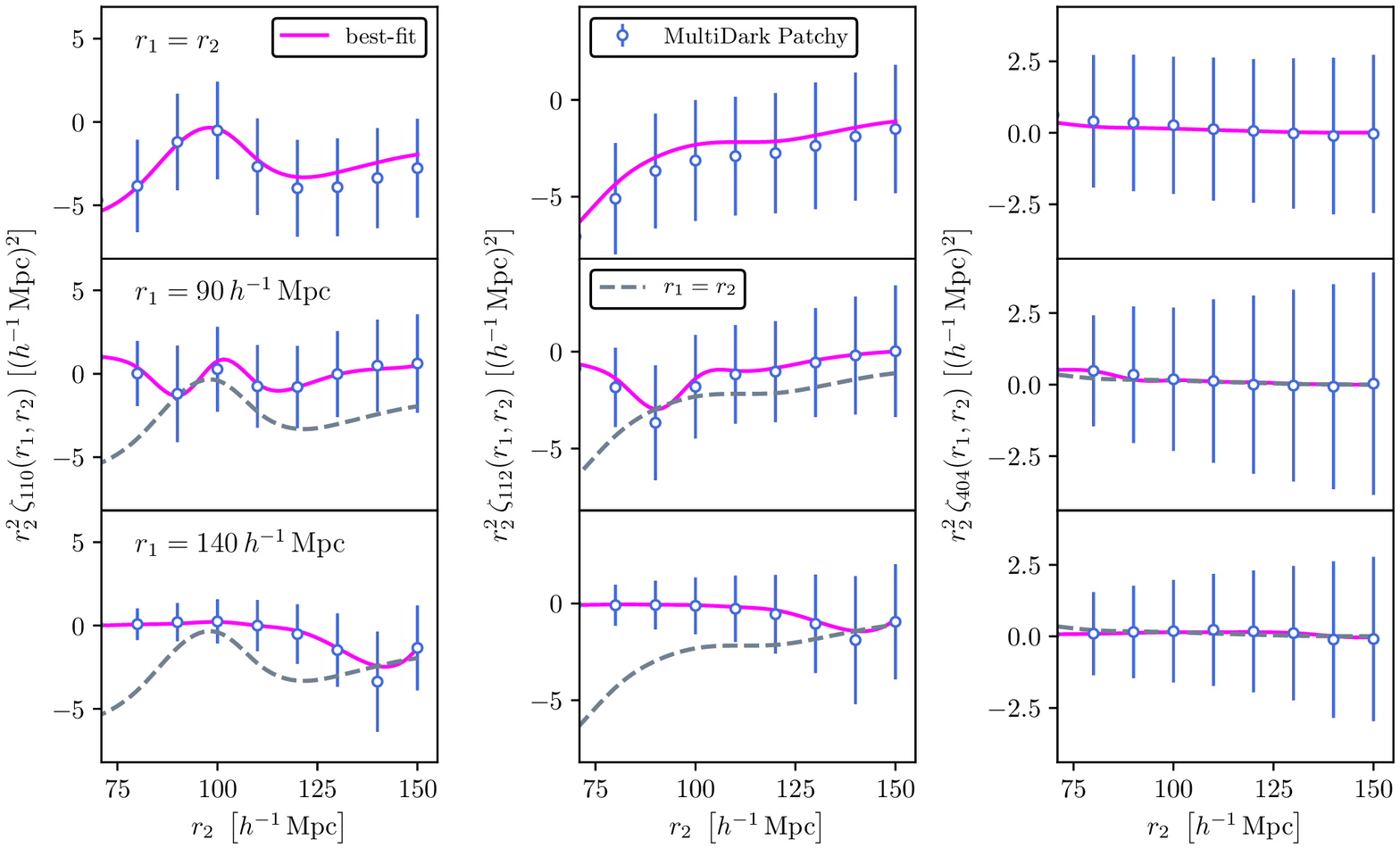}}
	\caption{
    Similar to Figure~\ref{fig:3PCFMeasured_000202}.
    The second terms of the monopole and quadrupole 3PCFs ($\zeta_{110}$ and $\zeta_{112}$) 
    and the first term of the hexadecapole 3CPF ($\zeta_{404}$) are shown from left to right.
	}
	\label{fig:3PCFMeasured_110112404}
\end{figure*}

One of the major problems in the 3PCF analysis is the high degree of freedom, i.e., the large number of data bins. To reduce the number of data bins, we choose $\Delta r = 10\hMpc$ for the 3PCF measurements in the same scale range, $80\leq r \leq 150\hMpc$, as used in the 2PCF analysis; the number of $r$ bins is then $n_{\rm b}=8$. For our decomposition formalism of the 3PCF, which is characterized by $\zeta_{\ell_1\ell_2\ell}(r_1,r_2)$, the number of data bins is $n_{\rm b}(n_{\rm b}+1)/2=36$ for $\ell_1=\ell_2$ and $n_{\rm b}^2=64$ for $\ell_1\neq \ell_2$.

Since we focus only on $r\geq 80\hMpc$, we can directly use our template model given in Eq.~(\ref{Eq:MainResult}), which is based on the tree-level solution with the nonlinear damping of BAO. We fix $\Sigma_{\perp}=6.5\hMpc$ and $\Sigma_{\parallel}=11.3\hMpc$ to the same values used in the pre-reconstruction 2PCF analysis. We use two separate nonlinear bias parameters in the NGC and SGC samples: $b_{2}^{\rm NGC}$, $b_{2}^{\rm SGC}$, $b_{\rm K^2}^{\rm NGC}$ and $b_{\rm K^2}^{\rm SGC}$. Thus, our 3PCF analysis uses $10$ fitting parameters as follows:
\begin{eqnarray}
    &\Big\{\alpha_{\perp}, \alpha_{\parallel}, f\sigma_8, \sigma_8, (b_1^{\rm NGC}\sigma_8), (b_1^{\rm SGC}\sigma_8), \nonumber \\
    & (b_2^{\rm NGC}\sigma_8^2), (b_2^{\rm SGC}\sigma_8^2), (b_{\rm K^2}^{\rm NGC}\sigma_8^2), (b_{\rm K^2}^{\rm SGC}\sigma_8^2)\Big\}.
\end{eqnarray}

We employ the FFT-based estimator of the 3PCF suggested by~\citet{Scoccimarro:2015bla,Slepian:2016qwa,Sugiyama:2018yzo}. The setting of the FFT grids for the 3PCF measurements is the same as used for the 2PCF measurements. Specifically, the estimator of $\zeta_{\ell_1\ell_2\ell}$ is given by~\citep[Section~4.3 in][]{Sugiyama:2018yzo}
\begin{eqnarray}
	\widehat{\zeta}_{\ell_1\ell_2\ell}(r_1,r_2)
	\hspace{-0.25cm}&=&\hspace{-0.25cm} 
    \frac{(4\pi)^2 h_{\ell_1\ell_2\ell}}{V} \sum_{m_1m_2m}	\left( \begin{smallmatrix} \ell_1 & \ell_2 & \ell \\ m_1 & m_2 & m \end{smallmatrix}  \right)  \nonumber \\
	\hspace{-0.25cm} &\times&\hspace{-0.25cm}
    \int d^3x F_{\ell_1}^{m_1}(\VEC{x};r_1) F_{\ell_2}^{m_2}(\VEC{x};r_2) G_{\ell}^{m}(\VEC{x}),
	\label{Eq:reduced_xi_estimator}
\end{eqnarray}
where
\begin{eqnarray}
	F_{\ell}^{\, m}(\VEC{x};r)
    \hspace{-0.25cm}&=& \hspace{-0.25cm}
	i^{\ell}\int \frac{d^3k}{(2\pi)^3} 
    e^{i\VEC{k}\cdot\VEC{x}} j_{\ell}(rk) Y_{\ell}^{m*}(\hat{k}) \frac{\widetilde{\delta}_{ {\rm obs}}(\VEC{k})}{W_{\rm mass}(\VEC{k})} \nonumber \\
    G_{\ell}^{m}(\VEC{x}) 
    \hspace{-0.25cm}&=& \hspace{-0.25cm}
    \int \frac{d^3k}{(2\pi)^3} \,
    e^{i\VEC{k}\cdot\VEC{x}} \, \frac{\widetilde{\delta}_{ {\rm obs}, \ell m}(\VEC{k})}{W_{\rm mass}(\VEC{k})}.
\end{eqnarray}
Similar to the 2PCF case, we subtract the shot-noise term~\citep[Eq.~(51) in ][]{Sugiyama:2018yzo} from the above estimator.

We have described in Section~\ref{Sec:WindowFunctionEffects} how to compute a theoretical model of 3PCF that takes into account the effects of the window function. 

We follow the same manner to compute the mean and covariance matrix of $2048$ mock 3PCF multipoles as used for the 2PCF analysis. Figure~\ref{fig:coeff} displays the correlation coefficients between the 2PCF and 3PCF multipoles. Since the cross-covariance between the 2PCF and 3PCF arises from the non-linearity effect of gravity, we find that the correlation coefficient decrease as expected when reconstruction of the 2PCF is performed to partially remove the non-linear gravitational effect. Note that our reconstruction can mainly affect the isotropic component of the density field because it preserves linear RSDs (\ref{Eq:recon_delta}). Therefore, ${\rm cov}[\xi_0,\zeta_{000}]$ and ${\rm cov}[\xi_2,\zeta_{202}]$, which are dominated by monopole components, can be reduced, while ${\rm cov}[\xi_2,\zeta_{000}]$ and ${\rm cov}[\xi_0,\zeta_{202}]$, which are dominated by quadrupole components, are virtually unchanged before and after the 2PCF reconstruction.

\subsection{Results}
\label{Sec:Results_3PCF}

\begin{table*}
\begin{tabular}{lcccccc} 
 \hline 
\multicolumn{7}{c}{ Patchy mock ($0.4<z<0.6$) } \\
 \hline 
 & $1-\alpha_{\perp}$  & $\sigma_{\perp}$ & $1-\alpha_{\parallel}$ &  $\sigma_{\parallel}$  & $r_{\alpha_{\perp}\alpha_{\parallel}}$ & $M_2$\\ 
 \hline 
 \hline 
$\xi_{0,2}$                                                      & 0.003 & 0.026 &  0.000 & 0.061 & -0.49 & 1.012\\
 \hline 
(1)  $\xi_{0,2}+\zeta_{(000),(110)}$                             & 0.008 & 0.026 & -0.024 & 0.062 & -0.46 & 1.047 \\
(2)  $\xi_{0,2}+\zeta_{(000),(202)}$                             & 0.002 & 0.023 & -0.008 & 0.043 & -0.16  & 1.062\\
(3)  $\xi_{0,2}+\zeta_{(000),(202),(112)}$                       & 0.001 & 0.023 & -0.005 & 0.040 & -0.02 & 1.083\\
(4)  $\xi_{0,2}+\zeta_{(000),(202),(404)}$                       & 0.002 & 0.024 & -0.008 & 0.044 & -0.12 & 1.099\\
(5)  $\xi_{0,2}+\zeta_{(000),(202),(112),(404)}$                 & 0.004 & 0.022 &\ 0.003 & 0.037 & -0.05 & 1.121\\
\hline
 \hline 
     $\xi_{0,2,4}$                                               & 0.002 & 0.023 &\ 0.007 & 0.048 & -0.28 & 1.020 \\
 \hline 
(6)  $\xi_{0,2,4}+\zeta_{(000),(110)}$                           & 0.002 & 0.024 & -0.003 & 0.049 & -0.28 & 1.055\\
(7)  $\xi_{0,2,4}+\zeta_{(000),(202)}$                           & 0.001 & 0.023 & -0.003 & 0.040 & -0.06 & 1.071\\
(8)  $\xi_{0,2,4}+\zeta_{(000),(202),(112)}$                     & 0.001 & 0.023 & -0.002 & 0.038 &\ 0.07 & 1.091\\
(9)  $\xi_{0,2,4}+\zeta_{(000),(202),(404)}$                     & 0.000 & 0.024 &\ 0.000 & 0.040 & -0.03 & 1.108\\
(10) $\xi_{0,2,4}+\zeta_{(000),(202),(112),(404)}$               & 0.000 & 0.023 &\ 0.000 & 0.037 &\ 0.08 & 1.130\\
\hline
\hline 
     $\xi_{0,2}^{(\rm rec)}$                                   & 0.006 & 0.017 & 0.006  & 0.033 & -0.44 & 1.012\\
 \hline                                                                                                            
(11) $\xi_{0,2}^{(\rm rec)}+\zeta_{(000),(110)}$               & 0.007 & 0.017 & -0.002 & 0.033 & -0.38 & 1.047\\
(12) $\xi_{0,2}^{(\rm rec)}+\zeta_{(000),(202)}$               & 0.007 & 0.016 &  0.000 & 0.027 & -0.23 & 1.062\\
(13) $\xi_{0,2}^{(\rm rec)}+\zeta_{(000),(202),(112)}$         & 0.007 & 0.016 &  0.001 & 0.027 & -0.20 & 1.083\\
(14) $\xi_{0,2}^{(\rm rec)}+\zeta_{(000),(202),(404)}$         & 0.007 & 0.017 & -0.001 & 0.028 & -0.26 & 1.099\\
(15) $\xi_{0,2}^{(\rm rec)}+\zeta_{(000),(202),(112),(404)}$   & 0.007 & 0.017 &  0.003 & 0.028 & -0.23 & 1.121\\
\hline
 \hline 
     $\xi_{0,2,4}^{(\rm rec)}$                                 & 0.005 & 0.017 & 0.008 & 0.031 & -0.35 & 1.020\\
 \hline                                                                                                            
(16) $\xi_{0,2,4}^{(\rm rec)}+\zeta_{(000),(110)}$             & 0.007 & 0.017 & 0.003 & 0.032 & -0.30 & 1.055 \\
(17) $\xi_{0,2,4}^{(\rm rec)}+\zeta_{(000),(202)}$             & 0.006 & 0.016 & 0.003 & 0.027 & -0.18 & 1.071 \\
(18) $\xi_{0,2,4}^{(\rm rec)}+\zeta_{(000),(202),(112)}$       & 0.006 & 0.016 & 0.004 & 0.027 & -0.15 & 1.091 \\
(19) $\xi_{0,2,4}^{(\rm rec)}+\zeta_{(000),(202),(404)}$       & 0.006 & 0.016 & 0.005 & 0.027 & -0.18 & 1.108 \\
(20) $\xi_{0,2,4}^{(\rm rec)}+\zeta_{(000),(202),(112),(404)}$ & 0.006 & 0.017 & 0.003 & 0.027 & -0.16 & 1.130 \\
\hline 
\end{tabular} \\ 
\caption{
Results for $\alpha_{\perp}$ and $\alpha_{\parallel}$, obtained from the mean of the 2PCF and 3PCF measured from the MultiDark-Patchy mock catalogues for $0.4<z<0.6$. Since the MD-Patchy mock cosmology is used in the analysis, the values of $\alpha_{\perp}$ and $\alpha_{\parallel}$ should be consistent with 1. The $1$-$\sigma$ errors in $\alpha_{\perp}$ and $\alpha_{\parallel}$, and the correlation coefficient $r_{\alpha_{\perp}\alpha_{\parallel}}$ between those two parameters are \Mod{for a single mock, not for the mean of the mocks.} Also shown is the values of the $M_2$ correction factor (\ref{Eq:M2}), which is required by evaluating the covariance matrix from a finite number of mocks. The errors given here is the value after correction by $M_2$.}
\label{Table:APlimits}
\end{table*}

\begin{table*}
\begin{tabular}{lcccccc} 
 \hline 
\multicolumn{7}{c}{ Patchy mock ($0.4<z<0.6$) } \\
\hline
& $1-\langle \alpha_{\perp}\rangle$ & $\langle \sigma_{\perp}\rangle$ & $1-\langle \alpha_{\parallel}\rangle$ & $\langle\sigma_{\parallel}\rangle$ & $\langle r_{\alpha_{\perp}\alpha_{\parallel}}\rangle$ & $ \langle \chi_{\rm min}^2\rangle/{\rm dof}$ \\ 
 \hline 
$\xi_{0,2}$                     &  0.003 & 0.026$\pm$0.007 & -0.011 & 0.061$\pm$0.020 & -0.48$\pm$0.05 & (55.5$\pm$10.5)/55 \\                
$\xi_{0,2}+\zeta_{(000),(202)}$ & -0.001 & 0.023$\pm$0.006 & -0.014 & 0.043$\pm$0.010 & -0.16$\pm$0.19 & (239.9$\pm$23.3)/250\\                
 \hline 
$\xi_{0,2,4}$                     &  0.001 & 0.024$\pm$0.006 &  0.000 & 0.049$\pm$0.015 & -0.29$\pm$0.08 & (84.2$\pm$12.0)/85\\
$\xi_{0,2,4}+\zeta_{(000),(202)}$ & -0.002 & 0.022$\pm$0.006 & -0.008 & 0.039$\pm$0.009 & -0.08$\pm$0.19 & (265.4$\pm$23.2)/280 \\                
 \hline 
$\xi^{\rm (rec)}_{0,2}$                     & 0.004 & 0.018$\pm$0.004 &  0.000 & 0.035$\pm$0.008 & -0.45$\pm$0.037 & (53.2$\pm$10.1)/55 \\
$\xi_{0,2}^{\rm (rec)}+\zeta_{(000),(202)}$ & 0.004 & 0.017$\pm$0.004 & -0.001 & 0.028$\pm$0.007 & -0.28$\pm$0.121 & (238.9$\pm$22.8)/250 \\                
\hline 
$\xi^{\rm (rec)}_{0,2,4}$     & 0.002 & 0.017$\pm$0.003 & 0.004 & 0.032$\pm$0.006 & -0.36$\pm$0.070 & (83.8$\pm$13.0)/85  \\ 
$\xi_{0,2,4}^{\rm (rec)}+\zeta_{(000),(202)}$ & 0.003 & 0.017$\pm$0.004 & 0.002 & 0.027$\pm$0.006 & -0.23$\pm$0.16 & (266.4$\pm$23.0)/280 \\                
\hline 
 \end{tabular} \\ 
\caption{
Similar to Table~\ref{Table:APlimits}. The results shown here are the mean of the results obtained from each $100$ MD-Patchy mock catalogue.
}
\label{Table:APlimits_mean}
\end{table*}

\begin{figure*}
	\includegraphics[width=\textwidth]{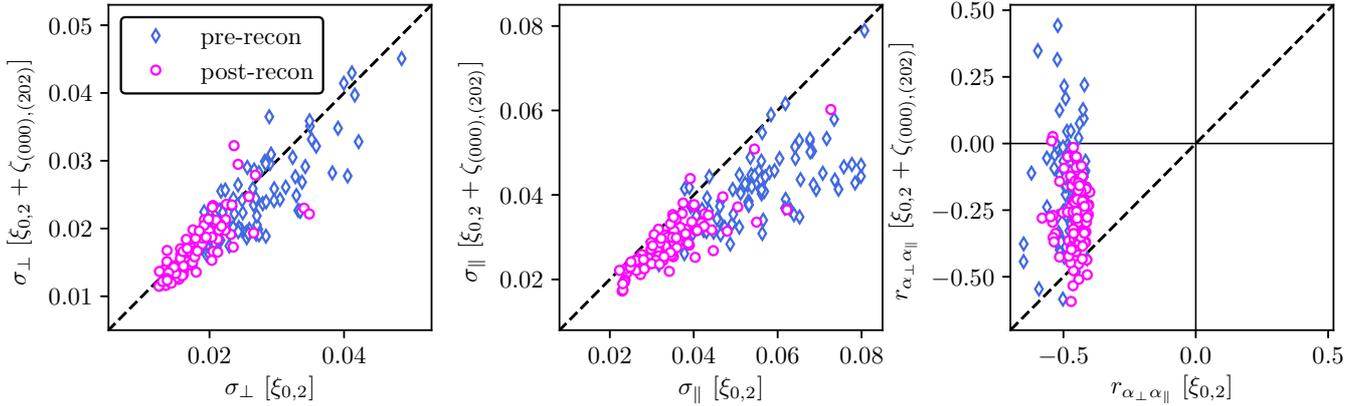}
	\caption{
    Uncertainties on $\alpha_{\perp}$ and $\alpha_{\parallel}$ before and after adding the anisotropic 3PCFs. 
    The blue diamonds and magenta circles show the results before and after reconstruction for 2PCFs.
    The possibility satisfying $\sigma[\xi+\zeta]<\sigma[\xi]$ is $70\%$ and $96\%$
    for $\sigma_{\perp}$ and $\sigma_{\parallel}$, respectively, in both cases before and after reconstruction of the 2PCF.
	}
	\label{fig:sigma_mocks}
\end{figure*}

Figures~\ref{fig:3PCFMeasured_000202} and~\ref{fig:3PCFMeasured_110112404} display $\zeta_{000}$, $\zeta_{202}$, $\zeta_{110}$, $\zeta_{112}$ and $\zeta_{404}$ measured from the MultiDark Patchy mock catalogues in the NGC for $0.4<z<0.6$. Since the 3PCF multipoles $\zeta_{\ell_1\ell_2\ell}$ depends on two scales, $r_1$ and $r_2$, for display purposes, we plot them as a function of $r_2$ when fixing $r_1$ to $r_1=r_2$, $r_1=90\hMpc$ and $r_1=140\hMpc$, where we also show the associated best-fit models. For the BAO signal and non-linear gravitational effects found in $\zeta_{000}$ in the case of real-space dark matter shown in Section~\ref{Sec:PredictionsOf3PCF}, a similar behavior is observed even in the case of redshift-space galaxies. In the case of quadrupole components such as $\zeta_{202}$ and $\zeta_{112}$, the BAO signal is largely suppressed by the non-linear effects of the bulk flow, so that only the effect of the trough at $r_1=r_2$ appears. Just to show how our best-fit model looks like at smaller scales, we leave Figures~\ref{fig:3PCFMeasured_000202_small} and~\ref{fig:3PCFMeasured_110112404_small} in Appendix; they are the same as Figures~\ref{fig:3PCFMeasured_000202} and~\ref{fig:3PCFMeasured_110112404}, but extend the multipole components of the 3PCF at scales smaller than $80\hMpc$. We find that when $r_1$ and $r_2$ are both smaller than $80\hMpc$, the best-fit models start to deviate from the measurements from the mocks.

We present the results fitted to the mean of the MD-Patchy 2PCF multipoles in Table~\ref{Table:APlimits} and illustrate the results in Figure~\ref{fig:joint_errors}. We build on four 2PCF-only analyses presented in Section~\ref{Sec:2PCFanalysis} and consider a joint analysis with the 3PCF multipoles in the following five cases. Hence in total we investigate $20$ cases, as labeled in Table~\ref{Table:APlimits} and Figure~\ref{fig:joint_errors}. First, let us consider the first and second terms of the monopole 3PCF. See cases (1), (6), (11) and (16). In all these four cases, we cannot find any improvement in the constraints on $\alpha_{\perp}$ and $\alpha_{\parallel}$ compared to the 2PCF-only analysis. Next, let us consider the first terms of the monopole and quadrupole 3PCFs. See cases (2), (7), (12) and (17). We then find an improvement in $\alpha_{\parallel}$ constraints of $30\%$, $20\%$, $20\%$ and $20\%$ compared to the 2PCF analyses using $\xi_{0,2}$, $\xi_{0,2,4}$, $\xi^{(\rm rec)}_{0,2}$ and $\xi^{(\rm rec)}_{0,2,4}$, respectively. With respect to $\alpha_{\perp}$ constraints, only in case (2), we find an $\sim10\%$ improvement compared to the corresponding 2PCF analysis. Note that after adding $\zeta_{000}$ and $\zeta_{202}$, the results with and without $\xi_4$ hardly change. Finally, we add the second term of the quadrupole 3PCFs ($\zeta_{112}$), the first term of the hexadecapole 3PCFs ($\zeta_{404}$), or both of them to $\zeta_{000}$ and $\zeta_{202}$. See cases (3)-(5), (8)-(10), (13)-(15) and (18)-(20). Even in all of these cases, we cannot find any significant change in the outcome. From these results, we conclude that the quadrupole 3PCFs are effective in constraining the Hubble parameter, and that a combination of data from $\xi_0$, $\xi_2$, $\zeta_{000}$ and $\zeta_{202}$ has the main information on the AP parameters. In the case of this minimum data set, the correction factor $M_2$ coming from the fact that we have a finite number of mocks to estimate the covariance matrix is $1.06$, indicating that our covariance estimate is conservative. The 2D contours in the $\alpha_{\perp}$-$\alpha_{\parallel}$ plane for the $\xi_{0,2}+\zeta_{(000),(202)}$ and $\xi^{\rm (rec)}_{0,2}+\zeta_{(000),(202)}$ analyses are shown in Figure~\ref{fig:AP2d}. 

%
%

We fit each of $100$ individual mock catalogue when we add $\zeta_{000}$ and $\zeta_{202}$ to the 2PCF analysis and summarize the mean and standard deviation of the results in Table~\ref{Table:APlimits_mean}. All of the four cases result in the $\langle \chi^2_{\rm min}\rangle /{\rm dof}$ values that are nearly unity and the $p$-values that are larger than $0.25$, indicating that our template model (\ref{Eq:MainResult}) can explain the measured $\zeta_{000}$ and $\zeta_{202}$ well at $r\geq80\hMpc$. Considering the offset between our mean results and the fiducial value, i.e., $1-\langle \alpha_{\perp}\rangle$ and $1-\langle \alpha_{\parallel}\rangle$, the largest one for $\alpha_{\perp}$ is $0.4\%$ in the $\xi^{(\rm rec)}_{0,2}+\zeta_{(000),(202)}$ analysis, less than $25\%$ of the standard deviation expected in these measurements. The bias in $\alpha_{\parallel}$ is larger than that in $\alpha_{\perp}$ for pre-reconstruction, which is $1-\langle \alpha_{\parallel}\rangle=1.4\%$ in the $\xi_{0,2}+\zeta_{(000),(202)}$ analysis. 2PCF reconstruction decreases the bias of $\alpha_{\parallel}$ and results in $1-\langle \alpha_{\parallel}\rangle=0.1\%$ in the $\xi^{(\rm rec)}_{0,2}+\zeta_{(000),(202)}$ analysis, which is less than $5\%$ of the standard deviation.

We find that the addition of the isotropic and anisotropic 3PCFs ($\zeta_{000}$ and $\zeta_{202}$) reduces the absolute value of $r_{\alpha_{\perp}\alpha_{\parallel}}$ by $70\%$, $70\%$, $40\%$ and $40\%$ compared to the original 2PCF analyses using $\xi_{0,2}$, $\xi_{0,2,4}$, $\xi_{0,2}^{(\rm rec)}$ and $\xi_{0,2,4}^{(\rm rec)}$, respectively, while the 2PCF reconstruction does not significantly reduce $r_{\alpha_{\perp}\alpha_{\parallel}}$, as shown in Table~\ref{Table:APlimits_mean}. In particular, before reconstruction, the value of $r_{\alpha_{\perp}\alpha_{\parallel}}$ is consistent with zero within the $1\sigma$ error. This fact means that we can treat the angular diameter distance and the Hubble parameter as mostly independent information to constrain cosmological parameters such as the equation of state of dark energy.

Figure~\ref{fig:sigma_mocks} displays the uncertainties in $\alpha_{\perp}$ and $\alpha_{\parallel}$ derived from the $\xi_{0,2}$-only analysis vs. those from the joint analysis with $\zeta_{000}$ and $\zeta_{202}$, recovered for each $100$ mock realization. This figure graphically confirms that the errors in $\alpha_{\perp}$ and $\alpha_{\parallel}$ and the correlation coefficient between them tend to be reduced with the 3PCF for both pre- and post-reconstruction of the 2PCF.
Specifically, the possibility that the information on $\zeta_{000}$ and $\zeta_{202}$ provides smaller errors in $\alpha_{\perp}$ and $\alpha_{\parallel}$ is $70\%$ and $96\%$, respectively.

In terms of derived parameters, we complement the results for $\alpha$ and $\varepsilon$ in Tables~\ref{Table:APlimits_derived} and~\ref{Table:APlimits_mean_derived}, and Figures~\ref{fig:AP2d_AE} and \ref{fig:sigma_mocks_AE}. While the addition of the anisotropic 3PCF information does not improve the constraints on $\alpha$, there are $40\%$, $25\%$, $20\%$ and $15\%$ improvements for $\varepsilon$ compared to the 2PCF analyses using $\xi_{0,2}$, $\xi_{0,2,4}$, $\xi_{0,2}^{(\rm rec)}$ and $\xi_{0,2,4}^{(\rm rec)}$, respectively. After 2PCF reconstruction, even $\varepsilon$ appears to show little improvement. Focusing on the correlation coefficient, $r_{\alpha\varepsilon}$, it is clearly less than half as much as the 2PCF results, similar to the case of $r_{\alpha_{\perp}\alpha_{\parallel}}$.


Finally, we show in Tables~\ref{Table:RSD} and~\ref{Table:RSD_mean} the results for $f\sigma_8$, $\sigma_8$ and the bias parameters. We do not find any improvement over the 2PCF results for $f\sigma_8$, and even for $\sigma_8$, the constraints are weak. There are two possible reasons for the large errors in $\sigma_8$ in particular: we focus only on large scales ($r\geq80\hMpc$), and $\sigma_8$ is degenerate with the nonlinear bias parameters such as $b_2$ and $b_{\rm K^2}$.

\subsection{Detection level of BAO}
\label{Sec:DetectionOfBAO}

\begin{table*}
\begin{tabular}{lccccccc} 
 \hline 
\multicolumn{8}{c}{ Patchy mock ($0.4<z<0.6$) } \\
\hline
& $1-\langle \alpha_{\perp}\rangle$ & $\langle \sigma_{\perp}\rangle$ & $1-\langle \alpha_{\parallel}\rangle$ & $\langle\sigma_{\parallel}\rangle$ & $\langle r_{\alpha_{\perp}\alpha_{\parallel}}\rangle$ & $ \langle \chi_{\rm min}^2\rangle/{\rm dof}$ & $\langle \Delta \chi^2_{\rm min} \rangle$\\ 
 \hline \vspace{0.07cm}    
(a) $\xi_{0,2}({\rm NoWiggle})$                                     & -0.001  & 0.065$\pm$0.017 & 0.007 & 0.126$\pm$0.040  & -0.10 $\pm$0.20 & (82.8$\pm$15.6)/55   & 27.3$\pm$10.6\\       
(b) $\xi_{0,2}+\zeta_{(000),(202)}({\rm NoWiggle})$                 &  0.003  & 0.025$\pm$0.007 & -0.014 & 0.048$\pm$0.013 & -0.19 $\pm$0.17 & (243.1$\pm$23.9)/250 & 3.28$\pm$4.05\\       
(c) $\xi_{0,2}({\rm NW})+\zeta_{(000),(202)}({\rm NW})$ & -0.022  & 0.061$\pm$0.009 & -0.054 & 0.096$\pm$0.014 &  0.54 $\pm$0.12 & (265.5$\pm$25.8)/250 & 25.7$\pm$11.1\\       
(d) $\xi_{0,2}+\zeta_{(000),(202)}({\rm Tree})$                     & -0.004 & 0.024$\pm$0.007 & -0.008 & 0.037$\pm$0.009 & -0.11$\pm$0.22 & (242.1$\pm$23.1)/250 & 2.27$\pm$3.33\\ 
 \hline
 \end{tabular} \\ 
\caption{
Repeat of the same analyses in Table~\ref{Table:APlimits_mean} using four different theoretical models: (a) the analysis using the no-wiggle 2PCF model, (b) the joint analysis using the non-linear BAO 2PCF model and the no-wiggle 3PCF model, (c) the joint analysis using the no-wiggle 2PCF and 3PCF models, and (d) the joint analysis using the non-linear BAO 2PCF model and the tree-level 3PCF model.
}
\label{Table:Models}
\end{table*}

\begin{figure*}
	\includegraphics[width=\textwidth]{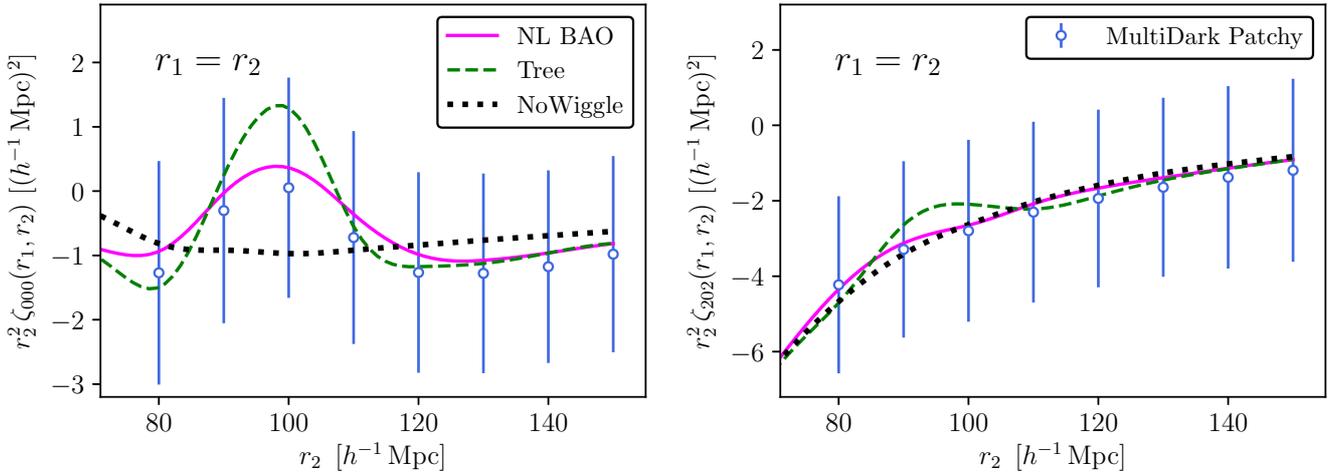}
	\caption{
    Same panels as the top left and top right panels of Figure~\ref{fig:3PCFMeasured_000202}.
    Additionally, the best-fit models of the no-wiggle (black dots) and tree-level (green dashed) solutions of the 3PCFs are shown.
	}
	\label{fig:zeta_BAO}
\end{figure*}

We discuss the detection level of the BAO signal. To do so, we analyze the dataset $\xi_{0,2}+\zeta_{(000),(202)}$ using each of the $100$ mock catalogues for the following three cases: (a) the 2PCF analysis using the no-wiggle 2PCF model, (b) the joint analysis using the non-linear BAO 2PCF model and the no-wiggle 3PCF model, and (c) the joint analysis using the no-wiggle 2PCF and 3PCF models. The results are summarized in Table~\ref{Table:Models}. Using the no-wiggle 2PCF model, the constraints of $\alpha_{\perp}$ and $\alpha_{\parallel}$ are $\langle \sigma_{\perp}\rangle=0.065$ and $\langle \sigma_{\parallel}\rangle=0.126$ in case (a), and $\langle \sigma_{\perp}\rangle=0.061$ and $\langle \sigma_{\parallel}\rangle=0.096$ in case (c), which are much weaker than in the results of Table~\ref{Table:APlimits_mean} for both cases. This fact suggests that the BAO signal on the 2PCF makes a significant contribution to the constraint of the AP parameters, as has been well studied in the past. In case (b), where the no-wiggle model is used only for the 3PCF, the error in $\alpha_{\parallel}$ increases to $\langle \sigma_{\parallel}\rangle=0.048$, indicating that we loose constraining power by $\sim10\%$ when we remove the BAO signal even on the 3PCF. In Figure~\ref{fig:zeta_BAO}, we find a clear difference between the no-wiggle 3PCF model and our non-linear BAO model. However, since case (b) results in a $p$-value of $0.61$, we cannot statistically reject the no-wiggle 3PCF model as long as we use the appropriate model for the 2PCF, which implies that we do not detect the BAO signal in the 3PCF with any high significance in our analysis.

To define the detection of the BAO signal, we calculate $\Delta \chi_{\rm min}$, which is the difference between the $\chi_{\rm min}$ computed in our main analysis presented in Table~\ref{Table:APlimits_mean} and those computed in cases (a), (b) and (c). Then, cases (a) to (c) results in statistical significances of $5.2\sigma$, $1.8\sigma$ and $5.1\sigma$, respectively, for the BAO detection. Our results show that the 3PCF contributes little to increasing the statistical significance for the BAO detection, and case (b) shows that the 3PCF alone can only yield a statistical significance of less than $2\sigma$. As a reminder, in Section~\ref{Sec:Results_3PCF} we showed that for BOSS, including higher-order multipoles such as $\zeta_{110}$ and $\zeta_{112}$ did not increase the S/N of the BAO detection. This however might change for other datasets at different redshift. In addition, if one increases the number of multipole components to be considered, the value of $M_2$ (\ref{Eq:M2}) will increase. See Section~\ref{Section:NumberOfMockRealizations} for a discussion of the impact of changes in the $M_2$ value on the results. From these results, we conclude that the reduction of $\alpha_{\parallel}$ (or $\varepsilon$) errors by adding the 3PCF is not due to the BAO signal in the 3PCF, but rather to the effect of the shape of the anisotropic 3PCF.

\subsection{Versus the tree-level solution}
\label{Sec:TheoreticalModels}

To demonstrate the validity of our template model (\ref{Eq:MainResult}), we compared our main result presented in Table~\ref{Table:APlimits_mean} with the results of the analysis using the tree-level model. We have performed a joint analysis using the non-linear BAO 2PCF model and the tree-level 3PCF model, and the results are summarized as case (d) in Table~\ref{Table:Models}. Figure~\ref{fig:zeta_BAO} displays the best-fit tree-level 3PCF model. $\Delta \chi_{\rm min}$ is calculated in the same way as in Section~\ref{Sec:DetectionOfBAO}. The results show that the tree level model reduces the error in $\alpha_{\parallel}$ by $\sim15\%$ compared to the results in Table~\ref{Table:APlimits_mean} due to the emphasis on the BAO signal. If we assume that $\Delta \chi_{\rm min}$ has a Gaussian distribution, the probability that our model gives a better fit, i.e., the probability that $\Delta \chi_{\rm min}$ has a positive value, is $75\%$. Based on these results, we conclude that it is safer to use our 3PCF model than using the tree-level solution, though as with the no-wiggle model, we cannot statistically reject the tree-level solution because the $p$-value in case (d) is $0.63$. 

\subsection{Number of mock realizations}
\label{Section:NumberOfMockRealizations}

\begin{table*}
\begin{tabular}{lcccccccc} 
 \hline 
\multicolumn{9}{c}{ Patchy mock ($0.4<z<0.6$) } \\
\hline   
& $N_{\rm s} $ & $\Delta r$ [$\hMpc$] & $1-\alpha_{\perp}$ & $\sigma_{\perp}$ & $1-\alpha_{\parallel}$ & $\sigma_{\parallel}$ & $ r_{\alpha_{\perp}\alpha_{\parallel}}$ & $M_2$ \\ 
 \hline   
 $\xi_{0,2}+\zeta_{(000),(202)}$ & 500  &  $10$ & 0.001 & 0.021 & -0.008 & 0.042 & -0.17 & 1.321 \\
 $\xi_{0,2}+\zeta_{(000),(202)}$ & 1000 &  $10$ & 0.004 & 0.022 & -0.005 & 0.042 & -0.09 & 1.137 \\
 $\xi_{0,2}+\zeta_{(000),(202)}$ & 1500 &  $10$ & 0.002 & 0.022 & -0.007 & 0.042 & -0.04 & 1.087 \\
 $\xi_{0,2}+\zeta_{(000),(202)}$ & 2048 &   $5$ & 0.007 & 0.023 & -0.011 & 0.040 & 0.053 & 1.218 \\
  \hline
  \end{tabular}
 \caption{
 Repeat of the same analyses in case (2) of Table~\ref{Table:APlimits} using three different numbers of the Patchy mock catalogues
 to estimate the covariance matrix and a different bin width $\Delta r$.
}
\label{Table:M2}
\end{table*}

We discuss the impact of changes in the number of the MD-Patchy mock catalogs on the results when estimating the covariance matrix. We compute three covariance matrices using $500$, $1000$, and $1500$ catalogs and then use them to analyze the data. We have performed an analysis using $\xi_{0,2}+\zeta_{(000),(202)}$ with $130$ bins, corresponding to case $(2)$ in Table~\ref{Table:APlimits}. The results are summarized in Table~\ref{Table:M2} and illustrated in Figure~\ref{fig:realizations}. In this case, which has $130$ data bins, we find that if we use a number of $500$ mock catalogues, the error in $\alpha_{\perp}$ is $\sim 10\%$ less than if we use $2048$ catalogues. Thus, even using about $4$ times the number of catalogues as the number of data bins, the estimated error results are not fully converged. This fact indicates that for a more conservative analysis, it is important to keep the value of $M_2$ as close to $1$ as possible, as in our main result corresponding to $M_2 = 1.06$~\footnote{\Mod{Instead of the $M_2$ parameter, a more improved method to correct for the effect of a finite number of simulations have been proposed by~\citep{Sellentin:2016}. This method requires fewer simulations and is expected to allow us to use higher multipole components safely.}}.

\subsection{A narrower bin size for 3PCFs}
\label{Sec:NarrowerBinSize}

So far in this paper, we have used a wider bin size, $\Delta r = 10\hMpc$, in our 3PCF analysis than $\Delta r=5\hMpc$ used in our 2PCF analysis. Here, we show in Table~\ref{Table:M2} the results for the 3PCF analysis when we used the same bin size, i.e., $\Delta r = 5\hMpc$, for the 3PCF analysis as for the 2PCF analysis. We have performed the same analysis as in case $(2)$ in Table~\ref{Table:APlimits}, where we have used $\xi_{0,2}+\zeta_{(000),(202)}$. In this case, the number of bins is $375$ and $M_2=1.22$. With respect to $\alpha_{\parallel}$, the error is reduced by about $7\%$ compared to the corresponding result in Table~\ref{Table:APlimits}. However, we cannot determine whether this reduction in the $\alpha_{\parallel}$ error is the result of using the narrow bin width or whether it is due to the fact that the number of the mock catalogues is too small compared to the number of data bins. In the analysis using $\Delta r = 10\hMpc$, the estimates of $\alpha_{\perp}$ and $\alpha_{\parallel}$ return sufficiently low bias values, and the value of $M_2$ is also kept to a small value of $1.06$. Therefore, to be conservative, we adopt the results in Tables~\ref{Table:APlimits} and \ref{Table:APlimits_mean} with $\Delta r = 10\hMpc$ as our main result. 

\subsection{Test of the AP effect}
\label{Sec:TestOfAP}

\begin{table*}
\begin{tabular}{lcccccc} 
 \hline 
\multicolumn{7}{c}{ Patchy mock ($0.4<z<0.6$) } \\ 
\hline
& $1.0575-\langle \alpha_{\perp}\rangle$ & $\langle \sigma_{\perp}\rangle$ & $1.1034-\langle \alpha_{\parallel}\rangle$ & $\langle\sigma_{\parallel}\rangle$ & $\langle r_{\alpha_{\perp}\alpha_{\parallel}}\rangle$ & $ \langle \chi_{\rm min}^2\rangle/{\rm dof}$\\ 
 \hline 
    $\xi_{0,2}$                                     & 0.0045  & 0.027$\pm$0.007 &  0.007 & 0.110$\pm$0.063 & -0.40 $\pm$0.08 & (54.4$\pm$10.6)/55   \\ 
    $\xi_{0,2}+\zeta_{(000),(202)}$                 & -0.0055 & 0.024$\pm$0.006 & -0.011 & 0.049$\pm$0.013 & -0.07 $\pm$0.17 & (232.5$\pm$20.2)/250 \\   
 \hline 
 $\xi^{(\rm rec)}_{0,2}$                            & 0.0031  & 0.020$\pm$0.004 & -0.003 & 0.042$\pm$0.014 & -0.43 $\pm$0.03 & (55.2$\pm$9.1)/55 \\   
 $\xi^{(\rm rec)}_{0,2}+\zeta_{(000),(202)}$        & -0.0024 & 0.019$\pm$0.004 &  0.001 & 0.034$\pm$0.008 & -0.21 $\pm$0.12 & (233.4$\pm$19.1)/250 \\   
 \hline
 \end{tabular} \\ 
\caption{
Similar to Table~\ref{Table:APlimits_mean}. 
The results are analyzed from the newly computed Patchy mock catalogues using $w_0=-0.7$ and $w_a=0.3$.
The expected values of the AP parameters are $\alpha_{\perp}=1.0575$ and $\alpha_{\parallel}=1.1034$.
}
\label{Table:TestAP}
\end{table*}

The previous analyses in this paper have verified that the estimated $\alpha_{\perp}$ and $\alpha_{\parallel}$ return unity in measurements where no AP effect is present, i.e., where $\alpha_{\perp}=\alpha_{\parallel}=1$. In this subsection, we test whether the parameter estimation can correctly yield the expected values of $\alpha_{\perp}$ and $\alpha_{\parallel}$ even when an AP effect is present, i.e., when $\alpha_{\perp}\neq 1$ and $\alpha_{\parallel}\neq 1$. This analysis is another validation of the approximate calculation of the anisotropy of the AP effect in the 3PCF given in Section~\ref{Sec:APeffects}.

We compute the radial distance to the galaxies using different cosmological parameters than those used to generate the Patchy mock catalogues and generate new 3D distributions of the galaxies. Specifically, we add the equation-of-state parameters for dark energy, $w_0=-0.7$ and $w_a=0.3$, to the fiducial parameters given in the introduction. In this case, the expected values of the AP parameters are $\alpha_{\perp}=1.0575$ and $\alpha_{\parallel}=1.1034$. For these newly created 3D galaxy catalogues, we repeat the same analysis as done in Table~\ref{Table:APlimits_mean} and show the results in Table~\ref{Table:TestAP}. The estimated biases of $\langle \alpha_{\perp}\rangle$ and $\langle \alpha_{\parallel}\rangle$ are sufficiently small, e.g., less than $\sim10\%$ of the standard deviation for the analysis using $\xi^{(\rm rec)}_{0,2}+\zeta_{(000),(202)}$. This result validates the approximation method given in Section~\ref{Sec:APeffects} to calculate the AP effect on the 3PCF. We therefore conclude that we can safely apply our analysis to actual observed galaxy data for which the AP parameters should not be 1.

\section{Conclusions}
\label{Sec:Conclusions}

We have established a joint analysis of anisotropic galaxy two- and three-point correlation functions (2PCF and 3PCF) in the range $80\leq r\leq 150\hMpc$. Our analysis is based on a decomposition formalism of the 3PCF using tri-polar spherical harmonics (\ref{Eq:TripoSH}). The reason why we have focused on the 3PCF instead of the bispectrum in Fourier space is that by looking only at scales larger than $80\hMpc$, it is possible to extract almost all of the BAO information that appears around $100\hMpc$. At such large scales, the non-linearity is weak, which simplifies the construction of the theoretical model of the 3PCF, and furthermore, the number of data bins required for analysis can be reduced. We apply density field reconstruction to the 2PCF measurement, but not in the 3PCF measurement. Specifically, we have (1) built a theoretical model of the bispectrum (3PCF) that incorporates the effects of BAO nonlinear damping (Section~\ref{Sec:Model}), (2) established an approximation to the fast computation of the theoretical model (Section~\ref{Sec:ThreePointCorrelationFunctions}), (3) applied our analysis method to the MultiDark Patchy mock catalogues to investigate which terms in the TripoSH expansion coefficients have the main cosmological information (Section~\ref{Sec:Results_3PCF}), and (4) examined various systematic errors in our analysis, \Mod{such as the comparison with the no-wiggle model (Section~\ref{Sec:DetectionOfBAO}), the comparison with the tree-level solution (Section~\ref{Sec:TheoreticalModels}), the change in the number of mock realizations to estimate the covariance matrix (Section~\ref{Section:NumberOfMockRealizations}), the change in data bin size of the measured 3    PCF (Section~\ref{Sec:NarrowerBinSize}), and the estimate of biases from the expected values of the AP parameters (Section~\ref{Sec:TestOfAP}). In particular, we highlight the result in Section~\ref{Sec:TestOfAP}, which confirms that the expected values of the AP parameters can be returned reliably even in the case of $\alpha_{\perp}\neq 1$ and $\alpha_{\parallel}\neq 1$, i.e., in the presence of the AP effect.} In conclusion, we find that, in addition to the monopole and quadrupole components of the 2PCF ($\xi_0(r)$ and $\xi_2(r)$), we can extract almost all the information for the angular diameter distance and the Hubble parameter by considering only the first terms of the monopole and quadrupole components of the 3PCF expanded using the TripoSH basis, i.e., $\zeta_{000}(r_1,r_2)$ and $\zeta_{202}(r_1,r_2)$; We have found the quadrupole component of the 3PCF to be particularly useful in constraining the Hubble parameter. 

Our bispectrum model (\ref{Eq:MainResult}) is an analogy to the power spectrum model widely used in BAO analysis given by~\citet{Eisenstein:2006nj}. The shape of the bispectrum model is described by a tree-level solution consisting of no-wiggle power spectra with no BAO, while including the nonlinear damping of the BAO signature. The following three steps are required to build our bispectrum model. The first step is to expand the bispectrum using the $\Gamma$-expansion according to~\citet{Bernardeau:2008fa} and classify it according to the structure of the mode-coupling integral (Section~\ref{Sec:GammaExpansion}). The second step is to focus on the long-wavelength (infra-red) mode of flow of galaxies (IR flow) and to describe the non-linearity from the IR flow using the $\Gamma$-expansion: in the limit where the IR flow is uncorrelated with the density perturbation, all the non-liniearities from the IR flow cancel each other out, resulting in the tree-level bispectrum (Sections~\ref{Sec:InfraRedFlow} and~\ref{Sec:Cancellation}). As the third step, we replace the linear power spectrum appearing in the mode-coupling terms of the bispectrum arising from the IR flow with no-wiggle power spectra without BAO (Section~\ref{Sec:TemplateModel}). These three operations yield the bispectrum model used in our analysis.

In describing the nonlinear terms arising from the IR flow using the $\Gamma$-expansion, we can estimate the scale at which the mode-coupling terms begin to dominate in the bispectrum. It turns out that the mode-coupling terms start to become more important at larger scales for higher order multipole components such as $\zeta_{110}$ and $\zeta_{112}$ than for $\zeta_{000}$ and $\zeta_{202}$, which are of particular interest to us in this paper (Figure~\ref{fig:bk_IR}). Thus, the model building would be much easier if one can show that it is sufficient to include $\zeta_{000}$ and $\zeta_{202}$ without the other higher multipole components.

A 2D Hankel transform (\ref{Eq:B_to_zeta}) of the bispectrum model expanded in the TripoSH basis allows us to compute a theoretical model of the corresponding multipole components of the 3PCF. The TripoSH-expanded 3PCF is more attenuated at $r_1 = r_2$ than at $r_1 \neq r_2$ (the right panel in Figure~\ref{fig:3PCF_RealSpace}). This is due to the terms called "shift" and "tidal force" in the second-order non-linear gravity (\ref{Eq:2ndOrder}). As expected, we find the BAO peak around $100\hMpc$ in $\zeta_{000}$ (top of the left panel in Figure~\ref{fig:3PCF_RealSpace}). Interestingly, when we fix $r_1=90\hMpc$ and plot $\zeta_{000}(r_1,r_2)$ as a function of $r_2$, the trough at $r_1=r_2$ (which is more obvious in the right panel) closes in on the BAO peak at $\sim 100\hMpc$ and the two partially cancel each other out and neither signal is clearly visible (middle of the left panel in Figure~\ref{fig:3PCF_RealSpace}).

Since we measure the 3PCF using an FFT estimator presented by~\citet{Sugiyama:2018yzo}, we need to consider the effect of the survey geometry in the theoretical model in order to explain the results of our measurements. In the TripoSH expansion, a method to correct for the effect of the survey geometry on the 3PCF is given in~\citet{Sugiyama:2018yzo}. Although this method, in principle, requires an infinite number of multipole components to account for a single observed multipole component, we have tested the convergence of the expansion formalism in Eq.~(\ref{Eq:zetaMask}) for BOSS in this paper. To do so, we have considered several multipole components of the 3PCF: $9$ for monopole, $15$ for quadrupole, $10$ for hexadecapole, and $7$ for tetrahexacontapole, for a total of $41$. For these $41$ components, we have calculated the theoretical model of the 3PCF and measured the window function 3PCF, respectively, to determine which multipole components contribute to the final result and to what extent. As a result, we have confirmed that the contribution tends to be smaller for higher order multipole components. In our analysis, we have decided to treat only the terms that contribute more than $0.5\%$ of these $41$ components (Tables~\ref{Table:zeta_window} and~\ref{Table:Q_window}).

In order to speed up the computation of the theoretical model of the 3PCF, we fix the shape of linear power spectra and the smoothing parameters, $\Sigma_{\perp}$ and $\Sigma_{\parallel}$ (\ref{Eq:Damping}), that describes the non-linear erasing of BAO. That is, we only vary parameters related to the amplitude of the 3PCF, such as $f\sigma_8$, $\sigma_8$, and the bias parameters ($b_1$, $b_2$, and $b_{\rm K^2}$). The 3PCF is then decomposed into $14$ terms (\ref{Eq:Bp}) that depend only on the linear power spectrum, $\Sigma_{\perp}$ and $\Sigma_{\parallel}$. By calculating them in advance before analyzing the data, we can create a table of the resulting data and simply load the table when we perform the data analysis.

In order to quickly calculate the dependence of the anisotropic components of the AP parameters parameterized in $\varepsilon$ (\ref{Eq:AAEE}) on the 3PCF, we expand the 3PCF with respect to $\varepsilon$ (\ref{Eq:zeta_approx_AP}). Then, by precomputing the terms that depend only on $\varepsilon$ and creating a table of the resulting data, we can calculate the $\varepsilon$-dependence on the 3PCF by simply reading that table. We have confirmed that this approximation method can calculate the 3PCF multipoles of interest with an accuracy of $2\%$ in the range of $-0.02<\varepsilon<0.04$ (Figure~\ref{fig:zeta_E}).

We analyze both the pre- and post-reconstruction 2PCFs in the range $80\leq r \leq 150\hMpc$. In this case, we use the template model of Eq.~(\ref{Eq:Eisenstein2007}) for the theoretical calculation of the 2PCF without any nuisance parameter. This means that our analysis is in principle not a BAO analysis, but an RSD analysis. In particular, even though we only focus on large scales ($r\geq 80\hMpc$), our analysis is the first rigorous RSD analysis to use the 2PCF after reconstruction. As for the constraints on the AP parameters (Table~\ref{Table:APlimits_mean}), our results using $\xi_0$ and $\xi_2$ measured from the MD-Patchy mock catalogues at $0.4<z<0.6$ have $\alpha_{\perp}=0.997\pm0.026$, $\alpha_{\parallel}=1.011\pm0.061$ and $r_{\alpha_{\perp}\alpha_{\parallel}}=-0.48$ for pre-reconstruction, and $\alpha_{\perp}=0.996\pm0.018$, $\alpha_{\parallel}=1.000\pm0.035$ and $r_{\alpha_{\perp}\alpha_{\parallel}}=-0.45$ for post-reconstruction, which are in good agreement with the previous studies of the BAO-only analysis~\citep{Ross:2016gvb,Beutler:2016ixs}. The results for $f\sigma_8$ (Table~\ref{Table:RSD_mean}) are $0.46\pm0.085$ for pre-reconstruction and $0.45\pm0.077$ for post-reconstruction, which are weaker than in the previous studies of the full shape clustering~\citep{Satpathy:2016tct,Beutler:2016arn}, since we have only used large scales above $80\hMpc$; after reconstruction, as expected, the error in $f\sigma_8$ is about $10\%$ smaller than the result before reconstruction.

We have measured five multipole components, $\zeta_{000}$, $\zeta_{110}$, $\zeta_{202}$, $\zeta_{112}$, and $\zeta_{404}$, and used them to perform a joint analysis with the 2PCF (Table~\ref{Table:APlimits}). In the analysis of the 3PCF, we pay particular attention to the value of $M_2$ (\ref{Eq:M2}), a factor that compensates for the uncertainty in the estimated errors in parameters arising from the fact that the number of simulations used to measure the covariance matrix is finite, and we believe that the closer $M_2$ is to $1$, the more conservative the analysis is. For this reason, we treat a minimum dataset $\xi_0+\xi_2+\zeta_{000}+\zeta_{202}$ as the main analysis of our paper, and we have confirmed that even adding other components such as $\zeta_{112}$ and $\zeta_{404}$ does not change the results significantly. In the main analysis, where the corresponding $M_2$ value is $1.06$, the error in the Hubble parameter is $\sim 30\%$ better before the 2PCF reconstruction (i.e., $\xi_{0,2}+\zeta_{(000),(202)}$) and $\sim 20\%$ better after the 2PCF reconstruction (i.e., $\xi_{0,2}^{\rm (rec)}+\zeta_{(000),(202)}$) than the results using the 2PCF alone. Furthermore, the absolute value of the anti-correlation coefficient ($r_{\alpha_{\perp}\alpha_{\parallel}}$) between the angular diameter distance and the Hubble parameter is reduced by $\sim 70\%$ before the 2PCF reconstruction and $\sim40\%$ after the 2PCF reconstruction. This is a major feature of adding the anisotropic 3PCF, considering that the $r_{\alpha_{\perp}\alpha_{\parallel}}$ with the pre- or post-reconstruction 2PCF alone remains near $-0.44\sim-0.49$ that is expected for the 2PCF BAO-only analysis. That is, adding 3PCF changes the degeneracy direction. Specifically, from the mean of the results from $100$ each independent mock catalogue (Table~\ref{Table:APlimits_mean}), we obtain $\alpha_{\perp}=1.001\pm0.023$, $\alpha_{\parallel}=1.014\pm0.043$ and $r_{\alpha_{\perp}\alpha_{\parallel}}=-0.16$ before the 2PCF reconstruction, and $\alpha_{\perp}=0.996\pm0.017$, $\alpha_{\parallel}=1.001\pm0.028$ and $r_{\alpha_{\perp}\alpha_{\parallel}}=-0.28$ after the 2PCF reconstruction. The constraint on the growth rate, $f\sigma_8=0.46\pm0.089$ (Table~\ref{Table:RSD_mean}), does not improve at all compared to the results of our analysis using only the 2PCF. This is due to the fact that we only use large scales ($r\geq 80\hMpc$). For similar reasons, the $\sigma_8$ constraint, $\sigma_8=0.96\pm0.55$, obtained from the 3PCF is also very weak.

In our main analysis using $\xi_{0,2}+\zeta_{(000),(202)}$, we have not detected the BAO signal in the 3PCF with any high significance: the statistical significance is $1.8\sigma$. Therefore, we conclude that the reduction of the error in the Hubble parameter in our analysis is not due to the BAO signal in the 3PCF, but rather to the changes in the shape of $\zeta_{000}$ and $\zeta_{202}$ due to the anisotropic AP effect (Section~\ref{Sec:DetectionOfBAO}). 

The possibility that our template model (\ref{Eq:MainResult}) gives a better fit to the measurement than the tree-level model is $75\%$; if we use the tree-level model, the error in the Hubble parameter is reduced compared to our main result because of the emphasis on the BAO signal. Therefore, it is safer to use our model for a conservative analysis (Section~\ref{Sec:TheoreticalModels}). \Mod{For future galaxy surveys, such as DESI, this difference in theoretical models should be more pronounced.}

Table~\ref{Table:M2} and Figure~\ref{fig:realizations} summarize the differences in results for different numbers of mock simulations to compute the covariance matrix. The number of data bins in our main analysis using $\xi_{0,2}+\zeta_{(000),(202)}$ is $130$. We have found that using a number of simulations of e.g. $500$ underestimates the error in the diameter distance by $\sim10\%$, even after accounting for the correction by the $M_2$ factor (\ref{Eq:M2}). Similarly, the same table and figure summarize the results when the bin width of the 3PCF measurement has been decreased to $\Delta r =5\hMpc$ and the number of data bins has been increased to $375$. Then, we have found a $7\%$ reduction in the Hubble parameter error when using $2048$ mock simulations compared to the results of our main analysis. We leave it to future work to find out if a similar reduction in the errors is seen when we increase the number of simulations beyond $2048$. As long as we only use $2048$ mock catalogues, for a conservative analysis we propose to continue to use $\Delta r = 10\hMpc$. 

Finally, we have tested whether our analysis can correctly estimate the expected values of the angular diameter distance and the Hubble parameter in the presence of an AP effect, i.e., if $\alpha_{\perp}$ and $\alpha_{\parallel}$ are not $1$. To do so, we have measured the distance to the mock galaxies using the equation-of-state parameters for dark energy, $w_0=-0.7$ and $w_a=0.3$, to create a new 3D mock galaxy catalogues. The expected value of the AP parameters are then $\alpha_{\perp}=1.0575$ and $\alpha_{\parallel}=1.1034$. Our results (Table~\ref{Table:TestAP}) show $\alpha_{\perp}=1.063\pm0.024$ and $\alpha_{\parallel}=1.1144\pm0.049$ for $\xi_{0,2}+\zeta_{(000),(202)}$, and $\alpha_{\perp}=1.0599\pm0.019$ and $\alpha_{\parallel}=1.1024\pm0.034$ for $\xi^{\rm (rec)}_{0,2}+\zeta_{(000),(202)}$. Thus, for the joint analysis of the post-reconstruction 2PCF and the pre-reconstruction 3PCF, the biases in $\alpha_{\perp}$ and $\alpha_{\parallel}$ are less than $10\%$ of the corresponding $1$-$\sigma$ errors. 
This test confirms the validity of applying our analysis method to real observed galaxy data, which should have $\alpha_{\perp}\neq1$ and $\alpha_{\parallel}\neq1$.

There are a variety of possible next research directions. The first and most obvious one is to apply our analysis method to actual observed galaxy data. The analysis of the 3PCF after reconstruction would also be straightforward. A conservative analysis of the 3PCF using up to small scales would require building a theoretical model of the 3PCF (or the bispectrum) to explain the observations and either increasing the number of mock simulations or calculating the covariance matrix of the 3PCF from an analytical approach~\citep{Sugiyama:2019ike,Philcox:2019xzt}. We expect that our 3PCF analysis will also play a major role in the probe of primordial non-Gaussianity~\citep{Shirasaki:2020vkk}. 

\section*{Acknowledgements}

NSS is grateful to Masato Shirasaki for very useful discussion. NSS acknowledges financial support from JSPS KAKENHI Grant Number 19K14703. Numerical computations were carried out on Cray XC50 at Center for Computational Astrophysics, National Astronomical Observatory of Japan. FB is a Royal Society University Research Fellow.

\section*{Data availability}
The data underlying this article are available at the SDSS data base (\url{https://www.sdss.org/dr12/}).

\bibliographystyle{mnras}
\bibliography{ms} 

\appendix

\section{Unbiased estimator of 3PCFs}
\label{Ap:Window}

The estimator of the 3PCF depending on $\VEC{r}_1$, $\VEC{r}_2$ and $\hat{n}$ is given by
\begin{eqnarray}
    \hspace{-0.7cm}&&\widehat{\zeta}(\VEC{r}_1,\VEC{r}_2,\hat{n}) \nonumber \\
    \hspace{-0.25cm}&=&\hspace{-0.25cm}
    \frac{1}{V}\int d^3x_1\int d^3x_2\int d^3x_3 \delta_{\rm D}\left( \VEC{r}_1-\VEC{x}_{13} \right)\left( \VEC{r}_2-\VEC{x}_{23} \right)
      \nonumber \\
    \hspace{-0.25cm}&\times&\hspace{-0.25cm}
    \delta_{\rm D}\left( \hat{n} - \hat{x}_3 \right)\, \delta_{\rm obs}(\VEC{x}_1)\, \delta_{\rm obs}(\VEC{x}_2)\, \delta_{\rm obs}(\VEC{x}_3),
    \label{Eq:zeta_rrn}
\end{eqnarray}
where $\VEC{x}_{13}=\VEC{x}_1-\VEC{x}_3$ and $\VEC{x}_{23}=\VEC{x}_2-\VEC{x}_3$, $V$ (\ref{Eq:V}) denotes the survey volume, $\delta_{\rm obs}$ (\ref{Eq:delta_obs}) is the observed density fluctuation, 
and $\hat{n}$ is the LOS direction under the assumption of the local plane-parallel approximation, i.e., $\hat{x}_1\approx\hat{x}_2\approx\hat{x}_3$.
Assuming that $\delta_{\rm obs}(\VEC{x})$ can be described by a product of the window function $W(\VEC{x})$ and the theoretically predicted $\delta(\VEC{x})$, where $W(\VEC{x})=V\, R(\VEC{x}) / \int d^3x R(\VEC{x})$ with $R(\VEC{x})$ being the density field of random particles, the ensemble average of $\widehat{\zeta}(\VEC{r}_1,\VEC{r}_2,\hat{n})$ is then calculated as
\begin{eqnarray}
    \langle \widehat{\zeta}(\VEC{r}_1,\VEC{r}_2,\hat{n}) \rangle
    = Q(\VEC{r}_1,\VEC{r}_2,\hat{n})\, \zeta(\VEC{r}_1,\VEC{r}_2,\hat{n}),
\end{eqnarray}
where the theoretical prediction of the 3PCF is
\begin{eqnarray}
    \zeta(\VEC{r}_1,\VEC{r}_2,\hat{n}) = \langle \delta(\VEC{x}_1)\delta(\VEC{x}_2)\delta(\VEC{x}_3) \rangle,
\end{eqnarray}
and the 3PCF of the window function is 
\begin{eqnarray}
    \hspace{-0.7cm}&&Q(\VEC{r}_1,\VEC{r}_2,\hat{n}) \nonumber \\
    \hspace{-0.25cm}&=&\hspace{-0.25cm}
    \frac{1}{V}\int d^3x_1\int d^3x_2\int d^3x_3 \delta_{\rm D}\left( \VEC{r}_1-\VEC{x}_{13} \right)\left( \VEC{r}_2-\VEC{x}_{23} \right)
      \nonumber \\
    \hspace{-0.25cm}&\times&\hspace{-0.25cm}
    \delta_{\rm D}\left( \hat{n} - \hat{x}_3 \right)\, W(\VEC{x}_1)\, W(\VEC{x}_2)\, W(\VEC{x}_3).
\end{eqnarray}
Therefore, the unbiased estimator of the 3PCF is obtained by dividing $\widehat{\zeta}(\VEC{r}_1,\VEC{r}_2,\hat{n})$ by $Q(\VEC{r}_1,\VEC{r}_2,\hat{n})$:
\begin{eqnarray}
    \left\langle \frac{\widehat{\zeta}(\VEC{r}_1,\VEC{r}_2,\hat{n}) }{Q(\VEC{r}_1,\VEC{r}_2,\hat{n})} \right\rangle = \zeta(\VEC{r}_1,\VEC{r}_2,\hat{n}).
    \label{Eq:zeta_unbiased}
\end{eqnarray}
We can measure such an unbiased 3PCF by direct pair-counting of galaxies, like the estimator of the 2PCF proposed by \citet{Landy:1993yu}.
However, we usually measure the multipole component of the 3PCF to increase the speed of our measurements. Specifically, 
substituting Eq.~(\ref{Eq:zeta_rrn}) into the formalism of the TripoSH expansion, given by
\begin{eqnarray}
	\zeta_{\ell_1\ell_2\ell}(r_1,r_2)
	\hspace{-0.25cm}&=&\hspace{-0.25cm}
    4\pi\, h_{\ell_1\ell_2\ell}^2
	\int \frac{d^2\hat{r}_1}{4\pi}\int \frac{d^2\hat{r}_2}{4\pi}\int \frac{d^2\hat{n}}{4\pi} \nonumber \\
	\hspace{-0.25cm}&\times&\hspace{-0.25cm}
	{\cal S}^*_{\ell_1\ell_2\ell}(\hat{r}_1,\hat{r}_2,\hat{n}) \zeta(\VEC{r}_1,\VEC{r}_2,\hat{n}),
	\label{Eq:zeta_multipole}
\end{eqnarray}
leads to Eq.~(\ref{Eq:zetaMask}).
Then, we can no longer treat the effects of the window function in such a simple way as in Eq.~(\ref{Eq:zeta_unbiased}), and we need to estimate the effect of the multipole component of the window 3PCF, as described in detail in Section~\ref{Sec:WindowFunctionEffects}.

\section{Supplemental tables and figures}

We summarize several supplemental tables and figures in this paper.

\begin{figure*}
    \scalebox{0.8}{\includegraphics[width=\textwidth]{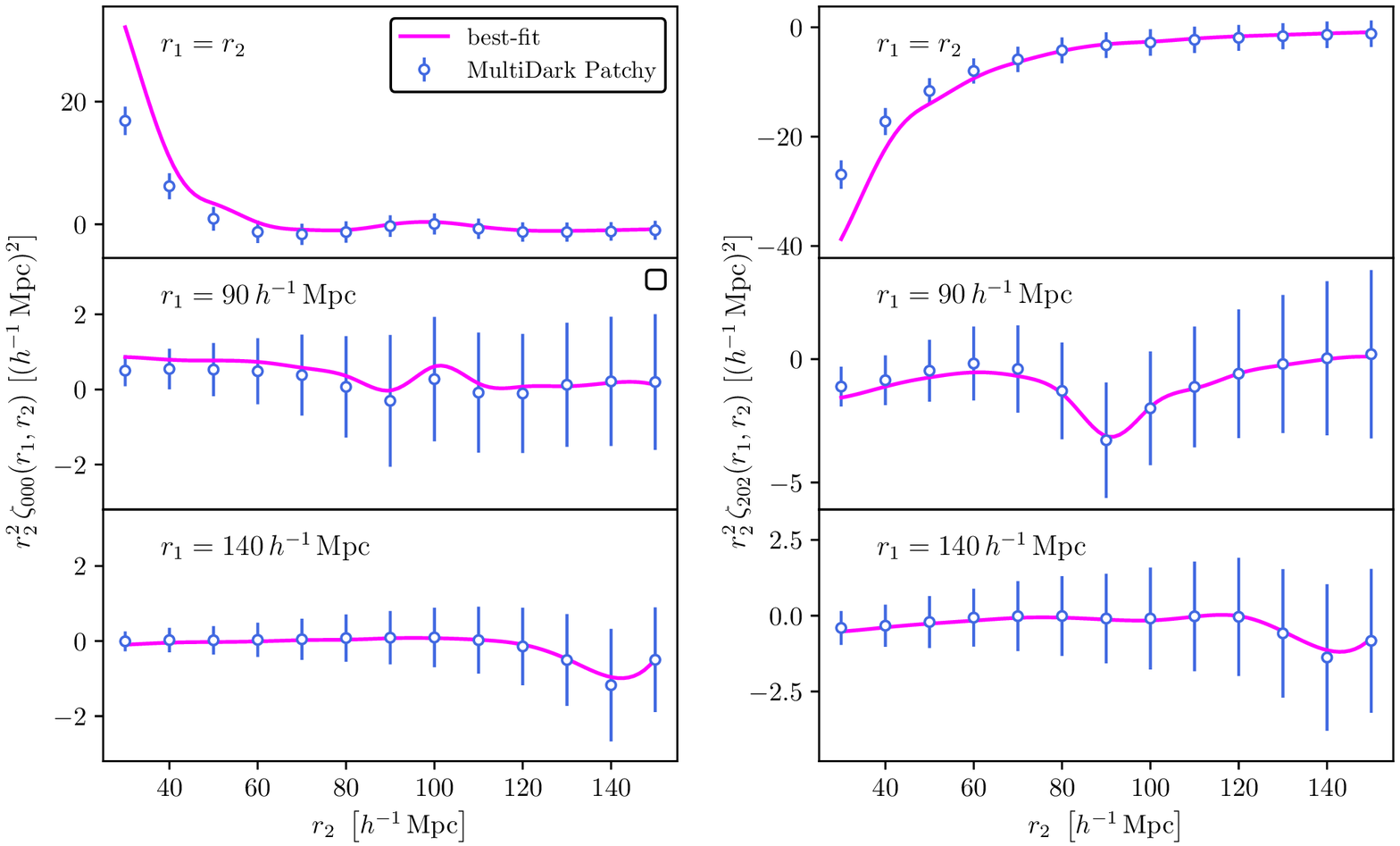}}
	\caption{
    Same as Figure~\ref{fig:3PCFMeasured_000202}, shown up to $r_2=30\hMpc$.
	}
	\label{fig:3PCFMeasured_000202_small}
    \scalebox{0.8}{\includegraphics[width=\textwidth]{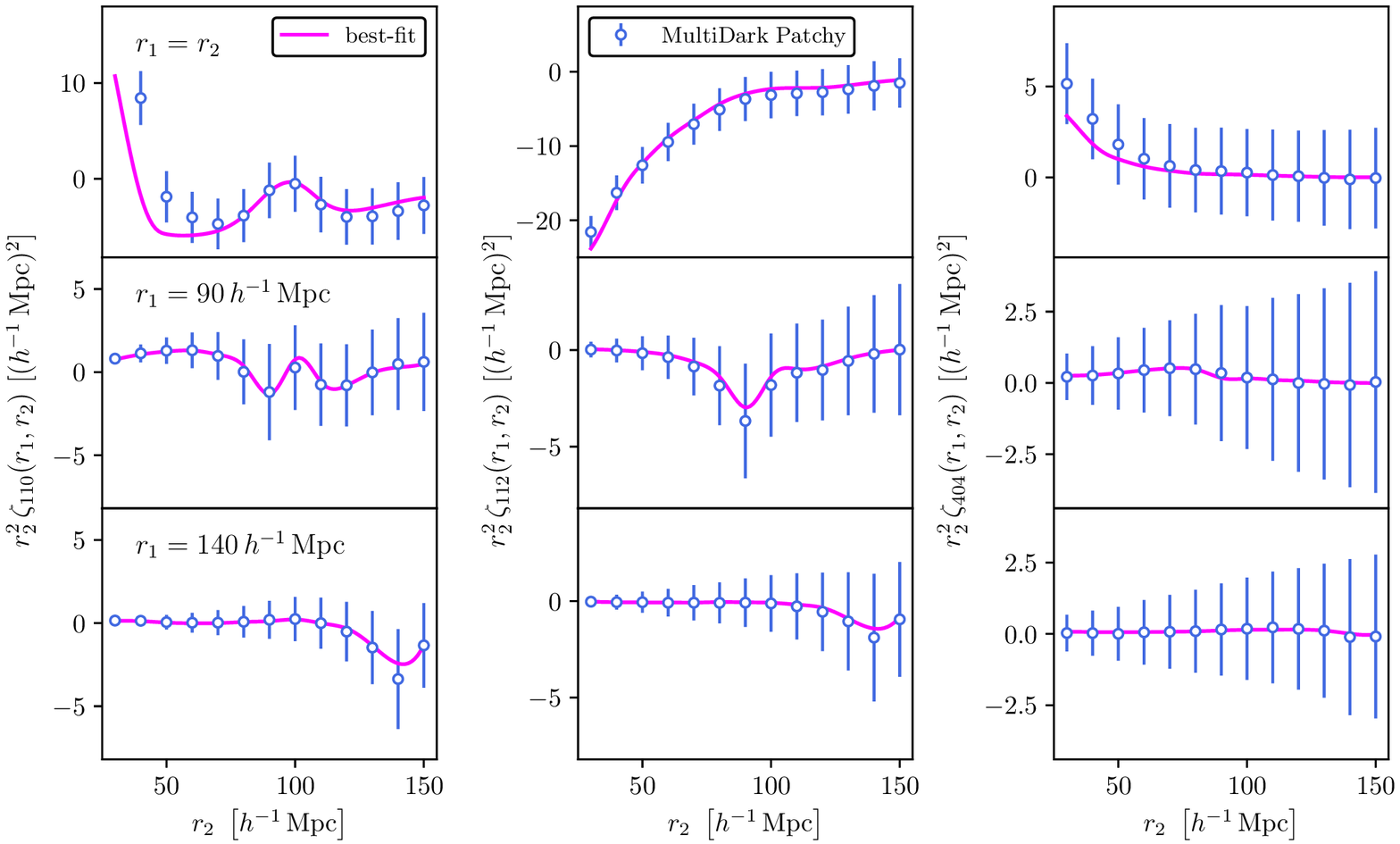}}
	\caption{
    Same as Figure~\ref{fig:3PCFMeasured_110112404}, shown up to $r_2=30\hMpc$.
	}
	\label{fig:3PCFMeasured_110112404_small}
\end{figure*}

\begin{figure*}
	\includegraphics[width=\textwidth]{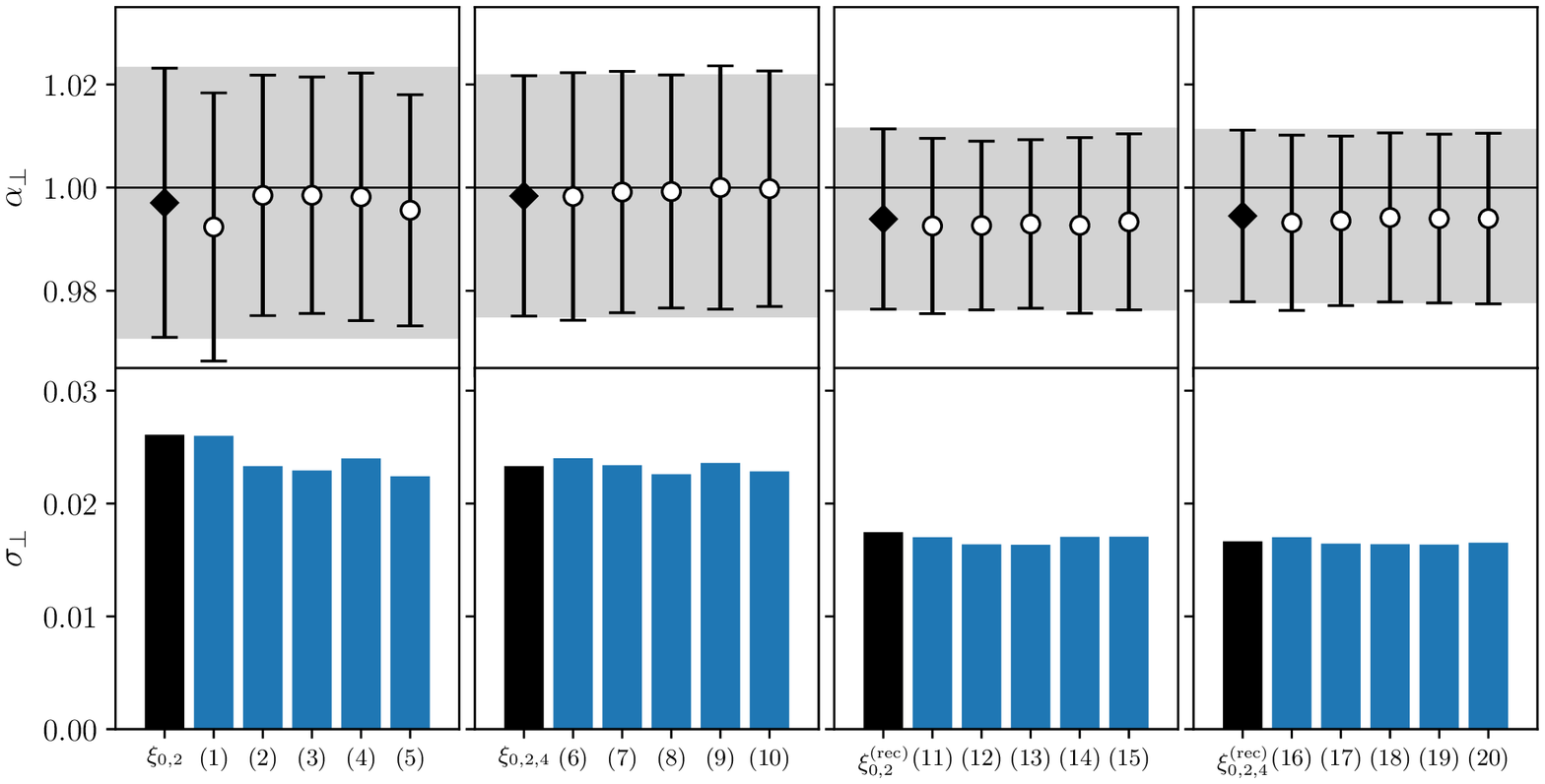}
	\includegraphics[width=\textwidth]{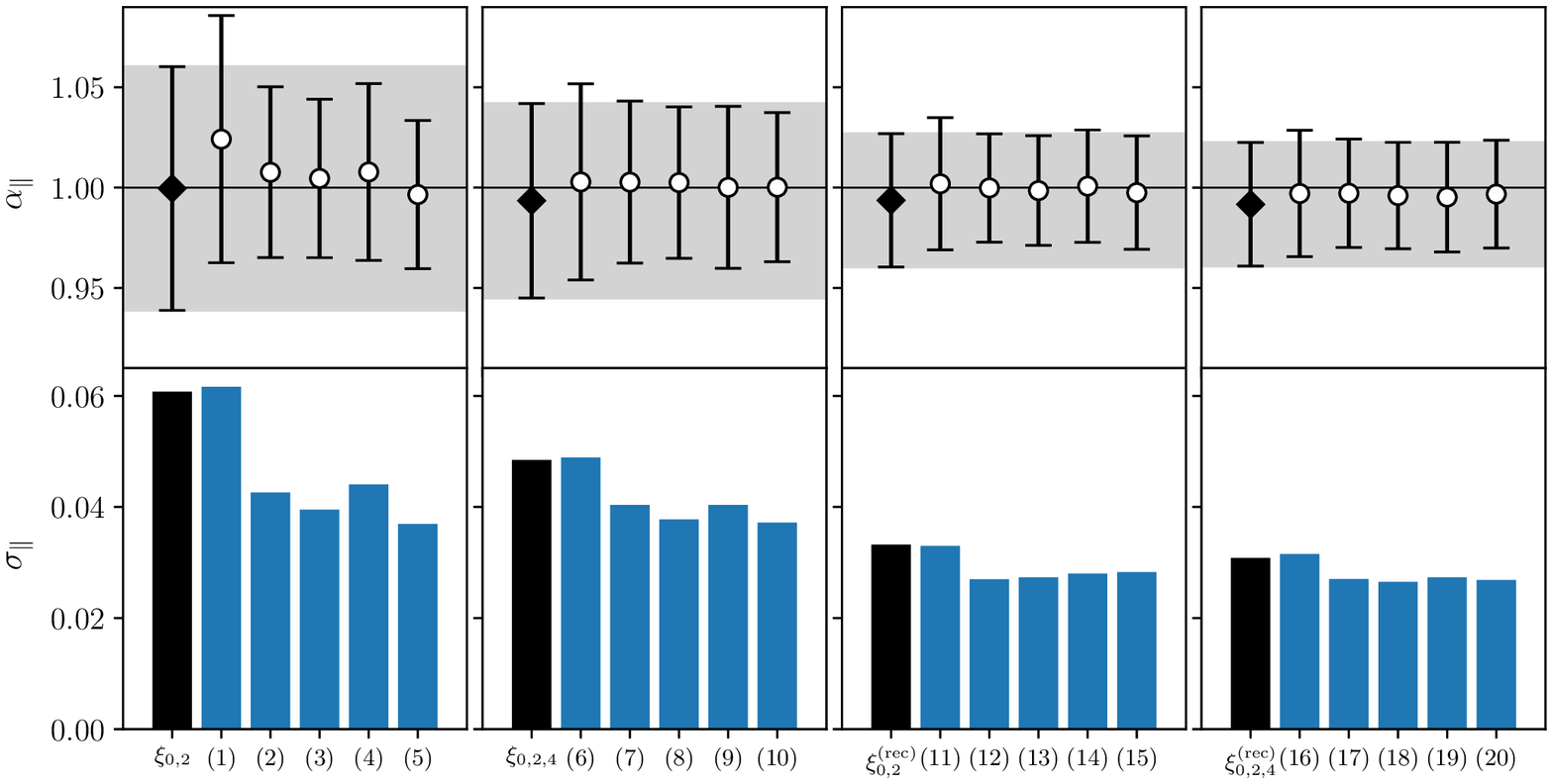}
	\caption{
    Illustration in Table~\ref{Table:APlimits}.
	}
	\label{fig:joint_errors}
\end{figure*}

\begin{figure*}
	\includegraphics[width=\columnwidth]{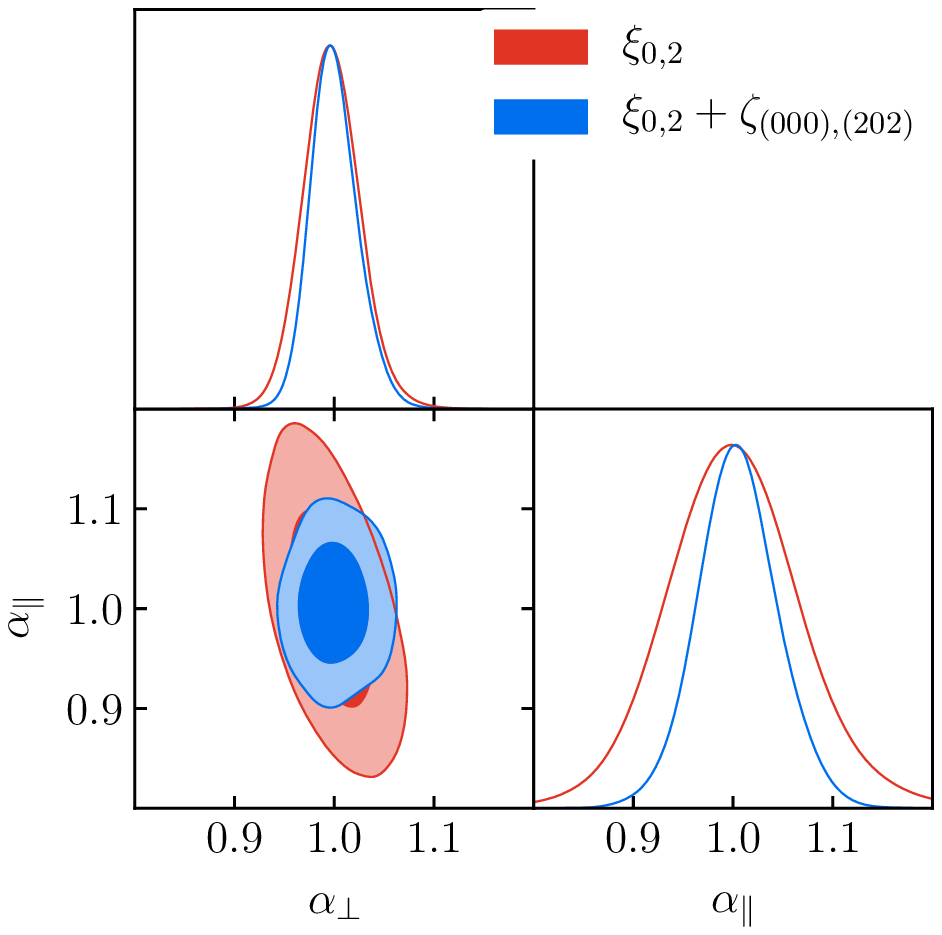}
	\includegraphics[width=\columnwidth]{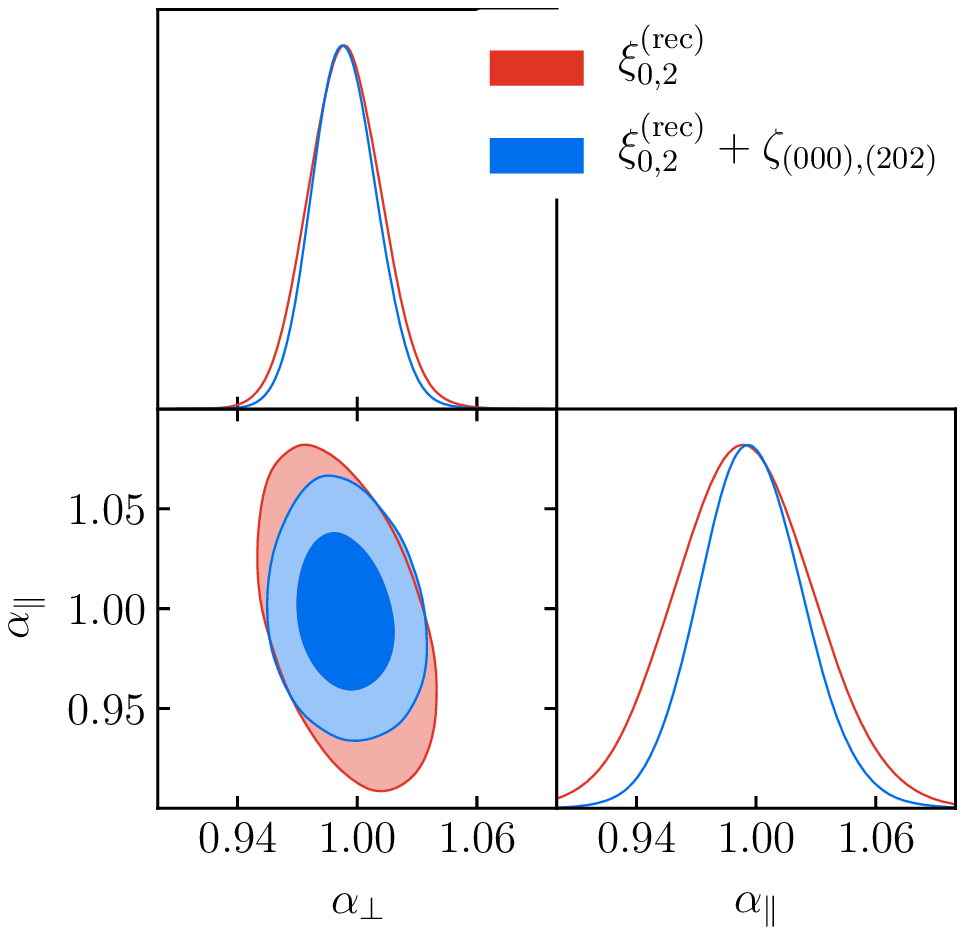}
	\caption{
    Likelihood distributions of $\alpha_{\perp}$ and $\alpha_{\parallel}$ for the MultiDark-Patchy mocks in the NGC and SGC for $0.4<z0.6$.
    The red contours only use the monopole and quadrupole 2PCFs before or after reconstruction,
    while the blue contours include the first terms of the monopole and quadrupole 3PCFs.
    We use the mean of 2048 mock measurements.
    The fitting range is $80\leq r \leq 150\hMpc$.
    The numerical values are summarized in Table~\ref{Table:APlimits}.
	}
	\label{fig:AP2d}
\end{figure*}
\begin{figure*}
	\includegraphics[width=\columnwidth]{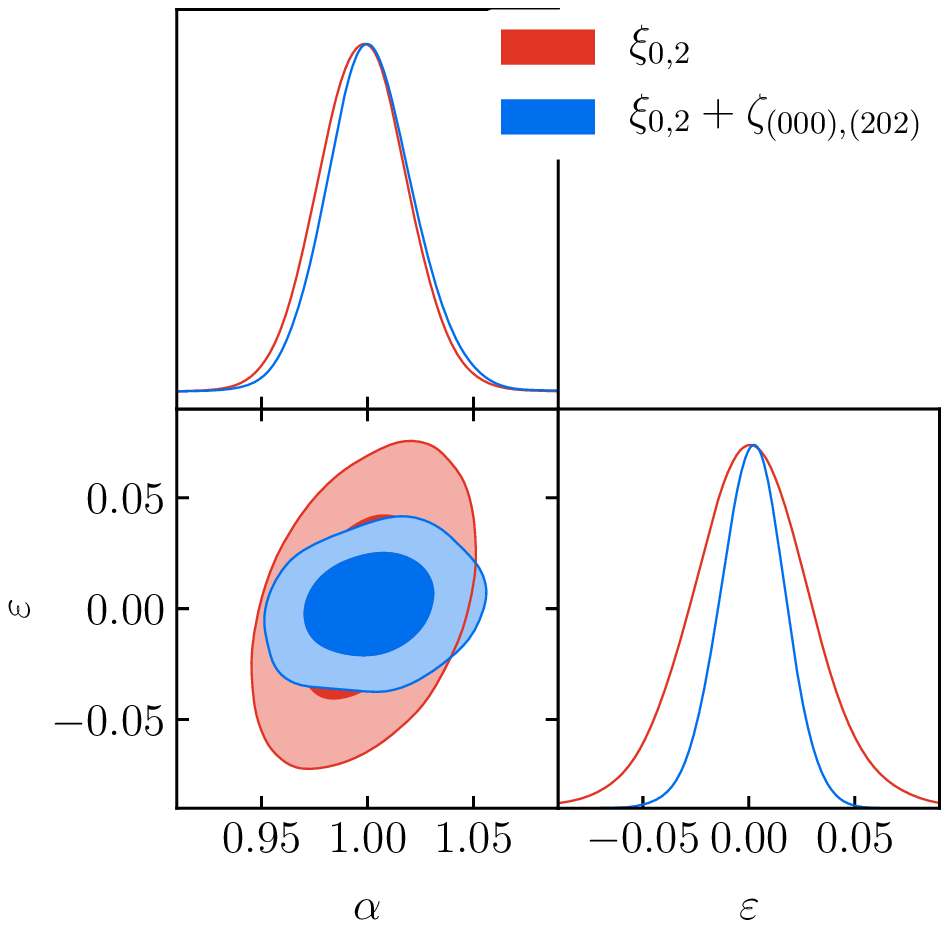}
	\includegraphics[width=\columnwidth]{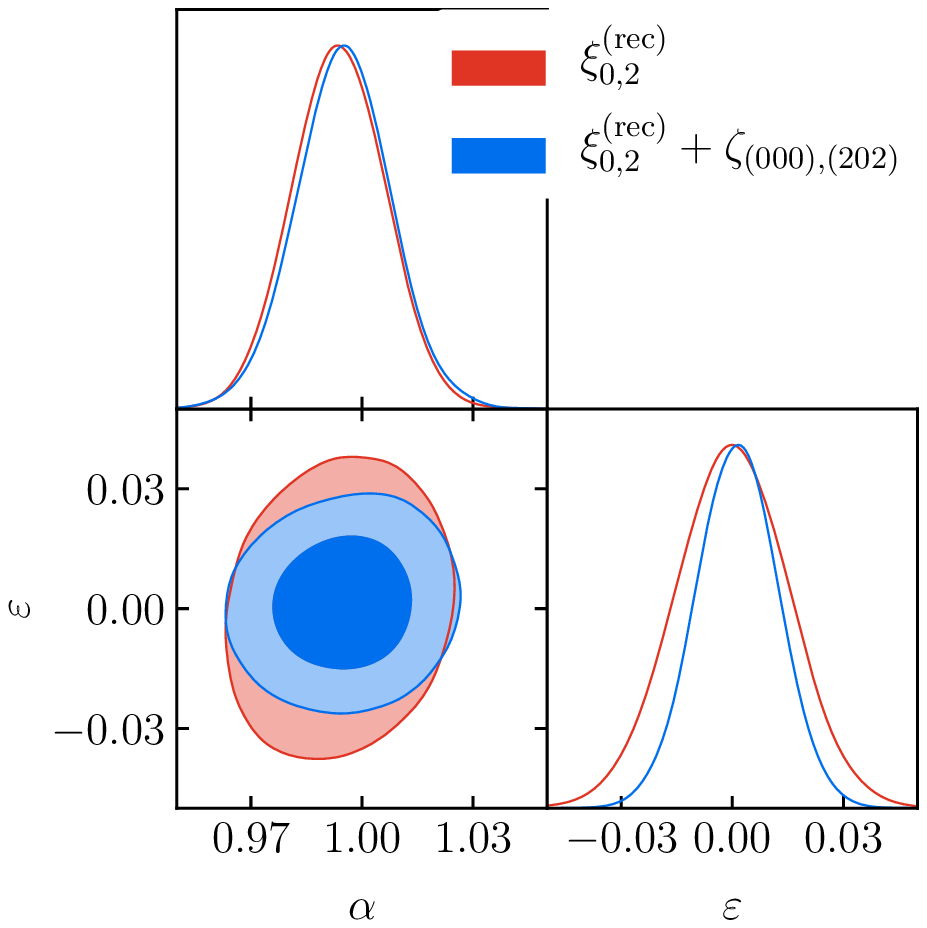}
	\caption{
    Similar to Figure~\ref{fig:AP2d}.
    The results for $\alpha$ and $\varepsilon$ are shown here, which are obtained as derived parameters.
	}
	\label{fig:AP2d_AE}
\end{figure*}

\begin{figure*}
	\includegraphics[width=\textwidth]{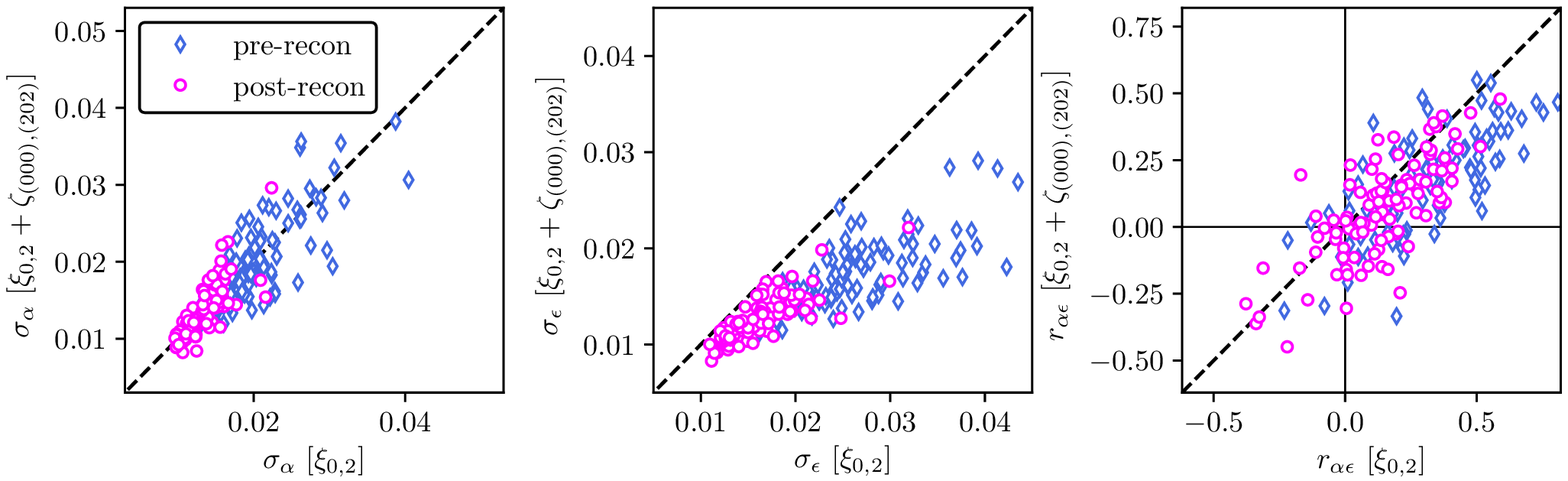}
	\caption{
    Similar to Figure~\ref{fig:sigma_mocks}.
    The results for $\alpha$ and $\varepsilon$ are shown here, which are obtained as derived parameters.
	}
	\label{fig:sigma_mocks_AE}
\end{figure*}

\begin{figure*}
	\includegraphics[width=\columnwidth]{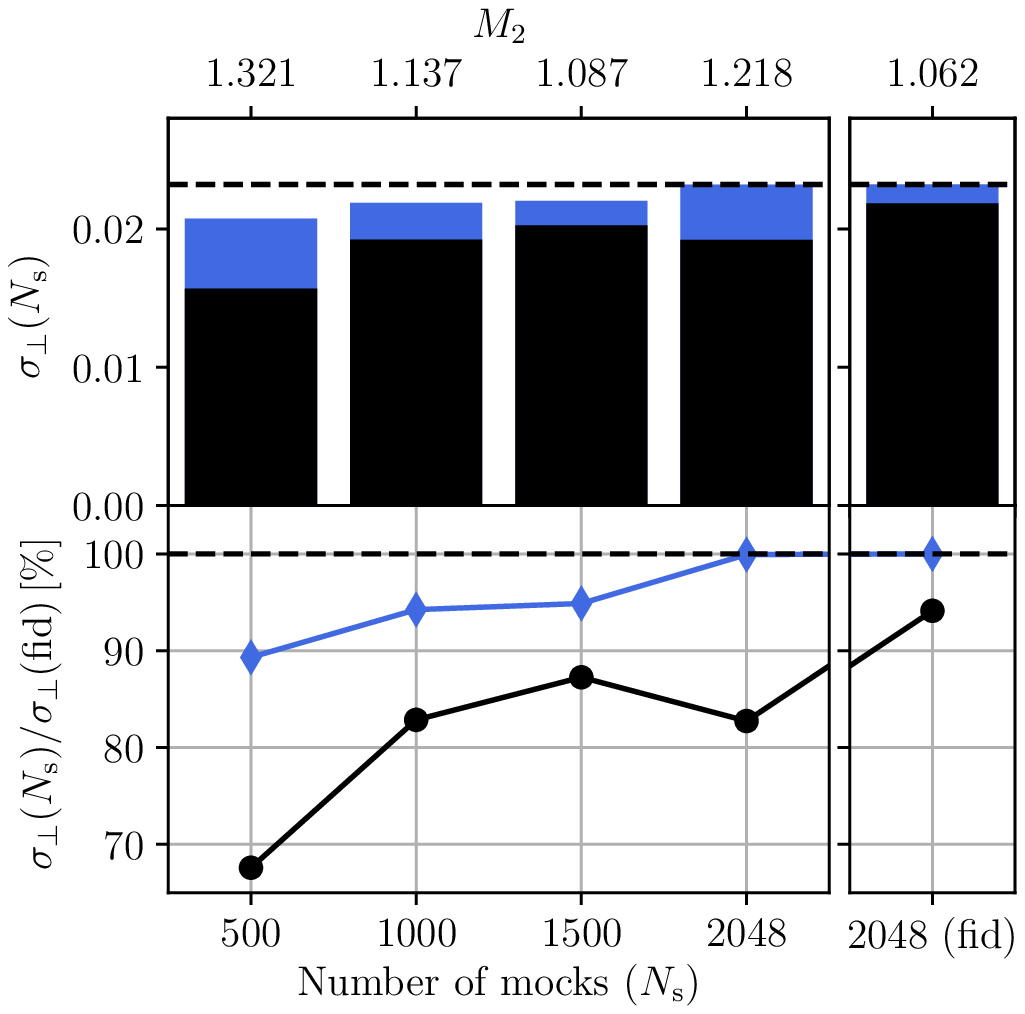}
	\includegraphics[width=\columnwidth]{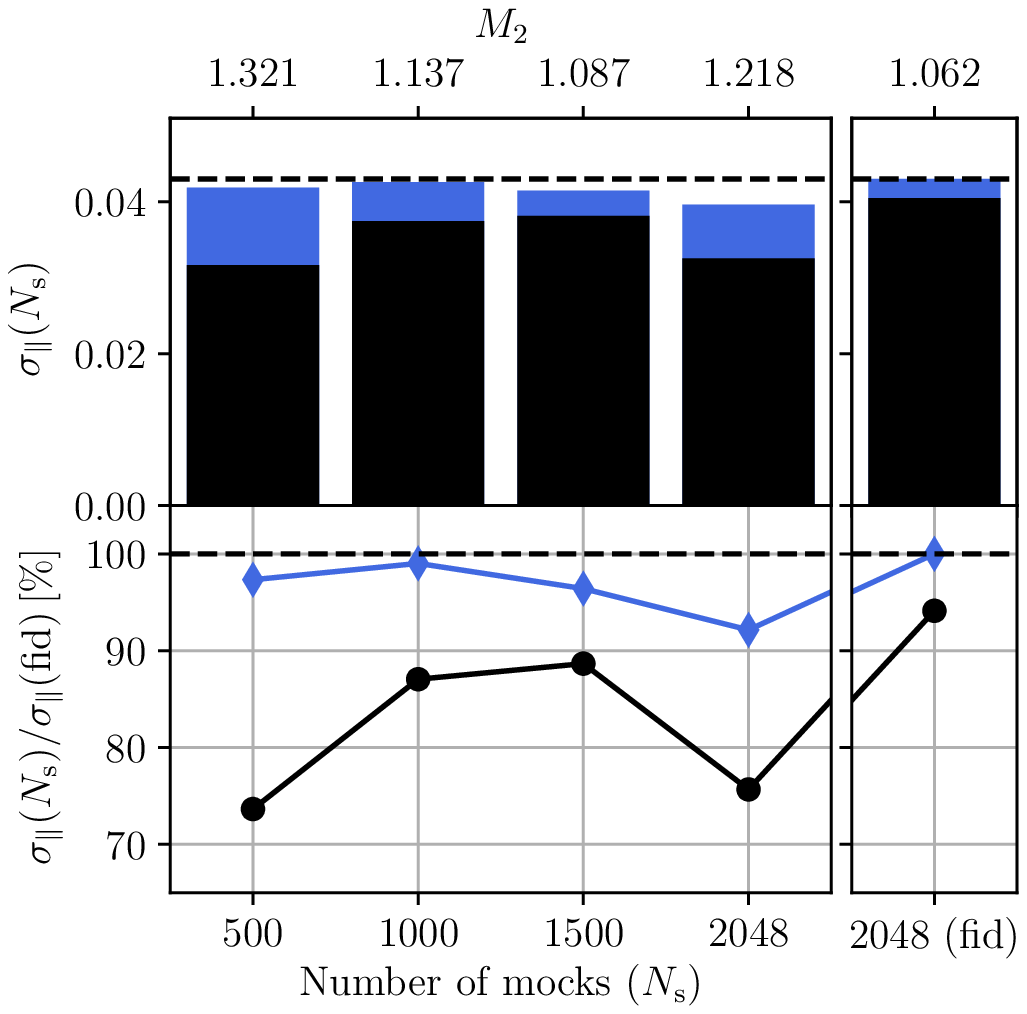}
	\caption{
    Illustration in Table~\ref{Table:M2}.
    The black and blue colors show the results before and after $M_2$ correction.
    The fiducial results is given in case (2) of Table~\ref{Table:APlimits}.
	}
	\label{fig:realizations}
\end{figure*}

\begin{table*}
\begin{tabular}{lccccc} 
 \hline 
\multicolumn{6}{c}{ Patchy mock ($0.4<z<0.6$) } \\
 \hline 
 & $1-\alpha$  & $\sigma_{\alpha}$ & $\varepsilon$ &  $\sigma_{\varepsilon}$  & $r_{\alpha\varepsilon}$\\ 
 \hline 
 \hline 
$\xi_{0,2}$                                                      &  0.0021 & 0.022 & 0.0008 & 0.029 & 0.38 \\
 \hline 
(1)  $\xi_{0,2}+\zeta_{(000),(110)}$                             & -0.0028 & 0.023 & 0.0113 & 0.029 & 0.21 \\
(2)  $\xi_{0,2}+\zeta_{(000),(202)}$                             & -0.0026 & 0.022 & 0.0033 & 0.018 & 0.18 \\
(3)  $\xi_{0,2}+\zeta_{(000),(202),(112)}$                       & -0.0016 & 0.022 & 0.0024 & 0.016 & 0.18 \\
(4)  $\xi_{0,2}+\zeta_{(000),(202),(404)}$                       & -0.0025 & 0.023 & 0.0034 & 0.019 & 0.19 \\
(5)  $\xi_{0,2}+\zeta_{(000),(202),(112),(404)}$                 &  0.0033 & 0.021 & 0.0006 & 0.016 & 0.12 \\
\hline
 \hline 
     $\xi_{0,2,4}$                                               &  0.0031 & 0.023 &-0.0019 & 0.021 & 0.27 \\
 \hline 
(6)  $\xi_{0,2,4}+\zeta_{(000),(110)}$                           &  0.0001 & 0.022 & 0.0014 & 0.022 & 0.15 \\
(7)  $\xi_{0,2,4}+\zeta_{(000),(202)}$                           & -0.0015 & 0.023 & 0.0014 & 0.017 & 0.17 \\
(8)  $\xi_{0,2,4}+\zeta_{(000),(202),(112)}$                     & -0.0012 & 0.022 & 0.0010 & 0.015 & 0.17 \\
(9)  $\xi_{0,2,4}+\zeta_{(000),(202),(404)}$                     & -0.0011 & 0.023 & 0.0009 & 0.017 & 0.17 \\
(10) $\xi_{0,2,4}+\zeta_{(000),(202),(112),(404)}$               & -0.0012 & 0.023 & 0.0005 & 0.015 & 0.13 \\
\hline
\hline 
     $\xi_{0,2}^{(\rm rec)}$                                   &  0.0062 & 0.014 &-0.0001 & 0.017 & 0.15 \\
 \hline                                                             
(11) $\xi_{0,2}^{(\rm rec)}+\zeta_{(000),(110)}$               &  0.0044 & 0.015 & 0.0034 & 0.016 & 0.09 \\
(12) $\xi_{0,2}^{(\rm rec)}+\zeta_{(000),(202)}$               &  0.0047 & 0.014 & 0.0026 & 0.013 & 0.11 \\
(13) $\xi_{0,2}^{(\rm rec)}+\zeta_{(000),(202),(112)}$         &  0.0049 & 0.014 & 0.0020 & 0.012 & 0.08 \\
(14) $\xi_{0,2}^{(\rm rec)}+\zeta_{(000),(202),(404)}$         &  0.0044 & 0.014 & 0.0029 & 0.014 & 0.10 \\
(15) $\xi_{0,2}^{(\rm rec)}+\zeta_{(000),(202),(112),(404)}$   &  0.0052 & 0.015 & 0.0017 & 0.013 & 0.10 \\
\hline
 \hline 
     $\xi_{0,2,4}^{(\rm rec)}$                                 &  0.0064 & 0.013 &-0.0011 & 0.014 & 0.14 \\
 \hline                                                             
(16) $\xi_{0,2,4}^{(\rm rec)}+\zeta_{(000),(110)}$             &  0.0055 & 0.014 & 0.0013 & 0.014 & 0.09 \\
(17) $\xi_{0,2,4}^{(\rm rec)}+\zeta_{(000),(202)}$             &  0.0049 & 0.014 & 0.0013 & 0.012 & 0.09 \\
(18) $\xi_{0,2,4}^{(\rm rec)}+\zeta_{(000),(202),(112)}$       &  0.0050 & 0.014 & 0.0007 & 0.012 & 0.08 \\
(19) $\xi_{0,2,4}^{(\rm rec)}+\zeta_{(000),(202),(404)}$       &  0.0053 & 0.014 & 0.0006 & 0.013 & 0.10 \\
(20) $\xi_{0,2,4}^{(\rm rec)}+\zeta_{(000),(202),(112),(404)}$ &  0.0048 & 0.015 & 0.0001 & 0.012 & 0.06 \\
\hline 
 \end{tabular} \\ 
\caption{
Similar to Table~\ref{Table:APlimits}. 
Here, the results for $\alpha$ and $\varepsilon$ are shown as derived parameters.
}
\label{Table:APlimits_derived}
\end{table*}

\begin{table*}
\begin{tabular}{lccccc} 
 \hline 
\multicolumn{6}{c}{ Patchy mock ($0.4<z<0.6$) } \\
\hline
& $1-\langle \alpha\rangle$ & $\langle \sigma_{\alpha} \rangle$ & $\langle \epsilon \rangle$ & $\langle\sigma_{\varepsilon}\rangle$ & $\langle r_{\alpha\varepsilon}\rangle$ \\ 
 \hline 
$\xi_{0,2}$                                   & -0.0008 & 0.021$\pm$0.0051 & 0.0042 & 0.028$\pm$0.0075 & 0.32$\pm$0.22 \\                
$\xi_{0,2}+\zeta_{(000),(202)}$               & -0.0060 & 0.021$\pm$0.0060 & 0.0051 & 0.018$\pm$0.0036 & 0.17$\pm$0.19 \\                
 \hline 
$\xi_{0,2,4}$                                 &  0.0008 & 0.021$\pm$0.0054 & 0.0002 & 0.021$\pm$0.0052 & 0.26$\pm$0.20 \\
$\xi_{0,2,4}+\zeta_{(000),(202)}$             & -0.0047 & 0.021$\pm$0.0058 & 0.0023 & 0.016$\pm$0.0031 & 0.14$\pm$0.18 \\                
 \hline 
$\xi^{\rm (rec)}_{0,2}$                       &  0.0033 & 0.013$\pm$0.0025 & 0.0011 & 0.016$\pm$0.0037 & 0.14$\pm$0.19 \\
$\xi_{0,2}^{\rm (rec)}+\zeta_{(000),(202)}$   &  0.0023 & 0.014$\pm$0.0035 & 0.0023 & 0.013$\pm$0.0024 & 0.07$\pm$0.19 \\                
\hline 
$\xi^{\rm (rec)}_{0,2,4}$                     &  0.0038 & 0.013$\pm$0.0023 &-0.0010 & 0.014$\pm$0.0024 & 0.12$\pm$0.16 \\ 
$\xi_{0,2,4}^{\rm (rec)}+\zeta_{(000),(202)}$ &  0.0028 & 0.013$\pm$0.0034 & 0.0004 & 0.012$\pm$0.0020 & 0.04$\pm$0.17 \\                
\hline 
 \end{tabular} \\ 
\caption{
Similar to Table~\ref{Table:APlimits_mean}. 
Here, the results for $\alpha$ and $\varepsilon$ are shown as derived parameters.
}
\label{Table:APlimits_mean_derived}
\end{table*}

\begin{table*}
\begin{tabular}{lcccccccc} 
 \hline 
\multicolumn{9}{c}{ Patchy mock ($0.4<z<0.6$) } \\
 \hline 
  & $f\sigma_8$ & $\sigma_8$ & $b_{1, \rm NGC}$ & $b_{1, \rm SGC}$ & $b_{2, \rm NGC}$ & $b_{2, \rm SGC}$ & $b_{\rm K^2, NGC}$ & $b_{\rm K^2, SGC}$ \\ 
 \hline 
 \hline 
$\xi_{0,2}$                                                      & 0.47$\pm$0.085 & -- & 1.21$\pm$0.15 & 1.19$\pm$0.24 & -- & -- & -- & --	\\
 \hline 
(1)  $\xi_{0,2}+\zeta_{(000),(110)}$                             & 0.52$\pm$0.100 & 0.98$\pm$0.57 & 1.03$\pm$0.19 & 0.91$\pm$0.24 &  0.24$\pm$2.06 & 0.07$\pm$4.10 & -0.17$\pm$0.64 & 0.53$\pm$1.10 \\
(2)  $\xi_{0,2}+\zeta_{(000),(202)}$                             & 0.47$\pm$0.089 & 0.88$\pm$0.54 & 1.13$\pm$0.16 & 1.02$\pm$0.25 &	 0.39$\pm$1.73 & 0.64$\pm$2.85 & -0.24$\pm$0.67 & 0.38$\pm$1.05 \\		
(3)  $\xi_{0,2}+\zeta_{(000),(202),(112)}$                       & 0.47$\pm$0.087 & 0.85$\pm$0.52 & 1.13$\pm$0.16 & 1.03$\pm$0.25 &	 0.48$\pm$1.62 & 0.55$\pm$2.55 & -0.21$\pm$0.65 & 0.37$\pm$1.01 \\
(4)  $\xi_{0,2}+\zeta_{(000),(202),(404)}$                       & 0.47$\pm$0.088 & 0.89$\pm$0.54 & 1.13$\pm$0.16 & 1.03$\pm$0.25 &  0.35$\pm$1.72 & 0.63$\pm$2.89 & -0.25$\pm$0.67 & 0.34$\pm$1.02 \\
(5)  $\xi_{0,2}+\zeta_{(000),(202),(112),(404)}$                 & 0.47$\pm$0.089 & 1.00$\pm$0.61 & 1.06$\pm$0.15 & 0.96$\pm$0.23 &  0.34$\pm$1.72 & 0.48$\pm$2.70 & -0.22$\pm$0.71 & 0.40$\pm$1.10 \\
\hline
\hline 
$\xi_{0,2,4}$                                                    & 0.46$\pm$0.082 & -- & 1.22$\pm$0.15 & 1.20$\pm$0.23 & -- & -- & -- & -- \\	
\hline 
(6)  $\xi_{0,2,4}+\zeta_{(000),(110)}$                           & 0.50$\pm$0.095 & 0.80$\pm$0.50 & 1.06$\pm$0.17 & 0.96$\pm$0.24 &  0.29$\pm$2.02 & 0.47$\pm$2.65 & -0.08$\pm$0.60 & 0.37$\pm$0.88 \\
(7)  $\xi_{0,2,4}+\zeta_{(000),(202)}$                           & 0.47$\pm$0.086 & 0.88$\pm$0.53 & 1.13$\pm$0.16 & 1.02$\pm$0.25 &  0.33$\pm$1.73 & 0.44$\pm$2.77 & -0.25$\pm$0.67 & 0.36$\pm$1.04 \\
(8)  $\xi_{0,2,4}+\zeta_{(000),(202),(112)}$                     & 0.46$\pm$0.082 & 0.89$\pm$0.55 & 1.14$\pm$0.16 & 1.03$\pm$0.25 &  0.30$\pm$1.72 & 0.57$\pm$2.80 & -0.26$\pm$0.67 & 0.38$\pm$1.04 \\
(9)  $\xi_{0,2,4}+\zeta_{(000),(202),(404)}$                     & 0.46$\pm$0.084 & 0.88$\pm$0.54 & 1.13$\pm$0.16 & 1.02$\pm$0.25 &  0.30$\pm$1.72 & 0.57$\pm$2.80 & -0.26$\pm$0.67 & 0.38$\pm$1.04 \\
(10) $\xi_{0,2,4}+\zeta_{(000),(202),(112),(404)}$               & 0.46$\pm$0.082 & 0.85$\pm$0.53 & 1.14$\pm$0.16 & 1.04$\pm$0.25 &  0.37$\pm$1.60 & 0.56$\pm$2.60 & -0.25$\pm$0.64 & 0.34$\pm$1.00 \\
\hline
\hline 
$\xi_{0,2}^{(\rm rec)}$                                          & 0.46$\pm$0.075 & -- & 1.26$\pm$0.13 & 1.24$\pm$0.20 & -- & -- & -- & -- \\	
\hline                                                              
(11) $\xi_{0,2}^{(\rm rec)}+\zeta_{(000),(110)}$                 & 0.49$\pm$0.085 & 0.62$\pm$0.37 & 1.14$\pm$0.14 & 1.03$\pm$0.23 & 0.09$\pm$1.54 & 0.02$\pm$3.27 & -0.14$\pm$0.48 & 0.35$\pm$0.82 \\
(12) $\xi_{0,2}^{(\rm rec)}+\zeta_{(000),(202)}$                 & 0.45$\pm$0.078 & 0.68$\pm$0.43 & 1.21$\pm$0.13 & 1.12$\pm$0.22 & 0.39$\pm$1.57 & 0.71$\pm$2.50 & -0.23$\pm$0.58 & 0.25$\pm$0.89 \\
(13) $\xi_{0,2}^{(\rm rec)}+\zeta_{(000),(202),(112)}$           & 0.45$\pm$0.077 & 0.66$\pm$0.41 & 1.21$\pm$0.13 & 1.11$\pm$0.20 & 0.47$\pm$1.46 & 0.66$\pm$2.42 & -0.20$\pm$0.56 & 0.27$\pm$0.84 \\
(14) $\xi_{0,2}^{(\rm rec)}+\zeta_{(000),(202),(404)}$           & 0.45$\pm$0.079 & 0.74$\pm$0.47 & 1.21$\pm$0.13 & 1.07$\pm$0.24 & 0.20$\pm$1.64 & 0.21$\pm$2.50 & -0.32$\pm$0.68 & 0.45$\pm$1.12 \\
(15) $\xi_{0,2}^{(\rm rec)}+\zeta_{(000),(202),(112),(404)}$     & 0.45$\pm$0.076 & 0.77$\pm$0.48 & 1.20$\pm$0.13 & 1.15$\pm$0.21 & 0.16$\pm$1.59 &-0.10$\pm$2.90 & -0.33$\pm$0.63 &-0.07$\pm$0.74  \\
\hline
\hline 
$\xi_{0,2,4}^{(\rm rec)}$                                        & 0.46$\pm$0.075 & -- & 1.26$\pm$0.13 & 1.25$\pm$0.20 & -- & -- & -- & --  \\
\hline                                                              
(16) $\xi_{0,2,4}^{(\rm rec)}+\zeta_{(000),(110)}$               & 0.48$\pm$0.079 & 0.51$\pm$0.31 & 1.17$\pm$0.14 & 1.09$\pm$0.21 & 0.85$\pm$1.40 & 0.81$\pm$2.12 &  0.01$\pm$0.48 & 0.35$\pm$0.63 \\
(17) $\xi_{0,2,4}^{(\rm rec)}+\zeta_{(000),(202)}$               & 0.46$\pm$0.077 & 0.66$\pm$0.41 & 1.21$\pm$0.13 & 1.11$\pm$0.22 & 0.37$\pm$1.58 & 0.73$\pm$2.40 & -0.21$\pm$0.57 & 0.31$\pm$0.87 \\
(18) $\xi_{0,2,4}^{(\rm rec)}+\zeta_{(000),(202),(112)}$         & 0.45$\pm$0.077 & 0.66$\pm$0.41 & 1.20$\pm$0.13 & 1.12$\pm$0.22 & 0.43$\pm$1.42 & 0.54$\pm$2.30 & -0.21$\pm$0.56 & 0.24$\pm$0.83 \\
(19) $\xi_{0,2,4}^{(\rm rec)}+\zeta_{(000),(202),(404)}$         & 0.45$\pm$0.078 & 0.79$\pm$0.50 & 1.21$\pm$0.13 & 1.11$\pm$0.22 &-0.05$\pm$1.80 & 0.30$\pm$2.43 & -0.38$\pm$0.67 & 0.08$\pm$0.92 \\
(20) $\xi_{0,2,4}^{(\rm rec)}+\zeta_{(000),(202),(112),(404)}$   & 0.45$\pm$0.076 & 0.68$\pm$0.42 & 1.21$\pm$0.13 & 1.12$\pm$0.23 & 0.37$\pm$1.52 & 0.82$\pm$2.30 & -0.23$\pm$0.58 & 0.26$\pm$0.84 \\
\hline 
 \end{tabular} \\ 
\caption{
Similar to Table~\ref{Table:APlimits}. 
The results for $f\sigma_8$, $\sigma_8$ and the bias parameters are shown.
}
\label{Table:RSD}
\end{table*}

\begin{table*}
\begin{tabular}{lcccccccc} 
 \hline 
\multicolumn{9}{c}{ Patchy mock ($0.4<z<0.6$) } \\
\hline
& $\langle f\sigma_8\rangle$ & $\langle \sigma_8\rangle$ & $\langle b_{1, \rm NGC}\rangle $ & $\langle b_{1, \rm SGC}\rangle $ & $\langle b_{2, \rm NGC}\rangle $ & $\langle b_{2, \rm SGC}\rangle$ & $\langle b_{\rm K^2, NGC}\rangle$ & $\langle b_{\rm K^2, SGC}\rangle$ \\ 
\hline 
$\xi_{0,2}$                                   & 0.46$\pm$0.085 & --            & 1.25$\pm$0.15 & 1.22$\pm$0.23 & -- & -- & -- & -- \\
$\xi_{0,2}+\zeta_{(000),(202)}$               & 0.46$\pm$0.089 & 0.96$\pm$0.55 & 1.15$\pm$0.16 & 1.04$\pm$0.23 & 0.29$\pm$1.77 & 0.29$\pm$3.01 & -0.26$\pm$0.69 & 0.36$\pm$1.09 \\                
\hline 
$\xi_{0,2,4}$                                 & 0.45$\pm$0.082 & --            & 1.25$\pm$0.15 & 1.22$\pm$0.23 & -- & -- & -- & -- \\
$\xi_{0,2,4}+\zeta_{(000),(202)}$             & 0.46$\pm$0.086 & 0.94$\pm$0.54 & 1.16$\pm$0.15 & 1.04$\pm$0.22 & 0.29$\pm$1.76 & 0.24$\pm$2.97 & -0.25$\pm$0.68 & 0.34$\pm$1.08 \\
\hline 
$\xi^{\rm (rec)}_{0,2}$                       & 0.45$\pm$0.077 & --            & 1.28$\pm$0.13 & 1.25$\pm$0.20 & -- & -- & -- & -- \\
$\xi_{0,2}^{\rm (rec)}+\zeta_{(000),(202)}$   & 0.45$\pm$0.079 & 0.74$\pm$0.43 & 1.22$\pm$0.13 & 1.12$\pm$0.21 & 0.32$\pm$1.58 & 0.38$\pm$2.70 & -0.24$\pm$0.59 & 0.25$\pm$0.92 \\
\hline 
$\xi^{\rm (rec)}_{0,2,4}$                     & 0.45$\pm$0.075 & --            & 1.28$\pm$0.13 & 1.26$\pm$0.20 & -- & -- & -- & -- \\ 
$\xi_{0,2,4}^{\rm (rec)}+\zeta_{(000),(202)}$ & 0.45$\pm$0.077 & 0.72$\pm$0.42 & 1.22$\pm$0.13 & 1.12$\pm$0.21 & 0.32$\pm$1.58 & 0.39$\pm$2.67 & -0.25$\pm$0.58 & 0.25$\pm$0.90 \\
\hline 
\end{tabular} \\ 
\caption{
Similar to Table~\ref{Table:APlimits_mean}. 
The results for $f\sigma_8$, $\sigma_8$ and the bias parameters are shown.
}
\label{Table:RSD_mean}
\end{table*}

\bsp	
\label{lastpage}

\end{document}